\documentclass[12pt,letterpaper,usenames,dvipsnames]{article}
\usepackage{jheppub}\usepackage{amsmath,amssymb,amsfonts,amsthm,amscd}
\usepackage{bm} %boldmath = \bm
\usepackage{bbm} %\mathbbm = doublestruck math for upper, lower, numbers, \mathbbmss=sans serif, \mathbbmtt=typewriter
\usepackage{eufrak} %gothic
\usepackage{mathrsfs} %caligraphic
\usepackage{lscape} 
\usepackage{subcaption} 
\usepackage{cancel}
\usepackage{enumitem}
\usepackage{tcolorbox}
\usepackage{multirow} % multiple row elements in a table
\usepackage{colortbl} % colored rows and columns in tables
\usepackage{array,hhline,arydshln,multirow}
\usepackage{floatrow}
\usepackage[export]{adjustbox}
\usepackage{rotating} % rotate text

\allowdisplaybreaks[2]  %allow page breaks in multi-line eqns, [1]=as little as poss; ... [4]=most permissive
\usepackage{pifont,dsfont}% http://ctan.org/pkg/pifont
\usepackage{bm} %boldmath

\usepackage{graphicx,tikz}
\usepackage[framemethod=TikZ]{mdframed} % fancy boxes
\usepackage{tikz-cd} %for complicated commutative diagrams

\usepackage{xcolor}
\def\red#1{{\color{red}{#1}}}
\def\blue#1{{\color{blue}{#1}}}
\def\green#1{{\color{black!30!green}{#1}}}

\def\yellow#1{{\color{black!25!yellow!70!red}{#1}}}

\definecolor{amaranth}{rgb}{0.9, 0.17, 0.31}
\definecolor{coolblack}{rgb}{0.0, 0.18, 0.39}
\definecolor{gold(web)(golden)}{rgb}{1.0, 0.84, 0.0}
\definecolor{deepcarmine}{rgb}{0.66, 0.13, 0.24}
\def\red#1{{\color{deepcarmine}{#1}}}
\def\green#1{{\color{green!70!black}{#1}}}
\def\Fgreen#1{{\color{ForestGreen}{#1}}}
\def\blue#1{{\color{blue}{#1}}}

\def\bred#1{{\color{BrickRed}{#1}}}

\def\H{\mathbb{H}}

\def\bM{\begin{matrix}}
\def\eM{\end{matrix}}
\newcommand{\bpm}{\begin{pmatrix}}
\newcommand{\epm}{\end{pmatrix}}
\newcommand{\bsm}{\begin{smallmatrix}}
\newcommand{\esm}{\end{smallmatrix}}
\newcommand{\bspm}{\left(\begin{smallmatrix}}
\newcommand{\espm}{\end{smallmatrix}\right)}
\newcommand{\beq}{\begin{equation}}
\newcommand{\eeq}{\end{equation}}
\def\bar{\overline}
\def\til{\widetilde}
\def\hat{\widehat}

\def\^{\wedge}

\def\so{\mathfrak{so}}

\def\C{\mathbbm{C}}

\def\H{\mathbbm{H}}

\def\cN{{\mathcal N}}
\def\cO{{\mathcal O}}

\def\cS{{\mathcal S}}

\def\cT{{\mathcal T}}

\def\V{\mathbbm{V}} 
\def\cV{{\mathcal V}}

\def\Z{\mathbbm{Z}}
\def\d{{\delta}}
\def\D{{\Delta}}

\def\L{{\Lambda}}

\def\S{{\Sigma}}

\usepackage{tikz}
\usetikzlibrary{arrows,snakes,shapes.arrows,decorations.markings}
     \tikzset{>=triangle 90}
     \tikzstyle{bbc}=[draw,circle,fill=black,scale=.75]
     \tikzstyle{rc}=[circle,fill=red,scale=.6]
     \tikzstyle{wc}=[draw,circle,scale=.75]

\def\blue#1{{\color{blue}{#1}}}

\def\green#1{{\color{black!25!green}{#1}}}

\def\yellow#1{{\color{black!25!yellow}{#1}}}
\def\red#1{{\color{red}{#1}}}

\def\yellow#1{{\color{yellow}{#1}}}

\def\bar{\overline}
\def\til{\widetilde}
\def\hat{\widehat}

\def\^{\wedge}

\def\d{{\delta}}

\def\D{{\Delta}}

\def\L{{\Lambda}}

\def\S{{\Sigma}}

\def\tQ{{\til Q}}

\def\af{\mathfrak{a}}
\def\bff{\mathfrak{b}}
\def\cf{\mathfrak{c}}
\def\df{\mathfrak{d}}
\def\ef{\mathfrak{e}}
\def\ff{\mathfrak{f}}
\def\gf{\mathfrak{g}}

\def\Rf{\mathfrak{R}}
\def\bRf{\bf \Rf}

\def\Sf{\mathfrak{S}}

\def\bSf{\blue{\mathfrak{S}}}

\def\sof{\mathfrak{so}}
\def\spf{\mathfrak{sp}}
\def\suf{\mathfrak{su}}

\def\bTf{\blue{\mathfrak{T}}}
\def\Tf{\mathfrak{T}}
\def\uf{\mathfrak{u}}

\def\cB{{\mathcal B}}

\def\cE{{\mathcal E}}

\def\cN{{\mathcal N}}
\def\cO{{\mathcal O}}

\def\cS{{\mathcal S}}

\def\cT{{\mathcal T}}

\def\cV{{\mathcal V}}

\def\CC{\mathbb{C}} 
 
\def\H{\mathbb{H}}

\def\V{\mathbb{V}} 
\def\Z{\mathbb{Z}} 

\def\beq{\begin{equation}}
\def\eeq{\end{equation}}

\newcommand{\bpmat}{\begin{pmatrix}}
\newcommand{\epmat}{\end{pmatrix}}
\newcommand{\bsmat}{\begin{smallmatrix}}
\newcommand{\esmat}{\end{smallmatrix}}
\newfloatcommand{capbtabbox}{table}[][\FBwidth]
\newtheorem{fact}{Criterion}

\def\rcb{\rowcolor{blue!07}}
\def\rcg{\rowcolor{green!07}}
\def\rcy{\rowcolor{black!25!yellow!10}}

\def\F{{\rm F}}
\def\S{{\rm S}}
\def\C{{\rm C}}
\def\V{{\rm V}}
\def\AS{{\rm AS}}

\title{On the compactification of 5d theories to 4d}
\author[1,2]{Mario Martone}
\author[3]{Gabi Zafrir}
\affiliation[1]{C.~N.~Yang Institute for Theoretical Physics,  Stony Brook University,Stony Brook, NY 11794-3840, USA}
\affiliation[2]{Simons Center for Geometry and Physics, Stony Brook University, Stony Brook, NY 11794-3840, USA}
\affiliation[3]{Dipartimento di Fisica, Università di Milano-Bicocca \& INFN, Sezione di Milano-Bicocca, I-20126 Milano, Italy}
\emailAdd{mmartone@scgp.stonybrook.edu,gabi.zafrir@unimib.it}

\abstract{We study general properties of the mapping between 5\emph{d} and 4\emph{d} superconformal field theories (SCFTs) under both twisted circle compactification and tuning of local relevant deformation and CB moduli. After elucidating in generality when a 5\emph{d} SCFT  reduces to a 4\emph{d} one, we identify nearly all $\cN=1$ 5\emph{d} SCFT parents of rank-2 4\emph{d} $\cN=2$ SCFTs. We then use this result to map out the mass deformation trajectories among the rank-2 theories in 4\emph{d}. This can be done by first understanding the mass deformations of the 5\emph{d} $\cN=1$ SCFTs and then map them to 4\emph{d}. The former task can be easily achieved by exploiting the fact that the 5\emph{d} parent theories can be obtained as the strong coupling limit of Lagrangian theories, and the latter by understanding the behavior under compactification. Finally we identify a set of general criteria that 4\emph{d} moduli spaces of vacua have to satisfy when the corresponding SCFTs are related by mass deformations and check that all our RG-flows satisfy them. Many of the mass deformations we find are not visible from the corresponding complex integrable systems.}

\begin{document}
\maketitle 

\section{Introduction and summary of the results}

In recent years there has been an incredible progress in understanding superconformal field theories (SCFTs) in various dimensions, particularly those with eight or more supercharges. These correspond to $\cN\geq4$, $\cN\geq2$, $\cN= 1$ and $\cN\geq (1,0)$ in, respectively, $d=3$, 4, 5 and 6. Here we focus on understanding the relations among $\cN=1$ SCFTs in $5d$ and $\cN=2$ SCFTs in 4d. In particular we will be concerned with two major questions: (1) which $5d$ SCFTs reduce, upon circle compactification and perhaps turning on some relevant operators and/or non-trivial holonomies, to 4$d$  SCFTs and which to IR-free theories? (2) is there anything we can learn about 4$d$  physics by exploring the physics in one dimension higher?

\begin{figure}[h!]
\centering
\begin{adjustbox}{center,max width=.9\textwidth}
\begin{tikzpicture}
[
auto,
good/.style={rectangle,rounded corners,fill=green!50,inner sep=2pt},
bad/.style={rectangle,rounded corners,fill=red!15,inner sep=2pt},
ugly/.style={rectangle,rounded corners,fill=blue!10,inner sep=2pt},
Marrow/.style={->,>=stealth[round],shorten >=1pt,line width=.4mm,coolblack},
IRarrow/.style={->,>=stealth[round],shorten >=1pt,line width=.4mm,deepcarmine,dashed}
]
\begin{scope}[yshift=.5cm]
\node at (4.5,0) {{\large {\fontfamily{qcs}\selectfont\textsc{RG-flows for 4$d$  $\cN=2$ rank-2 SCFTs}}}};
\end{scope}
%II^* line:
\begin{scope}[yshift=-.5cm,xshift=-2cm] %E8 and SO20 series
%\fill[color=gray!10, rounded corners] (0,1) rectangle (15.9,-5.4);
%\fill[color=amaranth!30, rounded corners] (9.4,.6) rectangle (14.3,-5);
\node (E824) at (1,0) [,align=center] {\scriptsize$[\ef_8]_{24}{\times}\suf(2)_{13}$};
\node (E716) at (1,-1) [,align=center] {\scriptsize$[\ef_7]_{16}{\times}\suf(2)_9$};
\node (E612) at (1,-2) [,align=center] {\scriptsize$[\ef_6]_{12}{\times}\suf(2)_7$};
\fill[color=yellow!40, rounded corners] (0,-2.7) rectangle (2,-3.3);
\node (SO88) at (1,-3) [,align=center] {\scriptsize$\sof(8)_8{\times}\suf(2)_5$};
\node (H2r2) at (1,-4) [,align=center] {\scriptsize$H_2$ rank-2};
\node (H1r2) at (1,-5) [,align=center] {\scriptsize$H_1$ rank-2};
\node (H0r2) at (1,-6) [,align=center] {\scriptsize$H_0$ rank-2};
\node (E820) at (-1.5,-1) [,align=center] {\scriptsize$[\ef_8]_{20}$};
\node (SO1410) at (3.5,-2) [,align=center] {\scriptsize$\sof(14)_{10}{\times}\uf(1)$};
\fill[color=yellow!40, rounded corners] (3,-2.7) rectangle (4,-3.3);
\node (U66) at (3.5,-3) [,align=center] {\scriptsize$\uf(6)_6$};
\node (D2A4) at (3.5,-4) [,align=center] {\scriptsize$D_2(\suf(5))$};
\node (A1D6) at (3.5,-5) [,align=center] {\scriptsize$(A_1,D_6)$};
\node (A1A5) at (4.25,-6) [,align=center] {\scriptsize$(A_1,A_5)$};
\node (A1A4) at (4.25,-7) [,align=center] {\scriptsize$(A_1,A_4)$};
\node (A1D5) at (2.75,-6) [,align=center] {\scriptsize$(A_1,D_5)$};
\fill[color=yellow!40, rounded corners] (-2.1,-2.7) rectangle (-.9,-3.3);
\node (SO128) at (-1.5,-3) [,align=center] {\scriptsize$\sof(12)_8$};
\node (SO1612) at (4,-1) [,align=center] {\scriptsize$\sof(16)_{12}{\times}\suf(2)_8$};
\node (SO2016) at (5,0) [,align=center] {\scriptsize$\sof(20)_{16}$};
\node (SU1010) at (6,-1) [,align=center] {\scriptsize$\suf(10)_{10}$};
\node (SU88) at (6,-2) [,align=center] {\scriptsize$\suf(2)_6{\times}\suf(8)_8$};
\fill[color=yellow!40, rounded corners] (5.4,-2.7) rectangle (6.6,-3.3);
\node (SU245) at (6,-3) [,align=center] {\scriptsize$\suf(2)_4^5$};
\end{scope}

\begin{scope}[yshift=-1cm,xshift=.5cm] %SU6 series%
\node (SU616) at (8,0) [,align=center] {\scriptsize$\suf(6)_{16}{\times}\suf(2)_9$};
\node (SU412) at (6.5,-1) [,align=center] {\scriptsize$\suf(4)_{12}{\times}\suf(2)_7{\times}\uf(1)$};
\node (SU3102) at (9.5,-1) [,align=center] {\scriptsize$\suf(3)_{10}{\times}\suf(3)_{10}{\times}\uf(1)$};
\node (SU310) at (6.5,-2) [,align=center] {\scriptsize$\suf(3)_{10}{\times}\suf(2)_6{\times}\uf(1)$};
\node (SU282) at (9.5,-2) [,align=center] {\scriptsize$\suf(2)_8{\times}\suf(2)_8{\times}\uf(1)^2$};
\fill[color=yellow!40, rounded corners] (7.2,-2.7) rectangle (8.8,-3.3);
\node (U12) at (8,-3) [,align=center] {\scriptsize$\uf(1){\times}\uf(1)$};
\end{scope}

\begin{scope}[yshift=-6cm,xshift=-1.5cm] %G2 series
\node (G28) at (8,0) [,align=center] {\scriptsize$[\gf_2]_8{\times}\suf(2)_{14}$};
\node (G2203) at (9.25,-1) [,align=center] {\scriptsize$[\gf_2]_{\frac{20}3}$};
\node (SU2163) at (6.75,-1) [,align=center] {\scriptsize$\suf(2)_{\frac{16}3}{\times}\suf(2)_{10}$};
\fill[color=blue!10, rounded corners] (6.15,-1.7) rectangle (7.35,-2.3);
\node (SU28) at (6.75,-2) [,align=center] {\scriptsize$\suf(2)_8$};
\end{scope}

\begin{scope}[yshift=-6cm,xshift=2.5cm] %SU326 series
\node (SU326) at (8,0) [,align=center] {\scriptsize$\suf(3)_{26}{\times}\uf(1)$};
\node (U12p) at (8,-1) [,align=center] {\scriptsize$\uf(1)^2$};
\fill[color=green!10, rounded corners] (6.9,-1.7) rectangle (9.1,-2.3);
\node (G312) at (8,-2) [,align=center] {\scriptsize$\cN=3\,G(3,1,2)$};
\end{scope}

\begin{scope}[yshift=-9.5cm,xshift=-11cm] %SU5 series
\node (SU510) at (8,0) [,align=center] {\scriptsize$\suf(5)_{10}$};
\node (SU312) at (8,-1) [,align=center] {\scriptsize$\suf(3)_{12}{\times}\uf(1)$};
\node (SU210U1) at (8,-2) [,align=center] {\scriptsize$\suf(2)_{10}{\times} \uf(1)$};
\end{scope}

\begin{scope}[yshift=-9.5cm,xshift=-8cm] %Sp128 series
\node (SP128) at (8,0) [,align=center] {\scriptsize$\spf(12)_8$};
\node (SP86p) at (8,-1) [,align=center] {\scriptsize$\spf(8)_6{\times}\suf(2)_8$};
\node (SP65p) at (8,-2) [,align=center] {\scriptsize$\spf(6)_5{\times}\uf(1)$};
\fill[color=yellow!40, rounded corners] (7.5,-2.7) rectangle (8.5,-3.3);
\node (SP44) at (8,-3) [,align=center] {\scriptsize$\spf(4)_4$};
\node (SP88) at (10.5,0) [,align=center] {\scriptsize$\spf(4)_7{\times}\spf(8)_8$};
\node (SP66p) at (10.5,-1) [,align=center] {\scriptsize$\suf(2)_5{\times}\spf(6)_6{\times}\uf(1)$};
\node (F412) at (13,0) [,align=center] {\scriptsize$[\ff_4]_{12}{\times}\suf(2)_7^2$};
\node (SO78) at (13,-1) [,align=center] {\scriptsize$\sof(7)_8{\times}\suf(2)_5^2$};
\node (F410) at (15.5,-1) [,align=center] {\scriptsize$[\ff_4]_{10}{\times}\uf(1)$};
\node (SU36) at (13,-2) [,align=center] {\scriptsize$\suf(3)_6{\times}\suf(2)_4^2$};
\fill[color=blue!10, rounded corners] (12.4,-2.7) rectangle (13.6,-3.3);
\node (SU26p) at (13,-3) [,align=center] {\scriptsize$\suf(2)_6$};
\end{scope}

\begin{scope}[yshift=-9.5cm,xshift=3cm] %Sp14 series
\node (SP149) at (8,0) [,align=center] {\scriptsize$\spf(14)_9$};
\node (SP107) at (6.75,-1) [,align=center] {\scriptsize$\spf(10)_7{\times}\suf(2)_8$};
\node (SP66) at (9.25,-1) [,align=center] {\scriptsize$\spf(8)_7{\times}\suf(2)_5$};
\node (SP86) at (8,-2) [,align=center] {\scriptsize$\spf(8)_6{\times} \uf(1)$};
\fill[color=yellow!40, rounded corners] (7.5,-2.7) rectangle (8.5,-3.3);
\node (SP65) at (8,-3) [,align=center] {\scriptsize$\spf(6)_5$};
\end{scope}

\begin{scope}[yshift=-14cm,xshift=-10cm] %SU216 series
\node (SU216) at (8,0) [,align=center] {\scriptsize$\suf(2)_{16}{\times}\uf(1)$};
\fill[color=green!10, rounded corners] (6.9,-.7) rectangle (9.1,-1.3);
\node (G412) at (8,-1) [,align=center] {\scriptsize$\cN=3\,G(4,1,2)$};
\end{scope}

\begin{scope}[yshift=-13.5cm,xshift=-4.5cm] %Sp8SU22 series
\node (SP813) at (8,0) [,align=center] {\scriptsize$\spf(8)_{13}{\times}\suf(2)_{26}$};
\node (SP49) at (8,-1) [,align=center] {\scriptsize$\spf(4)_9{\times}\suf(2)_{16}{\times}\suf(2)_{18}$};
\node (SU214) at (8,-2) [,align=center] {\scriptsize$\suf(2)_7{\times}\suf(2)_{14}{\times} \uf(1)\ \ \ \ $};
\fill[color=blue!10, rounded corners] (8.4,-2.7) rectangle (9.6,-3.3);
\node (SU210) at (9,-3) [,align=center] {\scriptsize$\suf(2)_{10}$};
\node (SU26) at (10,-2) [,align=center] {\ \ \ \ \scriptsize$\suf(2)_6{\times}\suf(2)_8$};
\node (SU25) at (11,-3) [,align=center] {\scriptsize$\suf(2)_5$};
\end{scope}

\begin{scope}[yshift=-14cm,xshift=2cm] %SP1211 series
\node (SP1211) at (8,0) [,align=center] {\scriptsize$\spf(12)_{11}$};
\fill[color=yellow!40, rounded corners] (8.4,-.7) rectangle (9.6,-1.3);
\node (SP87) at (9,-1) [,align=center] {\scriptsize$\spf(8)_7$};
\fill[color=yellow!40, rounded corners] (6,-.7) rectangle (8,-1.3);
\node (SP45) at (7,-1) [,align=center] {\scriptsize$\spf(4)_5{\times}\sof(4)_8$};
\end{scope}

\draw[Marrow] (E824) to (E716);
\draw[Marrow] (E824) to (E820);
\draw[IRarrow] (E820) to (SO128);
\draw[Marrow] (E824) to (SO1410);
\draw[IRarrow] (E716) to (SO128);
\draw[Marrow] (E716) to (E612);
\draw[IRarrow] (E612) to (SO88);
\draw[Marrow] (E612) to (U66);
\draw[Marrow] (SO88) to (H2r2);
\draw[Marrow] (H2r2) to (H1r2);
\draw[Marrow] (H1r2) to (H0r2);
\draw[Marrow] (SO2016) to (SO1612);
\draw[Marrow] (SO2016) to (SU1010);
\draw[Marrow] (SO1612) to (SO1410);
\draw[Marrow] (SU1010) to (SU88);
\draw[Marrow] (SO1612) to (SU88);
\draw[IRarrow] (SO1410) to (U66);
\draw[IRarrow] (SO1410) to (SO128);
\draw[IRarrow] (SU88) to (SU245);
\draw[IRarrow] (SU88) to (U66);
\draw[Marrow] (U66) to (D2A4);
\draw[Marrow] (D2A4) to (A1D6);
\draw[Marrow] (A1D6) to (A1D5);
\draw[Marrow] (A1D6) to (A1A5);
\draw[Marrow] (A1A5) to (A1A4);
\draw[Marrow] (SP813) to (SP49);
\draw[Marrow] (SP49) to (SU214);
\draw[IRarrow] (SU214) to (SU210);
\draw[Marrow] (SU616) to (SU3102);
\draw[Marrow] (SU616) to (SU412);
\draw[Marrow] (SU412) to (SU310);
\draw[Marrow] (SU412) to (SU282);
\draw[Marrow] (SU3102) to (SU282);
\draw[IRarrow] (SU282) to (U12);
\draw[IRarrow] (SU310) to (U12);
\draw[Marrow] (SP149) to (SP107);
\draw[Marrow] (SP107) to (SP86);
\draw[Marrow] (SP66) to (SP86);
\draw[IRarrow] (SP86) to (SP65);
\draw[Marrow] (SU326) to (U12p);
\draw[Marrow] (U12p) to (G312);
\draw[Marrow] (SU26) to (SU25);
\draw[Marrow] (SU26) to (SU210);
\draw[Marrow] (SU510) to (SU312);
\draw[Marrow] (SU312) to (SU210U1);
\draw[Marrow] (G28) to (G2203);
\draw[Marrow] (G28) to (SU2163);
\draw[Marrow] (SU2163) to (SU28);
\draw[Marrow] (SU216) to (G412);
\draw[IRarrow] (SP1211) to (SP45);
\draw[IRarrow] (SP1211) to (SP87);
\draw[Marrow] (SP128) to (SP86p);
\draw[Marrow] (SP86p) to (SP65p);
\draw[IRarrow] (SP65p) to (SP44);
\draw[Marrow] (SP88) to (SP86p);
\draw[Marrow] (SP88) to (SP66p);
\draw[Marrow] (SP66p) to (SP65p);
\draw[Marrow] (F412) to (SP66p);
\draw[Marrow] (F412) to (SO78);
\draw[Marrow] (F412) to (F410);
\draw[Marrow] (SO78) to (SU36);
\draw[IRarrow] (SU36) to (SU26p);
\draw[IRarrow] (SO78) to (SP44);
\draw[IRarrow] (F410) to (SP44);

%\draw[Marrow] (H1r2) to (H0r2);

%Legend
\begin{scope}[yshift=-17cm,xshift=-6.5cm]
\node (L1) at (5,-.75){};
\node (L2) at (6.1,-.75){};
\node (L3) at (5,-1.25){};
\node (L4) at (6.1,-1.25){};
\node[anchor=west] (Text2) at (6,-.75)  {\small{\textsc{: mass deformations involving two or more relevant parameters,}}};
\node[anchor=west] (Text3) at (6,-1.25) {\small{\textsc{: mass deformation among SCFTs.}}};
\draw[IRarrow] (L1) to (L2);
\draw[Marrow] (L3) to (L4);
\end{scope}
\end{tikzpicture}
\caption{Graphical depiction of the RG-relations among four dimensional $\cN=2$ SCFTs. We shade in \yellow{yellow} $\cN=2$ Lagrangian theories, in \green{green} $\cN=3$ theories and in \blue{blue} $\cN=4$ theories. All theories but two $\cN=3$ theories are labeled by their flavor symmetries.}
\label{MapR2}
\end{adjustbox}
\end{figure}
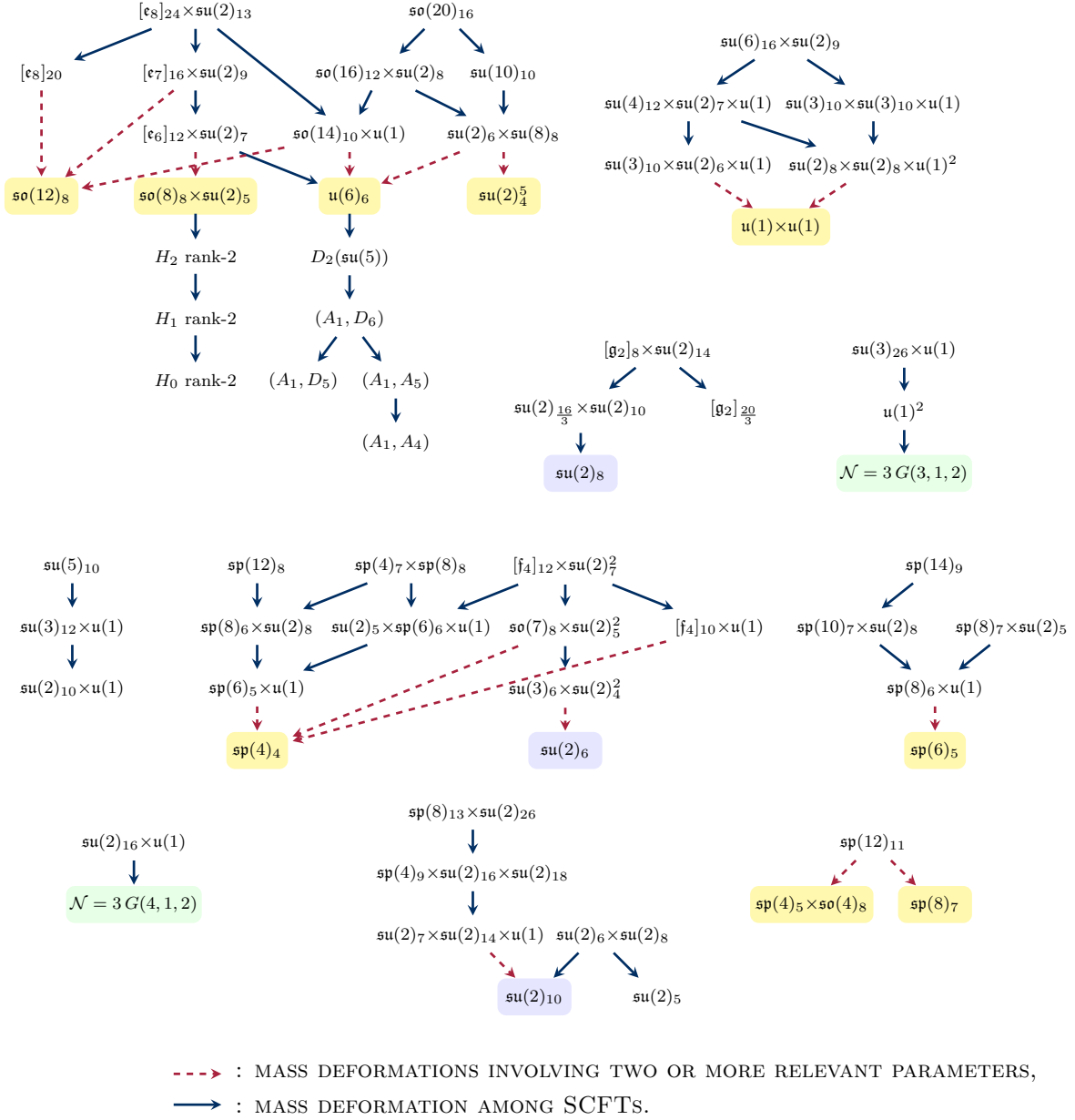

%Four dimensions is singled out for various reasons. Primarily, of course, because it is the dimensionality of our spacetime. Secondly, in four dimension the gauge coupling is marginal and, with eight or more supercharges, exactly marginal. In dimensions lower than four, the coupling is instead relevant and, perhaps not surprisingly, in dimension larger than four it is instead irrelevant. In the latter case, weakly coupled Lagrangian theories can be obtained from strongly coupled theories. 

To understand the relevance of these questions, it is useful to remind the reader of the major differences between 4 and 5 dimensions. Four dimensional SCFTs are for the most part inherintly strongly coupled, and no general strategy to understand their behaviour under relevant deformations is currently known. Many $\cN=1$ 5\emph{d} SCFTs arise instead as the infinite coupling limit of Lagrangian theories. In this limit the flavor symmetry generally enhances, including to exceptional algebras as it was shown famously in the rank-1 case \cite{Seiberg:1996bd,Morrison:1996xf,Intriligator:1997pq}. An advantage of the way SCFTs arise in five dimensions, is that it is possible to use the weakly coupled gauge theory description to extract insightful information about the behavior of the theory at the conformal point. Here we will exploit this fact to extract information about the relation of these theories under mass deformations.

With clarity on question (1), this observation readily implies a positive answer to question (2).  In fact if we understood which $\cN=1$ $5d$ SCFTs upon compactifying on a circle (with or without extra \emph{twisting}) reduce to a 4$d$  SCFT, we could map the mass deformations of four dimensional SCFTs gaining tremendous insights with relatively little effort by connecting an easily extractable quantity in five dimensions with what is instead very hard to analyze in four dimensions. 

A general answer to (1) is a challenging task, though. The compactification of one of the spatial dimensions introduces a scale breaking conformal invariance, thus in general the origin of the moduli space is no longer where interesting dynamics takes place. To obtain an interacting fixed point in 4d, we then usually have to tune other dimensionful quantities, \emph{e.g.} some moduli or relevant deformations, with result that the four dimensional theory obtained heavily depends on the details of how these limits are taken. Despite these challenges, in the absence of twist, we will be able to depict a somewhat coherent picture where specific types of $5d$ SCFT always reduce to $4d$ SCFTs. Unfortunately for twisted compactifications the analysis is far less straightforward and we will fall short in providing general conjectures.

A much more achievable goal is to try to provide an answer to (1) within a restricted set of theories, specifically restricted to $5d$ SCFTs which reduce to rank-2 $4d$ ones. In carrying out this analysis we will be far more successful. Using a variety of techniques, \emph{e.g.} matching moduli space of vacua and global symmetries or studying the behavior under gauging of various flavor symmetry factors, we identify the $\cN=1$ $5d$ parent of nearly all known rank-2 $4d$ $\cN=2$ SCFTs, the results are reported in table \ref{tab:r21}, \ref{tab:r22} and \ref{tab:r23}. The understanding of the $5d$ parents of Argyres-Douglas theories which arise on the Coulomb branch (CB) of Lagrangian theories is somewhat more challenging. Here, we specifically refer to $4d$ SCFTs containing Coulomb branch operators with dimensions $1< \Delta < 2$ \footnote{This class of theories has some interesting unique features, like the existence of relevant deformations unrelated to conserved currents, the so-called chiral deformations (see the discussion in section \ref{sec:4dsetup}). It is not clear whether these are related or not to the observations regarding their $5d$ origin.}. These are not observed to be the result of a direct reduction of $5d$ SCFTs, but rather arising by the tuning of various parameters involved in the compactification, similarly to how these arise on the Coulomb branch of $4d$ Lagrangian theories. It is not clear to us, at this point in time, whether this is a general feature or merely an artifact of the yet incomplete set of studied cases. We leave this study for future work, and as such, these theories are simply not included in table \ref{tab:r21}-\ref{tab:r23}.

A somewhat complete answer to (1) restricted to the set of $5d$ theories reducing to rank-2 $4d$ theories will allow us to map out the set of mass deformations among rank-2 $4d$ theories. We spend a significant part of our paper in carrying out this study and our results are delightfully summarized in figure \ref{MapR2}. The insights gained will allow us to also begin identifying criteria for how to read off the SCFTs behavior under mass deformations directly from 4$d$  in terms of their moduli spaces of vacua; we observe that we can track mass deformations via both the CB and the Higgs branch (HB) which together provide a somewhat constrained picture. We explicitly work out how this works in the case of 4$d$  rank-2 theories.

The paper is organized as follows. We begin in section \ref{sec:GL} with a general discussion on the compactification of $5d$ SCFTs. In section \ref{sec:4dsetup} we discuss mass deformations of $4d$ $\mathcal{N}=2$ SCFTs from the $4d$ perspective. We then move on to section \ref{sec:sum}, where we summarize the observational status of the study of the $4d$ compactifications of $5d$ SCFTs. Sections \ref{sec:def} and \ref{checks} are then devoted to the study of the rank-2 cases, with section \ref{sec:def} concentrating on the $5d$ side while section \ref{checks} concentrates on the $4d$ side. Appendix \ref{app:Higgs} contains somewhat technical results about HBs of $4d$ theories.
\vspace{1em}

\noindent\emph{A note on notation}: throughout the paper, capital letter will denote the gauge group of the various theories (though we won't be extremely careful on the global structure) \emph{e.g.} $SU(3)+N_f F$ will indicate a theory, either in 4 or 5 $d$, with gauge group $SU(3)$ and matter in the fundamental representation. Conversely lower case bold letter, will instead indicate the global flavor symmetry group (again we will be somewhat sloppy on questions regarding the global structure), \emph{e.g.} $\suf(3)$ will denote a theory in 5 or 4 $d$ with global flavor symmetry $\suf(3)$.

\section{General lesson on the compactification from 5d  to 4d}
\label{sec:GL}

Here we consider various general aspects of compactifications of $5d$ SCFTs to $4d$. This is a very complex subject and we shall try to summarize various features of it that have been previously observed. Our starting point is a $5d$ SCFT, that is a $5d$ theory enjoying $5d$ superconformal symmetry. There is only one superconformal group in $5d$, called $F(4)$, and contains the bosonic symmetries $\so(5,2)\times \suf(2)$, with the former being the $5d$ conformal group and the latter the R-symmetry group. The supersymmetry subgroup of the superconformal group contains eight conserved supercharges. Additionally, the $5d$ SCFT may contain flavor currents that commute with the superconformal group, corresponding to continuous flavor symmetries. The $5d$ SCFT may also posses various discrete global symmetries, both commuting and not with the supersymmetry algebra.   

We wish to compactify $5d$ SCFTs on a circle of radius $R$, leading to $4d$ theories at low-energies. Here, there are various options involved in the compactification process. Notably, we can turn on holonomies along the circle in the flavor symmetries of the $5d$ SCFT, both discrete and continuous. Here we shall concentrate on supersymmetry preserving compactifications, so that the resulting $4d$ theory will have $\mathcal{N}=2$ supersymmetry. This in turn rules out holonomies in the R-symmetry and in discrete symmetries that do not commute with the supersymmetry algebra.

In theories with eight supercharges vevs to scalars in background vector multiplets are expected to give mass deformations. As such we expect the holonomies associated with continuous symmetries to lead to mass deformations of the associated flavor symmetries\footnote{To be precise, this gives one of the components of the mas deformation. The other comes from a vev to the single real scalar in the $5d$ background vector multiplet. Together, these form the single complex scalar in the $4d$ background vector multiplet.}. However, holonomies in discrete symmetries are expected to lead to different $4d$ theories. 

There are various different ways of performing the reduction. One common option is to take the limit $R\rightarrow 0$, while keeping the holonomies, or more correctly the associated mass deformations, finite. This is expected to lead to some theory in $4d$ deformed by the mass deformations associated with the holonomies. In this case, the theory of most interest is the one when the compactification is done without the holonomies. An interesting feature of this type of compactifications is that the $4d$ theory should inherit the full flavor symmetry of the $5d$ SCFT, although as we shall soon discuss, this expectation may fail in some cases.

Another option for compactification that is commonly employed is to take a scaling limit with some holonomy. Specifically, take a mass deformation $M$, associated with a specific holonomy, and consider the compactification in the limit $R\rightarrow 0$, $M\rightarrow \infty$, $R M$ finite. This type of compactifications break the symmetry to that which commutes with the global symmetry element associated with $M$, and as such the resulting $4d$ theory only manifests this subset of the symmetry of the $5d$ SCFT. In many cases the resulting $4d$ theories are equal to the low-energy limit of the $4d$ theories you get without taking it, deformed by the mass deformation $M$. However, that is not necessarily true, that is the limits may not commute. We shall see various examples of this later on. It is also possible to take such scaling limits for Coulomb branch vevs, which are vevs to scalars in dynamical rather than background gauge fields.

\begin{table}[ht]
\begin{adjustbox}{center,max width=.9\textwidth}
$\def\arraystretch{1.0}
\begin{array}{r|c|c:cc|c:cc|c:c:c:c}
\multicolumn{12}{c}{\Large\textsc{Table\ of\ Rank-2\ theories\ I}}\\
\hline
\hline
&&\multicolumn{3}{l|}{\quad \text{Moduli Space}} &
\multicolumn{3}{l|}{\quad \text{Flavor and central charges}} &
\multicolumn{4}{c}{\quad \text{5$d$ Gauge theory realization}}\\[1mm]
\multirow{-2}{4mm}{\#}&
& \D_{u,v}&\ \ d_{\text{HB}}\ \ &\ \  h\ \  
&\quad \ff\quad &\ \ 24a\ \ & 12c 
&[ SU(3) ]_k&USp(4)&[SU(2)^\theta,SU(2)]&G_2
\\[1.5mm]
%%%%%%%%%%%%%%%% [E8 SO(20) series]
\hline\hline
\multicolumn{12}{c}{\ef_8-\sof(20)\ \text{series}}\\
\hline
1.&
& \{6,12\}
& 59 & 1
& [\ef_8]_{24}\times \suf(2)_{13} & 263 & 161 
&\ [8\F]_2\ &\ 1\AS+7\F\ &\ [1\F,5\F]\ &\ -\ \\
2.&
& \{6,8\}
& 46 & 0
&\sof(20)_{16} & 202 & 124  
&\ [9\F]_{\frac12}\ &\ 9\F\ &\ [3\F,4\F]\ &\ - \ \\
3.&
& \{4,10\}
 & 46 & 0
&[\ef_8]_{20} & 202 & 124 
&\ - \ &\ -\ &[0^0,5\F]\ &\ - \ \\
4.&
& \{4,8\}
& 35 & 1
& [\ef_7]_{16}\times \suf(2)_{9} & 167 & 101
&\ [7\F]_{\frac52}\ &\ 1\AS+6\F\ &\ [0^{\pi},5\F]\ &\ -\ \\
5.&
& \{4,6\}
&30 & 0
&\suf(2)_8 \times \sof(16)_{12} & 138 & 84 
&\ [8\F]_1\ &\ 8\F\ &\ [2\F,4\F]\ &\ -\ \\
6.&
& \{4,5\}
&  26 & 0
&\suf(10)_{10} & 122  & 74 
&\ [8\F]_0\ &\ -\ &\ [3\F,3\F]\ &\ -\ \\
7.&
& \{3,6\}
&  23 & 1
& [\ef_6]_{12}\times \suf(2)_{7} & 119 & 71 
&\ [6\F]_{3}\ &\ 1\AS+5\F\ &\ -\ &\ -\ \\
8.&
& \{3,5\}
&  22 & 0
&\sof(14)_{10}\times \uf(1) & 106  & 64 
&\ [7\F]_{\frac32}\ &\ 7\F\ &\ [1\F,4\F]\ &\ -\ \\
9.&
& \{3,4\}
&18 & 0
&\suf(2)_6\times \suf(8)_8 & 90  & 54  
&\ [7\F]_{\frac12}\ &\ -\ &\ [2\F,3\F]\ &\ -\ \\
\rcy10.&
& \{2,4\}
& 14 & 0 
& \sof(12)_8 & 74 & 44 
&\multicolumn{4}{c}{USp(4)+6\F\ \text{GT}}\\
\rcy 11.&
& \{2,4\} 
& 11 & 1 
& \sof(8)_8\times\suf(2)_5 & 75 & 42  
&\multicolumn{4}{c}{USp(4)+1AS+4\F\ \text{GT}}\\
\rcy12.&
& \{2,3\} 
& 10 & 0
& \uf(6)_6 & 58 & 34
&\multicolumn{4}{c}{SU(3)_{k}+6\F\ \text{GT}}\\
\rcy13.&
& \{2,2\}
&  6 & 0
& \suf(2)_4^5 & 42 & 24
&\multicolumn{4}{c}{2\F+SU(2)\times SU(2)+2\F\ \text{GT}}\\

%%%%%%%%%%%%%%%% [Sp(12) Sp(8) F4 series]

\cdashline{1-12}
\multicolumn{12}{c}{\spf(12)-\spf(8)-\ff_4\ \text{series}}\\
\hline
22.&
& \{4,6\} 
&  22 & 0
& \spf(12)_8 & 130 & 76  
&\multicolumn{4}{c}{SU(4)_0+10\F\  \Z_2\ {\rm twisted}} \\
23.&
& \{4,6\}
& 20 & 2
&\spf(4)_7 \times \spf(8)_8 & 128 & 74 
&\multicolumn{4}{c}{SU(4)_0+1\AS+8\F\  \Z_2\ {\rm twisted}} \\
24.&
&\{6,6\} 
&  24 & 2
& \suf(2)_7^2\times [\ff_4]_{12} & 156 & 90     
&\multicolumn{4}{c}{SU(4)_0+2\AS+6\F\  \Z_2\ {\rm twisted}} \\
25.&
& \{3,4\}
&12 & 0
&\suf(2)_8\times \spf(8)_6 & 84  & 48 
&\multicolumn{4}{c}{SU(4)_0+8\F\  \Z_2\ {\rm twisted}} \\
26.&
& \{3,4\}
& 11 & 1
&\suf(2)_5\times \spf(6)_6\times \uf(1) & 83  & 47 
&\multicolumn{4}{c}{SU(4)_0+1\AS+6\F\  \Z_2\ {\rm twisted}} \\
27.&
&\{4,4\} 
& 12 & 2 
& \suf(2)_5^2\times \sof(7)_8 & 96 & 54    
&\multicolumn{4}{c}{SU(4)_0+2\AS+4\F\  \Z_2\ {\rm twisted}} \\
28.&
& \{4,5\}
& 16 & 0
&[\ff_4]_{10}\times \uf(1) & 112  & 64
&\multicolumn{4}{c}{SU(2)_0-[SU(2)+2\F]-SU(2)_0\  \Z_2\ {\rm twisted}} \\
29.&
& \{\frac52,3\}
&7 & 0
& \spf(6)_5\times \uf(1) & 61 & 34
&\multicolumn{4}{c}{SU(4)_0+6\F\  \Z_2\ {\rm twisted}} \\
30.&
& \{3,3\}
& 6 & 2
& \suf(3)_6\times\suf(2)^2_4 & 66 & 36  
&\multicolumn{4}{c}{SU(4)_0+2\AS+2\F\  \Z_2\ {\rm twisted}} \\
\rcy31.&
& \{2,2\}
& 3 & 0
& \spf(4)_4 & 38 & 20 
&\multicolumn{4}{c}{SU(4)_0+4\F\ \text{GT}\ \Z_2\ {\rm twisted}}\\
\rcb32.&
& \{2,2\}
&2 & 2
& \suf(2)_6 & 36 & 18    
&\multicolumn{4}{c}{SU(4)_0+2\AS\ \text{GT}\ \Z_2\ {\rm twisted}}\\
\hline\hline
\end{array}$
\caption{\footnotesize{\label{tab:r21}List of rank-2 4$d$ theories with their corresponding CFT data and 5\emph{d} origin. The last column in the table gives the 5$d$ gauge theory description of
the 5$d$ SCFT whose compactification gives the 4$d$ SCFT. Here we shall
usually give only one gauge theory description although others may
exist. For most Lagrangian 4$d$ SCFTs, the 5$d$ parent is given by the
compactification of a 5$d$ gauge theory, rather than a 5$d$ SCFT. This is
denoted in the table by the label `GT'. When the compactification
involves twisting, the discrete symmetry group being twisted is
specified. In these cases this group is $\mathbb{Z}_2$ and it is manifested in the gauge theory as the combination
of quiver reflection and charge conjugation. Notice that we use the convention $\spf(2n)$ where $n$ is rank. Also, compared to similar tables in \cite{Martone:2021ixp}, we won't report the SCFTs which can be obtained by mass-deforming the Lagrangian theories. Finally we shade in \yellow{yellow} $\cN=2$ Lagrangian theories and in \blue{blue} $\cN=4$ theories.}}
\end{adjustbox}
\end{table}

Following the reduction we end up with a $4d$ $\mathcal{N}=2$ theory. A question of particular interest to us is whether said theory is a $4d$ SCFT or not. This is a rather difficult question to answer. To understand why, it is enough to consider the symmetries expected from the $5d$ theory. Specifically, as we mentioned we expect the $4d$ theory to inherit $\mathcal{N}=2$ supersymmetry from the $5d$ theory. Additionally, it should also inherit the $\suf(2)$ R-symmetry. However, that is not sufficient for $4d$ $\mathcal{N}=2$ superconformal symmetry, which also requires a $\uf(1)$ R-symmetry. Such a symmetry may arise accidentally, leading to a $4d$ $\mathcal{N}=2$ SCFT, or it may not. In general, determining whether there is or isn't an accidental enhancement of symmetry is a complicated problem\footnote{Most of what is said in this paragraph is also true for the compactification of $6d$ $(1,0)$ SCFTs on a torus.}. We shall see examples of both behaviors later. 

In spite of what was said so far, there have been many studies on the compactifications of $5d$ SCFT. These uncovered various patterns suggesting that some specific classes of $5d$ SCFTs reduce to $4d$ SCFTs, while others don't. We shall summarize these mostly experimental results and list some of their implications in section \ref{sec:sum}. Before that though, we wish to have a somewhat less technical discussion illustrating some of the phenomena that can happen in the RG flow from $5d$ to $4d$. We shall see various examples of these phenomena in the more technical summary that will come in later sections.

As we previously mentioned we consider here the compactification of $5d$ SCFTs to $4d$. For simplicity, we shall assume the compactification is done without a discrete symmetry twist. The $5d$ SCFTs have several field theory data that would be useful in analyzing the reduction. First, they have a moduli space, that is split into the Coulomb and Higgs branch. They also have a space of mass deformations. These initiate RG flows leading to either new $5d$ SCFTs or to an IR free gauge theory.  

The Coulomb branch should reduce to the $4d$ Coulomb branch, although as it can receive quantum corrections, the only thing we can determine about the $4d$ Coulomb branch from the $5d$ one is its dimension\footnote{Note that the $5d$ Coulomb branch is a real space, while the $4d$ Coulomb branch is a complex space. As such, when referring to the dimension of the $5d$ ($4d$) Coulomb branch we mean the real (complex) dimension respectively. We further note that a $5d$ SCFT with a $d$ dimensional Coulomb branch reduces to a $4d$ $\mathcal{N}=2$ theory with a $d$ dimensional Coulomb branch.}. Similarly, the Higgs branch should reduce to the $4d$ Higgs branch. However, unlike the Coulomb branch, the Higgs branch is invariant under quantum corrections. As such knowledge of the $5d$ Higgs branch translates directly to knowledge of the $4d$ Higgs branch. This makes the Higgs branch a very useful object in the study of dimensional reductions, and much of the tests performed in the study of dimensional reduction involve it in one way or another.

Finally, we consider mass deformations. These are associated with vevs to scalar fields in background vector multiplets associated with flavor symmetries. As such these are expected to reduce to mass deformations in $4d$ of the associated symmetries. There are several interesting questions involving mass deformations that we shall next consider.

\begin{table}[ht]
\begin{adjustbox}{center,max width=.61\textwidth}
$\def\arraystretch{1.0}
\begin{array}{r|c|c:cc|c:cc|c:c:c:c}
\multicolumn{12}{c}{\Large\textsc{Table\ of\ Rank-2\ theories\ II}}\\
\hline
\hline
&&\multicolumn{3}{l|}{\quad \text{Moduli Space}} &
\multicolumn{3}{l|}{\quad \text{Flavor and central charges}} &
\multicolumn{4}{c}{\quad \text{5$d$ Gauge theory realization}}\\[1mm]
\multirow{-2}{4mm}{\#}&
& \D_{u,v}&\ \ d_{\text{HB}}\ \ &\ \  h\ \  
&\quad \ff\quad &\ \ 24a\ \ & 12c 
&[ SU(3) ]_k&USp(4)&[SU(2)^\theta,SU(2)]&G_2
\\[1.5mm]
\hline\hline
\multicolumn{12}{c}{\suf(6)\ \text{series}}\\
\hline
33.& 
& \{6,8\} 
 & 23 &1  &
\suf(6)_{16}{\times}\suf(2)_9 & 179 & 101 
&\multicolumn{4}{c}{4\F+SU(3)_{\frac12}\times SU(3)_{-\frac12}+4\F\ \Z_2\ {\rm twisted}}\\
34.&
& \{4,6\}
& 13 & 1 &
\suf(4)_{12}{\times} \suf(2)_7{\times}\uf(1) & 121 & 67 
&\multicolumn{4}{c}{3\F+SU(3)_1\times SU(3)_{-1}+3\F\ \Z_2\ {\rm twisted}}\\
35.&
& \{4,5\}
& 11 & 0 &
\suf(3)_{10}{\times} \suf(3)_{10}{\times}\uf(1) & 107 & 59  
&\multicolumn{4}{c}{3\F+SU(3)_0\times SU(3)_0+3\F\ \Z_2\ {\rm twisted}}\\
36.&
& \{3,5\}
& 8 & 1 &
\suf(3)_{10}{\times} \suf(2)_6{\times}\uf(1) & 92 & 50  
&\multicolumn{4}{c}{2\F+SU(3)_{\frac32}\times SU(3)_{-\frac32}+2\F\ \Z_2\ {\rm twisted}}\\
37.&
& \{3,4\}
& 6 & 0 &
\suf(2)_8{\times} \suf(2)_8{\times}\uf(1)^2 & 78 & 42 
&\multicolumn{4}{c}{2\F+SU(3)_{\frac12}\times SU(3)_{-\frac12}+2\F\ \Z_2\ {\rm twisted}}\\
\rcy38.&
& \{2,3\} 
& 2 & 0
& \uf(1)\times \uf(1) & 49 & 25  
&\multicolumn{4}{c}{1\F+SU(3)_{k}\times SU(3)_{-k}+1\F\ \text{GT}\ \Z_2\ {\rm twisted}}\\

\cdashline{1-12}
\multicolumn{12}{c}{\spf(14)\ \text{series}}\\
\hline
39.&
& \{6,8\}
&  29 & 7
&\spf(14)_9 & 185 & 107 
&\multicolumn{4}{c}{SU(5)_0+12\F\ \Z_2\ {\rm twisted}}\\
40.&
& \{4,6\}
& 17 & 5
&\suf(2)_8 \times\spf(10)_7& 125  & 71 
&\multicolumn{4}{c}{SU(5)_0+10\F\ \Z_2\ {\rm twisted}}\\
41.&
& \{4,6\}
& 15 & 5
&\suf(2)_5 \times \spf(8)_7 & 123 & 69 
&\multicolumn{4}{c}{\red{\textsc {no\ known\ 5$d$  \ construction}}}\\
42.&
& \{3,5\}
& 11 & 4
&\spf(8)_6\times \uf(1) & 95  & 53 
&\multicolumn{4}{c}{SU(5)_0+8\F\ \Z_2\ {\rm twisted}}\\
\rcy43.&
& \{2,4\}
& 6 & 3
& \spf(6)_5 & 65 & 35 
&\multicolumn{4}{c}{SU(5)_0+6\F\ \text{GT}\ \Z_2\ {\rm twisted}}\\

\cdashline{1-12}
\multicolumn{12}{c}{\suf(5)\ \text{series}}\\
\hline

44.&
& \{6,8\}
& 19 & 0
&\suf(5)_{16} & 170 & 92  
&\multicolumn{4}{c}{5\F+SU(4)_0 \times SU(3)_0 \times SU(2)+2\F\  \Z_3\ {\rm twisted}}\\
45.&
& \{4,6\}
& 6 & 0 &
\suf(3)_{12}{\times} \uf(1) & 114 & 60  
&\multicolumn{4}{c}{3\F+SU(4)_0 \times SU(3)_0 \times SU(2)+1\F\  \Z_3\ {\rm twisted}}\\
46.&
& \{3,5\}
& 3 & 0 &
\suf(2)_{10}{\times} \uf(1) & 86 & 44  
&\multicolumn{4}{c}{\text{5$d$   SCFT (no known Lagrangian)}\ \Z_3\ {\rm twisted}}\\

\cdashline{1-12}
\multicolumn{12}{c}{\spf(12)\ \text{series}}\\
\hline
47.&
& \{4,10\}
& 32 & 6
&\spf(12)_{11} & 188 & 110
&\ [5\F]_{\frac92}\ &\ 2\AS+4\F\ &\ -\ &\ 5\F\ \\
\rcy48.&
& \{2,4\}
& 8 & 2
&\spf(4)_5 \times\sof(4)_8 & 68 & 38
&\multicolumn{4}{c}{USp(4)+2\AS+2\F\ \text{GT}}\\
\rcy49.&
& \{2,6\}
& 14 & 4
& \spf(8)_7 & 98 & 56
&\multicolumn{4}{c}{G_2+4\F\ \text{GT}}\\

\cdashline{1-12}
\multicolumn{12}{c}{\spf(8)-\suf(2)^2\ \text{series}}\\
\hline
51.&
& \{6,12\}
& 28 & 6
& \spf(8)_{13}\times \suf(2)_{26} & 232 & 130  
&\multicolumn{4}{c}{SU(5)_0+2\AS+6\F\ \Z_2\ {\rm twisted}}\\
52.&
& \{4,8\}
& 14 & 4
& \spf(4)_{9}{\times} \suf(2)_{16} {\times}\suf(2)_{18}& 146 & 80  
&\multicolumn{4}{c}{SU(5)_0+2\AS+4\F\ \Z_2\ {\rm twisted}}\\
53.&
& \{3,6\}
&  7 & 3
& \suf(2)_{7}\times \suf(2)_{14}\times \uf(1) & 103 & 55  
&\multicolumn{4}{c}{SU(5)_0+2\AS+2\F\ \Z_2\ {\rm twisted}}\\
54.&
&\{3,6\}   
& 6 & 2
& \suf(2)_6\times \suf(2)_8 & 102 & 54     
&\multicolumn{4}{c}{1\F+SU(2)\times [SU(4)_0+2\F]\times SU(2)+1\F\ \Z_4\ {\rm twisted}}\\
55.&
& \{\frac52,4\}
& 2 & 0
&\suf(2)_5 & 67 & 34 
&\multicolumn{4}{c}{SU(3)_0\times SU(2)_{\pi}\times SU(3)_0 \ \Z_4\ {\rm twisted}}\\
\rcb56.&
& \{2,4\}
&  2 & 
& \suf(2)_{10} & 60 & 30    
&\multicolumn{4}{c}{SU(2)_0\times SU(4)_0\times SU(2)_0 \ \Z_4\ {\rm twisted}}\\
\cdashline{1-12}
\multicolumn{12}{c}{\gf_2\ \text{series}}\\
\hline
57.&
&\{4,6\} 
& 12 & 2
& [\gf_2]_8\times \suf(2)_{14} & 120 & 66    
&\multicolumn{4}{c}{4\F+1\AS+SU(4)_0\times SU(2)+1\F \ \Z_3\ {\rm twisted}}\\
58.&
&\{\frac83,4\} 
& 4 & 2
& \suf(2)_{\frac{16}3}\times \suf(2)_{10}  & 72 & 38   
&\multicolumn{4}{c}{2\F+1\AS+SU(4)_0\times SU(2)_0 \ \Z_3\ {\rm twisted}}\\
59.&
& \{\frac{10}3,4\}
& 6& 0
&[\gf_2]_{\frac{20}3} & 82  & 44
&\multicolumn{4}{c}{SU(3)_0\times [SU(2)+1\F]\times SU(2)_0 \ \Z_3\ {\rm twisted}}\\
\rcb60.&
& \{2,3\}
& 2 & 2
& \suf(2)_8 & 48 & 24  
&\multicolumn{4}{c}{\text{5$d$   SCFT (no known Lagrangian)}\ \Z_3\ {\rm twisted}}\\

\cdashline{1-12}
\multicolumn{12}{c}{\suf(3)\ \text{series}}\\
\hline

61.&
& \{6,12\}
& 15 & 5
& \suf(3)_{26}\times \uf(1) & 219 & 117  
&\multicolumn{4}{c}{3\F+1\AS+SU(5)_0\times USp(4)+2\F \ \Z_3\ {\rm twisted}}\\
62.&
& \{4,8\}
& 5 & 3
& \uf(1)^2 & 137 & 71   
&\multicolumn{4}{c}{1\F+1\AS+SU(5)_0\times USp(4)+1\F \ \Z_3\ {\rm twisted}}\\
\rcg63.&
& \{3,6\}
&  2 & 2
& \uf(1) & 96 & 48  
&\multicolumn{4}{c}{\text{5$d$   SCFT (no known Lagrangian)}\ \Z_3\ {\rm twisted}}\\

\cdashline{1-12}
\multicolumn{12}{c}{\suf(2)\ \text{series}}\\
\hline
64.& 
& \{6,12\}
&  8 & 4
& \suf(2)_{16}\times \uf(1) & 212&  110  
&\multicolumn{4}{c}{2\F+1\AS+SU(5)_0\times SU(5)_0+1\AS+2\F \ \Z_4\ {\rm twisted}}\\
\rcg65.&
& \{4,8\}
&  2 & 2
& \uf(1) & 132 & 66  
&\multicolumn{4}{c}{1\AS+SU(5)_0\times SU(5)_0+1\AS \ \Z_4\ {\rm twisted}}\\
\hline\hline
\end{array}$
\caption{\label{tab:r22}\footnotesize{Continuation of rank-2 $4$d$  $ theories with their corresponding CFT data and 5\emph{d} origin. Again the last entry in the table gives the 5$d$   gauge theory description of the 5$d$ SCFT whose compactification gives the 4$d$ SCFT. When the compactification involves twisting, the discrete symmetry group being twisted is specified and again when it is $\mathbb{Z}_2$, the discrete symmetry is manifested in the gauge theory as the combination of quiver reflection and charge conjugation. For other groups, the discrete symmetry is generally broken by the gauge deformation. For some cases there is a 5$d$   SCFT parent, but we are not aware of any Lagrangian description for it. This is specified in the table by `5$d$   SCFT (no known Lagrangian)'. Finally, the entry `no known 5$d$  
construction' implies that we do not know the 5$d$ origin of said 4$d$  
SCFT. As before, we shade in \yellow{yellow} $\cN=2$ Lagrangian theories, in \blue{blue} $\cN=4$ theories and in \green{green} $\cN=3$ theories.}}
\end{adjustbox}
\end{table}

One interesting question is whether mass deformations commute with dimensional reduction or not. That is consider a mass deformation of a $5d$ SCFT, leading to a new $5d$ SCFT or to an IR free gauge theory. Now, consider the $4d$ theory resulting from the compactification of the $5d$ SCFT. As we mentioned it should inherit the mass deformation from the $5d$ SCFT, and we can inquire whether the result of such a mass deformation is the $4d$ theory resulting from the compactification of the low-energy $5d$ theory resulting from the mass deformation of the $5d$ SCFT, be it an SCFT or IR free gauge theory. Naively, we would suspect this to be true. The argument is that both the mass deformation and dimensional reduction causes some states to become massive. These are expected to be all integrated out of the low-energy $4d$ theory, regardless of which of them is heavier, as long as there is no tuning between the two deformations that causes some of them to be extremely light. However, this argument is too naive, as we shall next explain.

To illustrate this consider a mass deformation sending the $5d$ SCFT to a $5d$ gauge theory, potentially interacting with other stuff. As $5d$ gauge theories are always IR free, we expect them to reduce to analogous $4d$ gauge theories. However, $4d$ gauge theories have a coupling constant, which while classically scaleless, generically acquires a scale via dimensional transmutation. As such, the resulting $4d$ theory may still have a scale even after we integrated out the massive matter expected from the two deformations, and it is possible that this will be sensitive to the order in which the reduction is done. Here the behavior is expected to be different depending on whether the resulting $4d$ gauge theory is asymptotically free or IR free. In the latter case, the strong coupling scale is associated with the behavior in the UV, as such we expect the matter whose mass is proportional to it to have already been integrated out, and so our argument should still be valid. However, if the $4d$ gauge theory is asymptotically free then the strong coupling scale is associated with the behavior in the IR, and we expect the presence of massive matter whose mass is proportional to this scale to still survive in the low-energy theory. This invalidates our argument. As such, we expect mass deformations to $5d$ SCFTs and $5d$ gauge theories, whose $4d$ analogs are conformal or IR free, to commute with dimensional reduction while mass deformations to $5d$ gauge theories, whose $4d$ analogs are asymptotically free, to not commute. As we shall see, this matches our experimental observations.

We have noted that the space of mass deformations is spanned by the Cartan subalgebra of the flavor symmetry group. We have also noted that the $4d$ theory should inherit the mass deformations of the $5d$ theory. However, we can ask whether this must indeed be true. Specifically, is it possible for the $4d$ theory to have more (or less) mass deformations than the $5d$ theory? Due to the relation between mass deformations and global symmetries, this is equivalent to the problem of whether the $4d$ theory must have the same continuous flavor symmetries as the $5d$ theory.

One argument why this should be the case is as follows. The conserved current multiplets, for both the $5d$ $\mathcal{N}=1$ and $4d$ $\mathcal{N}=2$ SCFTs, belong to a special chiral ring of operators associated with the Higgs branch. Specifically, these are related to coordinate functions on this space. Recall that the Higgs branch is invariant under dimensional reduction. As such any conserved current multiplets possessed by the $5d$ SCFT must also be present in the $4d$ SCFT, as these are realized as functions on the Higgs branch which is the same for both theories. 

While this argument is usually valid, there is one interesting loophole that we next consider\footnote{There can also be potential subtleties that would arise if the multiplets containing the conserved currents are nilpotent. We shall for the most part ignore this possibility here.}. This loophole arises in cases where the $5d$ Higgs branch has multiple components, that is the Higgs branch is a union of several disconnected spaces. These spaces all intersect with one another and the Coulomb branch at the SCFT point. This is necessarily the case because of conformal invariance. In fact, say this was not the case, the distance between the points where the different components intersect the Coulomb branch would constitute a scale, thus violating the conformal invariance.

\begin{table}[t!]
\begin{adjustbox}{center,max width=.9\textwidth}
$\def\arraystretch{1.0}
\begin{array}{r|c|c:cc|c:cc|c:c:c:c}
\multicolumn{12}{c}{\Large\textsc{Table\ of\ Rank-2\ theories\ III - Isolated theories}}\\
\hline
\hline
&&\multicolumn{3}{l|}{\quad \text{Moduli Space}} &
\multicolumn{3}{l|}{\quad \text{Flavor and central charges}} &
\multicolumn{4}{c}{\quad \text{5$d$ Gauge theory realization}}\\[1mm]
\multirow{-2}{4mm}{\#}&
& \D_{u,v}&\ \ d_{\text{HB}}\ \ &\ \  h\ \  
&\quad \ff\quad &\ \ 24a\ \ & 12c 
&[ SU(3) ]_k&USp(4)&[SU(2)^\theta,SU(2)]&G_2
\\[1.5mm]
\hline\hline
%%%%%%%%%%%%%%%% [I_1 series]
66.&
& \{4,6\}
&  10 & 4
& \spf(4)_{14}\times\suf(2)_8 & 118 & 64   
&\multicolumn{4}{c}{SO(8)+2\V+2\S+2{\rm C}\ \Z_3\ {\rm twisted}}\\
67.&
& \{\frac{12}5,6\}
&  2 & 2
& \suf(2)_{14} & \frac{456}5 & \frac{234}5   
&\multicolumn{4}{c}{\red{\textsc{no\ known\ 5d\ construction}}}\\
\rcb68.&
& \{2,6\}
& 2 & 2
& \suf(2)_{14} & 84 & 42    
&\multicolumn{4}{c}{2F+SU(3)_0\times SU(4)_0\times SU(3)_0+2F\ \Z_6\ {\rm twisted}}\\
\rcy 69.&
& \{2,4\}
&  0 & 0
& \varnothing & 58 & 28  
&\multicolumn{4}{c}{\red{\textsc{no\ known\ 5d\ construction}}}\\
\hline\hline
\end{array}$
\caption{\label{tab:r23}List of isolated rank-2 $4d$ theories with their corresponding CFT data and 5\emph{d} origin. The column in the table gives the 5$d$  gauge theory description of
the 5$d$  SCFT whose compactification gives the 4$d$ SCFT. Here all compactifications involve twisting and since the discrete symmetry group being twisted is not $\mathbb{Z}_2$, the discrete symmetry is generally broken by the gauge deformation. The entry `no known 5$d$  construction' implies that we do not know the 5$d$  origin of said 4$d$ SCFT. As before, we shade in \yellow{yellow} $\cN=2$ Lagrangian theories and in \blue{blue} $\cN=4$ theories.}
\end{adjustbox}
\end{table}

What makes this case interesting is that while the Higgs branch should be invariant under quantum corrections, and so also its dimensional reduction, the property that all the different components intersect the Coulomb branch at a point is not. Specifically, there are examples, notably the case of the $4d$ $\mathcal{N}=2$ $SU(2)$ gauge theory with two fundamental hypermultiplets, where the Higgs branch is made of multiple components that classically intersect the Coulomb branch at the same point, but get separated by quantum corrections\cite{Seiberg:1994aj}. Similar phenomenon can also take place upon dimensional reduction. In fact, the only reason the components all intersect at a point is due to the $5d$ conformal invariance. Once that is broken, we expect quantum corrections to cause these to separate with a distance proportional to the scale associated with the dimensional reduction. When this scale is taken to infinity, giving the limit of the $4d$ theory, the components are expected to become infinitely separated, and we expect only one to be accessible. As such in this case we expect to be able to get different interesting $4d$ theories, each arising at a point where a component of the Higgs branch intersects the Coulomb branch. 

When we compactify on a circle of non-zero radius then we should be able to continuously interpolate between the different theories by moving along the Coulomb branch. However, in the limit where the radius of compactification goes to zero, the distance between all these theories becomes infinite, and we can only access one of them, depending on choices made during the compactification. As such, in these cases the Higgs branch of the resulting $4d$ theory is merely a single component of the $5d$ Higgs branch and, consequently, the flavor symmetry of the $4d$ theory is in general a subgroup of the $5d$ flavor symmetry, \emph{i.e.} the one associated with functions defined on this component. Therefore, we see that the $5d$ and $4d$ theories should have the same global symmetry if the Higgs branch has a single component. However, the $4d$ theory may only inherit a subgroup of the symmetries of the $5d$ SCFT in cases where the Higgs branch has multiple components. As we shall see, this matches the experimental observations. 

We have seen that properties of the Higgs branch, notably its invariance under quantum corrections, makes it an interesting object to consider in the study of dimensional reduction. A further interesting property of the Higgs branch of $\mathcal{N}=2$ SCFTs, not unrelated to the first, is that it preserves the $\uf(1)$ part of the superconformal R-symmetry. This implies that $4d$ SCFTs are Higgsed to other $4d$ SCFTs. Note that the converse is not necessarily true, specifically, it is possible for non-SCFT $4d$ $\mathcal{N}=2$ theories to be Higgsed to an SCFT, in which case the $\uf(1)$ part of its superconformal R-symmetry arises accidentally. 

Combining these two observations on the Higgs branch leads to an interesting approach to study the reduction of higher dimensional theories to $4d$ $\mathcal{N}=2$ theories. Specifically, say we can determine the resulting $4d$ theories for a special class of relatively simple higher dimensional theories. These are assumed to be such that they can either only be Higgsed to free hypers or alternatively have no Higgs branch. Some of these will reduce to $4d$ SCFTs and some will reduce to $4d$ theories that are not SCFTs. Now consider other more complicated higher dimensional theories that are related to the previous ones by a Higgs branch flow. Due to what we mentioned previously, we expect that any higher dimensional theory that is related via Higgs branch flows to a theory reducing to a $4d$ IR free theory, cannot itself reduce to a $4d$ SCFT. As such, the understanding of the reduction for simple theories, that are in a sense the bottom non-trivial theories in any Higgs branch flow, can help us map the space of theories reducing to $4d$ SCFTs. Specifically, these can only be those theories that are only Higgsed to theories reducing to $4d$ SCFTs. As such, the space of higher dimensional theories can usually be grouped into families with members of the family related via Higgs branch flows, and these families share similar behavior under dimensional reduction. We shall see many illustration of this behavior in section \ref{sec:sum}, where we have the more technical discussion.

\section{Mass deformation of rank-2 $\mathcal{N}=2$ SCFTs}\label{sec:4dsetup}

Let's leave the question of dimensional reduction behind for a minute and concentrate on the understanding of mass deformations directly from $4d$. In a recent paper \cite{Martone:2021ixp}, one of the authors compiled a catalogue of currently known rank-2 $4d$ $\cN=2$ SCFTs surveying the many different results scattered across the literature and, for most of the discussion below, we will restrict our attention to this case. The outcome of this study is that the theories are largely organized in series which are interconnected by renormalization group flows and theories in each series share similar moduli space properties.

It might be helpful to first remind the reader how the study in \cite{Martone:2021ixp} was performed. Given the complication of the study of rank-2 CB geometries, the study was not exhaustive. In fact the techniques used largely leverage an intricate set of consistency conditions which lead to tight predictions for a detailed analysis of the moduli space structure, namely its CB and HB stratification. This has provided new insights on the structure of these theories with relatively little effort, but a shortcoming of this method, is that it is not yet as systematic as for example the analysis of rank-1 theories performed in \cite{Argyres:2015ffa,Argyres:2015gha,Argyres:2016xmc,Argyres:2016xua}. Here we will make full use of this new information to highlight even more the structure of the space of four dimensional $\cN=2$ SCFTs with two complex dimensional CBs. In fact, this detailed data will allow us to quite straightforwardly identify the 5\emph{d} SCFT which dimensionally reduce to one of the entry in the catalogue in \cite{Martone:2021ixp}. The 5\emph{d} picture will allow us to more easily identify RG-trajectories among the rank-2 theories. We perform this study in the next section.

Rather than introducing a large set of technical details regarding the stratification of Higgs and Coulomb branches, which can be found in for example in \cite{Bourget:2019aer,Argyres:2020wmq,Martone:2021ixp}, we will highlight the evidence for the lack of completeness of our current understanding of rank-2 theories thus motivating the more careful analysis performed below. First of all there are a set of theories, reported in table \ref{tab:r23} which are not connected by a known mass deformation to any other $\cN=2$ SCFT. This is at least suspicious since some of these theories have a large flavor symmetry suggesting that perhaps we are missing some theories. Also the study of mass deformations of some of the Lagrangian SCFTs has not been performed systematically and new superconformal field theories might be obtained by tuning their masses as well as their CB moduli. For example a new fixed point on the CB of the theory with $G_2$ gauge groups and hypermultiplets in the fundamentals was recently found by one of the authors \cite{Kaidi:2021tgr}. Finally it remains an open question whether we have an exhaustive understanding of theories with $\cN=3$ supersymmetry. The moduli space of vacua for new candidate theories of rank-2 was put forth in \cite{Argyres:2019ngz} but no direct construction of these theories has been so far found. If these new entries corresponded to new physical $\cN=3$ theories, it is reasonable that each belongs to a whole new series of $\cN=2$ theories.

This state of the art justifies thinking more systematically about the structure of currently known theories to identify patterns which allow us to fill up the gaps of our understanding. An interesting route to pursue is understanding from the bottom up the way the stratified moduli space of vacua of $\cN=2$ theories is deformed by mass deformations, this is what we are going to do next.

\subsection{Analysis of relevant deformations}

The possible local $\cN=2$ supersymmetric deformations of an SCFT are built as integrals over space-time of some Lorentz scalar descendant, $\d$, of a superconformal primary field $\cO$.  To deform the theory it should not be a total derivative, so $\d$ can only be a descendant formed by acting with some combination of the eight $\cN=2$ supercharges, $Q$ and $\tQ$, on $\cO$,
\begin{align}\label{desc}
\d_{n,\ell}\sim Q^n \tQ^\ell \cO
\end{align}
for $n,m\in\{0,1,\ldots,4\}$.  To be $\cN=2$ supersymmetric, $Q$ or $\tQ$ acting on it must annihilate it or give a total derivative.  This can happen either by virtue of the supersymmetry algebra or, if the primary $\cO$ satisfies a null state condition, by virtue of it being in a shortened representation of the superconformal algebra. In order for the deformation to be relevant we have
\beq
\D_{Q^n\tQ^\ell \cO}<4
\eeq
where $\D$ indicates the scaling dimension of the superconformal descendent in \eqref{desc}. The analysis of the allowed $\cN=2$ relevant deformations is then reduced to a straightforward exercise of superconformal representation theory which was carried out in \cite{Argyres:2015ffa,Cordova:2016xhm}. 

The result of this analysis is that there are two type of relevant deformations which we will call \emph{mass} and \emph{chiral} deformations respectively. In both cases $\cO$ is a Lorentz scalar but the quantum numbers under the R-symmetry group are different. 

\paragraph{Chiral deformations} In the case of chiral deformations, the superconformal primary, which we call $\cO_\chi$, is only charged under the $\uf(1)_r$ symmetry and satisfies the shortening condition that its scaling dimension is proportional to its charge with the constant of proportionality fixed by the normalization of the $\uf(1)_r$\footnote{For those who are familiar with 4\emph{d} $\cN=2$ superconformal representation theory, $\cO_\chi$ is the superconformal primary of the $\cE_r$ or $L\bar{B}_1[0;0]^{(0,r)}_{\tfrac{r}2}$ multiplet in the notation of  \cite{Dolan:2002zh} and \cite{Cordova:2016emh} respectively.}. In this case the $\cN=2$ preserving deformation is associated to the following descendent:
\beq
{\rm chiral\ deformation}\ \sim\ \tQ^4\cO_\chi
\eeq
From that it follows that chiral deformations are relevant only if CB operators have $1<\D<2$. In this paper we won't have much to say regarding how to track chiral deformations so let us move now to talk about the other possibility.

\paragraph{Mass deformations} In the case of mass deformations the superconformal primary, which we will label $\cO_m$, is an $\suf(2)_R$ triplet, neutral instead under the $\uf(1)_r$, and satisfying the shortening condition that its scaling dimension is proportional to the $\suf(2)_R$ charge\footnote{In this case, $\cO_m$ is the superconformal primary of the $\hat{\cB}_1$ or $B_1\bar{B}_1[0;0]^{(2,0)}_2$ multiplet in the notation of  \cite{Dolan:2002zh} and \cite{Cordova:2016emh} respectively.}. 
In this case the $\cN=2$ preserving deformation is associated to the following descendent:
\beq
{\rm mass\ deformation}\ \sim\ \tQ^2\cO_m
\eeq
It can be shown that the flavor current has to necessarily belong to the same multiplet as $\cO_m$ which therefore implies that $\cO_m$, and the associated $\cN=2$ deformation, have to transform in the adjoint representation of the flavor symmetry algebra $\ff$. We will indicate this with an upper superscript, $\cO_m^A$, $A=1,...,{\rm dim}(\ff)$. 

The general mass deformation has the form:
\beq
m_A\int d^4x\tQ^2 \cO^A_m+{\rm c.c.}
\eeq
with $m_A\in \CC$ and carrying dimension of mass. Although there are in principle dim$(\ff)$ parameters $m^A$, many are related by global symmetry transformations and give rise to equivalent deformations. All in all an SCFT has rank$(\ff)$ inequivalent mass deformations.

As explained in \cite{Argyres:2015ffa}, understanding the RG-flows of a theory with a moduli space of vacua is a non-trivial operation as the continuous allowed set of vacua maps every single relevant deformations into a family of RG-flow. When an SCFT with a moduli space of vacua is deformed, the moduli space can either be partially lifted or deformed. The HB provides an example of the former phenomenon while the CB an example of the latter. We will study both in turn.

\subsubsection{Coulomb branch}

From the point of view of the low energy theory on the CB, $\cN=2$ supersymmetry disallows any scalar potential for the $U(1)$ vector multiplets. In fact we have just shown above that superconformal representation theory forbids $\cN=2$ FI terms \cite{Antoniadis:1995vb} which would instead have the effect of lifting some CB directions. This observation, along with complex analyticity, implies that the CB cannot be lifted but only deformed under mass deformations.

This property was key in the complete classification of rank-1 SCFTs \cite{Argyres:2015ffa,Argyres:2015gha,Argyres:2016xmc,Argyres:2016xua}. In the rank-1 case, the CB is one complex dimensional and the singular locus $\cV$ zero dimensional. By the absence of scales in the theory, one can easily conclude that $\cV$ corresponds to a single point, namely the origin of the CB which we will also call \emph{the superconformal vacuum}. In these circumstances, it is possible to argue that the only possible effect of a mass deformation on the CB vacuum is to split the singularity. The analysis of higher rank case is more involved. 

The singular locus of the CB of rank-2 SCFTs is in general the union of multiple irreducible one complex dimensional components which could intersect along complex co-dimension two loci. Using scale invariance, it is possible to argue that there is a single co-dimension two locus, which corresponds to the superconformal vacuum, but it is not possible to constrain the number of co-dimension one components\footnote{At rank-2, scale invariance still constraints the topology, or rather the algebraic form, of the complex co-dimension one loci. For more see \cite{Argyres:2018zay}.} which, by the previous constraint, have to all intersect at the origin of the moduli space. 

Before turning on any mass deformation, each of the irreducible complex co-dimension one loci stemming out of the origin supports charged massless matter which is effectively described by a rank-1 theory \cite{Argyres:2020wmq}. The transverse complex line to each singular stratum has then the approximate interpretation of being the CB of the theory supported on each locus. This approximation is valid locally near the singular locus. Using the \emph{UV-IR simple flavor condition} \cite{Martone:2020nsy}  the rank-2 SCFT mass deformations can be understood as deformations of the rank-1 theories supported on the complex dimension one components. This observation greatly simplifies the problem of understanding rank-2 mass deformations by reducing it to that of a specific set of rank-1 theories, \emph{i.e.} those supported on the complex co-dimension one loci, and allows us to understand how the asymptotics of rank-2 CBs (this argument is really valid for any rank $r$) behave under mass deformation. Let's see how this works.

As we mentioned above, turning on masses splits the rank-1 superconformal vacuum. This splitting of rank-1 singularities translates into a local transverse splitting of the complex co-dimension one loci. % see figure \ref{SplitCB}. 
By complex analyticity, the splitting can be extended to the asymptotics of the moduli space and it implies that the far end of each co-dimension one locus is separated in multiple components as dictated by the rank-1 mass deformations.\footnote{This behavior differs majorly from rank-1 where the asymptotics of the CB are simply undeformed.} The latter is characterized by the \emph{deformation pattern} \cite{Argyres:2015ffa}. Say the rank-1 theory is supported on the $\mathfrak{i}$-th stratum, if the initial singularity splits into $n_\mathfrak{i}$ singularities, the deformation pattern contains $n_\mathfrak{i}$ entries which specify the low energy effective theory at each of the resulting singular loci. These are severely constrained by the rules outlined in \cite{Argyres:2015ffa,Argyres:2015gha,Argyres:2016xmc} and can be worked out explicitly for any chosen mass deformation. It then follows that, correspondingly, the $\mathfrak{i}$-th complex co-dimension one locus splits into $n_\mathfrak{i}$ loci each supporting a rank-1 low energy theory as dictated by the deformation pattern. 

It might be useful to work out an example, and in section \ref{checks} below we will report many more. Consider $SU(3)+6F$ which has flavor symmetry $\uf(6)$ which has a two complex dimensional CB. Call $\{u,v\}$ the global coordinates which describe such a space. The CB stratification of this theory includes a $v=0$ stratum with $\red{\Tf}_v=[I_6,\suf(6)]$ and two knotted strata with $\red{\Tf}_{u^3+v^2}=[I_1,\varnothing]$. Here we are following the notation of \cite{Martone:2021ixp} where $\Tf$ represents a generic SCFT and $\red{\Tf}_{\star}$ an SCFT describing the low-energy effective theory on a specific CB stratum identified by $\star$. The simple factor of the flavor symmetry is realized on the $v=0$ stratum and therefore a mass deformation will split the $v=0$ to a number of complex dimension one loci. Since the possible deformation patterns of the $I_6$ are:
\beq\label{DefSU3}
I_6\to\{I_{n_1},...,I_{n_\ell}\}
\eeq
where $(n_1,...,n_\ell)$ is a partition of six, the $v=0$ locus splits in $n_\ell$ different components supporting $\{[I_{n_1},\suf(n_1)],...,[I_{n_\ell},\suf(n_\ell)]\}$. Since the knotted strata don't support any flavor symmetry, they simply go along with the ride undeformed.

This analysis is powerful but it is unfortunately only valid in the asymptotic region of the CB. The relevant scales in this problem are the distance from the origin along the complex co-dimension one singular locus, the distance along the transverse direction, \emph{i.e.} the rank-1 CB direction, and the size of the mass deformation which determines the distance of the splitting on the transverse complex line. So long as the latter is much smaller than the other two, the analysis performed above holds. We call this region, $(u,v)\gg m$, the \emph{outer} region. For the \emph{inner} region, \emph{i.e.} the region at distance roughly of the size of the mass from the origin of the CB, we can no longer treat each rank-1 theory independently. As mentioned in passing, rank-2 fixed points correspond to co-dimension two singular loci which arise at the intersection of co-dimension one loci. Our analysis above showed that no such intersection can appear in the outer region. Thus the inner region is the one relevant for the appearance of superconformal fixed points. Let's turn to this then.

The analysis of the inner region is extremely cumbersome, to have a sense of what can happen, it might be useful to work out an a explicit example. Consider $SU(3)+5F$. This theory is asymptotically free, thus it's Seiberg Witten curve depends on a scale $\Lambda$ which corresponds to the scale at which the coupling becomes strong. The anlaysis below can also be performed by starting with an SCFT, namely $SU(3)+6F$, but we then need to turn on an extra mass on top of what we describe below, which corresponds to decoupling one flavor. 

The $SU(3)+5F$ has a $\uf(5)$ flavor symmetry and the simplest mass deformation to analyze is the one which leaves an $\suf(5)$ flavor symmetry unbroken, this corresponds to giving a common mass $m$ to all hypermultiplets \cite{Eguchi:1996vu}. When all the masses are turned off the Seiberg Witten curve is given by \cite{Hanany:1995na}:
\beq
y^2=\left(x^3+u x+v+\frac{\L} 4 x^2\right)^2-\L x^5
\eeq
taking its quantum discriminant \cite{Martone:2020nsy}, we obtain the structure of the singular locus to be:
\beq
{\rm Dis}\sim v^5\left((4 u^3+27 v^2)^2+\cO(\L)\right)
\eeq
thus asymptotically the stratification is made of two strata. Now turning on the mass deformation the curve is deformed to:
\beq\label{D2su5}
y^2=\left(x^3+u x+v+\frac{\L} 4 (x^2+5 xm + 10 m^2)\right)^2-\L (x+m)^5
\eeq
The discriminant is too complicated to analyze and, following \cite{Eguchi:1996vu}, we will instead focus on a special one complex dimensional subspace of the $(u,v,m)$ hyperplane which has higher criticality. For generic values of the masses, the leading behavior near the critical point of the discriminant is:
\beq\label{U2po}
{\rm Dis}_{\rm crit}\sim \d v^5(\d u^2+\d v)^2
\eeq
where the zeros of the $\d v$ and $\d u$ are at the critical point that we are analyzing, which is where we expect to find a rank-2 fixed point ($\d v=\d u=0$). Because of the topology of the knotted stratum, we can immediately infer that $\D_{\d v}=2\D_{\d u}$. 

The physical interpretation of this fixed point is a $U(2)+5F$ theory, which is obviously IR-free. The Landau pole of this theory, $\tilde{\L}$, depends on the only two scales in the problem $m$ and $\L$. It is then possible to tune $m$ in such a way that $\tilde{\L}\to0$. This point corresponds to the point of maximal criticality of the mass deformed curve \eqref{D2su5} which give rise to the $D_2(\suf(5))$ Argyres-Douglas theory \cite{Eguchi:1996vu}. 

We can perform again the analysis of the leading behavior of the quantum discriminant for this special value of the mass finding the surprising result:
\beq
{\rm Dis}_{\rm crit}\sim \d v^5(\d u^5+\d v^3)
\eeq
which differs greatly from \ref{U2po} and precisely reproduces the stratification corresponding to the $D_2(\suf(5))$ SCFT \cite{Martone:2021ixp}. Notice in particular that by tuning the mass $m$ to a special value, \emph{i.e.} the one for which $\tilde{\L}\to0$, a (2,1) torus link has merged into a single (3,5) torus knot! As previously anticipated, this analysis confirms that the behavior of the inner region is quite involved. Below we instead state a simple, necessary, criterion that CB stratification have to satisfy to be connected by mass deformation. It is quite remarkable that we can find any such thing and which applies to all cases we have analyzed.

Before we state it, let's introduce the concept of a decoration and a deformation of a CB stratification. Recall that a CB stratification for a rank-2 theory is specified by two data, a set of complex dimensional one strata and a set of rank-1 theories which describe the low-energy effective theory on each stratum. Henceforth we will call the latter the \emph{decoration} of the CB stratification and will indicate the set of theories within angle brackets:
\beq\label{deco}
\langle \red{\Tf}_1,..., \red{\Tf}_n\rangle
\eeq
Then a CB stratification ($I$) is said to be \emph{a deformation} of a CB stratification ($II$) if there exists a deformation pattern of the decoration of ($II$) which includes the decoration of ($I$). This should also account for the multiple $[I_1,\varnothing]$. Now we can introduce the first criterion to test whether two \emph{4d} $\cN=2$ SCFTs are related by mass deformation:\vspace{1em}

\begin{tcolorbox}
\begin{fact}\label{knot}
If a four dimensional $\cN=2$ SCFT $\Tf_{\rm UV}$ can be mass deformed to another SCFT $\Tf_{\rm IR}$, then the CB stratification of $\Tf_{\rm IR}$ should be a deformation of the CB stratification of $\Tf_{\rm UV}$.
\end{fact}
\end{tcolorbox}\vspace{0.5em}

We conjecture this to be a general property of mass deformation of CBs of $\cN=2$ SCFTs. Let's see how this works in explicit examples. The stratification of $SU(3)+6F$ has decoration:
\beq
\langle[I_6,\suf(6)],[I_1,\varnothing]^2\rangle
\eeq
The two $I_1$s are not deformable, while the allowed deformation of the $I_6$ can are discussed around \eqref{DefSU3}. The CB stratification of $D_2(\suf(5))$, which as we discussed above can be reached by mass deforming the $SU(3)$ theory, has decoration \cite{Argyres:2020wmq}
\beq
\langle[I_5,\suf(5)],[I_1,\varnothing]\rangle
\eeq
from which it follows immediately that indeed the two CB stratifications satisfy criterion \ref{knot}. Any theory with a decoration including any non $I_n$ theory or multiple $I_{n_i}s$ such that the $n_i$s add up to nine or more, would instead not satisfy criterion \ref{knot} and could therefore not be obtained as a mass deformation of the $SU(3)$ theory.

\subsubsection{Higgs branch}

Let's now turn our attention to the HB. The technical mathematical definitions which will follow are necessary for completeness but not to grasp the main ideas of this subsection, thus can be safely skipped at a first reading. With this warning behind us, we observe that the spontaneous breaking of the $\suf(2)_R$ symmetry equips the HB with a hyperkahler structure which in turn implies that HB is holomorphic symplecitc. The holomorphic symplectic structure on the HB is in general singular, thus the HB is in fact a symplectic singularity \cite{brieskorn1970singular,slodowy1980simple,beauville2000symplectic}. 

As in the case of the CB, the singular loci have the physical interpretation of corresponding to where extra charged massless states appear in the theory. On these special loci the low-energy effective theory is therefore no longer described by a set of free hypermultiplets but rather, because the HB is a hyperkhaler cone, by (possibly lower rank) interacting superconformal field theories. A symplectic singularity is naturally partitioned into symplectic manifolds of possibly different dimensions, called its \emph{symplectic leaves}. The operation of inclusion of their closures, which we will also call \emph{symplectic strata}, makes symplectic leaves of a symplectic singularity a partial order set. Their disjoint union thus give rise to a stratification of the HB as a holomorphic symplectic variety, which we will also call the \emph{HB stratification}, and symplectic leaves can be graphically depicted as a Hasse diagram, for more details see for example \cite{Bourget:2019aer}. 

By scale invariance, the smallest symplectic stratum is always the zero dimensional superconformal vacuum where the SCFT we want to study is supported. Next we call the symplectic strata which only include the origin the \emph{bottom symplectic strata} and the SCFTs supported on these strata, or more correctly leaves, the \emph{bottom SCFTs}. In our discussion we will mainly focus on the study of these. In fact the structure of the HB of the bottom theories determines the rest of the structure and it is usually the case that this structure is known.

Following a standard terminology, we call the scalar operators whose vacuum expectation values form the coordinate ring of the HB, the HB operators. They are $\uf(1)_r$ singlets, and in general are charged under both the $\suf(2)_R$ and the flavor symmetry. A mass deformation can be also thought of as a background field configuration of a vector multiplet associated to a weakly gauged flavor symmetry. As such, it is equivalent to an explicit breaking of the theory's flavor symmetry to the commutant of the particular background configuration. This implies that after a mass deformation we only expect the HB operators charged under the unbroken flavor symmetry to survive and thus the HB is in general partially lifted. 

This behavior under mass deformation differentiates the HB from the CB considerably. In fact, it is no longer true that all the mass deformations of the theory at the superconformal vacuum map to those supported on various loci of the moduli space. This follows straightforwardly from the fact that different regions of the HB can be mapped to patterns of spontaneous flavor symmetry breaking and thus some mass deformations are simply no longer available as we explore the HB in less and less singular loci. What it is surprising is that, despite all of these shortcomings, the HB of two fixed points, at the start and at the end of an RG-flow initiated by a mass deformations, are closely related. To understand how, we have to clarify the way mass deformations of the SCFTs appearing on various symplectic leaves are related to those of the theory at the superconformal vacuum. %We will focus almost entirely on discussing the relation between the mass deformations of the superconformal vacuum and the bottom theories.

To make any progress on this we need to introduce a bit of notation. Let's consider an SCFT with flavor symmetry $\ff_{\rm UV}$ (not necessarily simple) and a spontaneous breaking of a simple factor in $\ff_{\rm UV}$. Let us denote the unbroken symmetry along this particular HB direction by $\ff^\natural$. Denote the flavor symmetry (level) of the theory at the end of this HB flow by $\ff_{\mathrm{IR}} \,\,(k_{\mathrm{IR}})$. Furthermore, say that upon spontaneously breaking $\mathfrak{f}_{\rm UV}$ to $\ff^\natural$ one produces Goldstone bosons in a representation $\mathfrak{R}$. In terms of this data, the flavor level of the original theory with symmetry $\ff_{\rm UV}$ is given by \cite{Beem:2019tfp,Beem:2019snk,Giacomelli:2020jel}
\begin{align}\label{kmatch}
k_{\ff_{\rm UV}} = {T_2(\mathfrak{R}) +  k_{\mathrm{IR}} I_{\ff^\natural\hookrightarrow\ff_{\rm IR}}  \over I_{{\ff^\natural}\hookrightarrow\ff_{\rm UV}}}
\end{align}
where $T_2(\mathfrak{R})$ is the Dynkin index of the representation $\mathfrak{R}$ and $I_{\mathfrak{g}\hookrightarrow \ff}$ is the embedding index of $\mathfrak{g}$ into $\ff$. Henceforth we will normalize the Dynkin index in such a way that the fundamental representation of $\suf(n)$ has $T_2({\bf{n}})=1$. Before we turn to discussing an example to familiarize on how \eqref{kmatch} works, a few remarks are in order.

Our main goal in this subsection is to be able to come up with clear criteria for when HBs of two SCFTs are compatible with the two SCFTs being related via a mass deformation. For that, the main datum we are after is a way to track how the mass deformations of a superconformal theory act on its HB. In particular we want to understand how the mass deformations at the superconformal vacuum map to mass deformations of SCFTs supported elsewhere on the HB, \emph{e.g.} the bottom theories. In terms of the notation just introduced, the mass deformations of the former and the latter are determined by adjoint breakings of $\ff_{\rm UV}$ and $\ff_{\rm IR}$ respectively. Because of the spontaneous breaking $\ff_{\rm UV}\to \ff^\natural$ it is clear that some mass deformations simply do not act. The rest can be determined by identifying a series of groups embedding, namely $\ff^\natural\hookrightarrow\ff_{\rm UV}$ and $\ff^\natural\hookrightarrow\ff_{\rm IR}$. And that's where \eqref{kmatch} comes handy as with a bit arithmetic allows to determine $I_{\ff^\natural\hookrightarrow\ff_{\rm UV}}$ and $I_{\ff^\natural\hookrightarrow\ff_{\rm IR}}$.

Let's now discuss an example to see how all of this works. Consider our beloved $SU(3)+6F$ with flavor symmetry $\uf(6)_6$. It is a straightforward group theory exercise to show that spontaneously breaking the simple factor $\suf(6)_6\to \suf(4)_6{\times} \uf(1)$, where we are keeping track of the level for later convenience, Higgses the $SU(3)+6F\to SU(2)+4F$, see for example \cite{Bourget:2019aer}. The $SU(2)+4F$ has a $\sof(8)_4$ flavor symmetry. Now we want to use this knowledge to understand how the mass deformations of $SU(3)+6F$ act on the $SU(2)+4F$. Clearly the one corresponding to the commutant of $\suf(4){\times}\uf(1)$ inside $\uf(6)$, which corresponds to an $\suf(2)$ associated to the highest root of $\suf(6)$, does not act, and in fact they simply lift altogether the symplectic leaf supporting the $SU(2)+4F$. To instead understand how the mass deformations associated to the $\suf(4){\times}\uf(1)$ factor are mapped we need to understand how this flavor symmetry embeds in $\sof(8)$. In this case there is a single embedding of $\suf(4){\times}\uf(1)\hookrightarrow\sof(8)$ with index of embedding one, but for completeness let's check how \eqref{kmatch} can be used to derive a constraining picture.

The spontaneous breaking pattern which we are considering gives rise to goldstone bosons in the ${\bf 4}\oplus \bar{\bf 4}$, thus $T_2(\mathfrak{R})=2$ in this case, as well as $I_{\suf(4){\times}\uf(1)\hookrightarrow\suf(6)}=1$. Using the fact that $k_{\ff_{\rm UV}}=6$ and $k_{\ff_{\rm IR}}=4$, we readily obtain that indeed $I_{\suf(4){\times}\uf(1)\hookrightarrow\sof(8)}=1$ confirming our previous reasoning. In this case \eqref{kmatch} provided little information that we didn't know already, though considerably less trivial examples will be discussed below in section \ref{checks} where \eqref{kmatch} will be key to be able to fully understand how things work. 

Now that we clarified how to map mass deformations along the HB, we can work our way towards the second criterion for two moduli spaces to be compatible of being connected by mass deformations. As mentioned already, the low energy effective theories on the singular loci of a HB can only be described by SCFTs; no IR-free theory is allowed. It then follows that the only mass deformations which could possibly deform a given SCFT to another SCFT are those which act non trivially on the theories supported on the various symplectic leaves and deform those to other SCFTs as well. This is very constraining. 

Let's consider our previous example. We can ask whether there is any mass deformation of the $SU(3)+6F$ which deforms it to another SCFT. Because of the aforementioned properties, this SCFT, if it exists, has to Higgs to an SCFT obtainable as a mass deformation of the $SU(2)+4F$. But the mass deformations of the latter are known, specifically $SU(2)+4F$ can be mass deformed to three rank-1 Argyres-Douglas theories, namely $H_2$, $H_1$ and $H_0$ or, in more modern notation, $\cT^{(1)}_{A_2,1}$, $\cT^{(1)}_{A_1,1}$ and $\cT^{(1)}_{\varnothing,1}$. We then expect three different mass deformations of the $SU(3)+6F$, all adjoint breakings of the $\suf(4){\times}\uf(1)\subset \uf(6)$, which map to the appropriate mass deformations of $\sof(8)$ and give rise to three mass deformations from the $SU(3)+6F$ to SCFTs. These are precisely those depicted in figure \ref{MapR2} linking $\uf(6)_6$ to $D_2(\suf(5))$, $(A_1,D_6)$ and $(A_1,A_5)$\footnote{The reader might wonder why in figure \ref{MapR2} there two more SCFTs reacheable from $SU(3)+6F$. These involve also turning on chiral deformations and therefore we will not discuss them here.}.

We can now turn the previous discussion around which leads to our second criterion to identify moduli spaces of SCFTs which can be mass-deformed into one another:\vspace{1em}

\begin{tcolorbox}
\begin{fact}\label{Higgs}
If a four dimensional $\cN=2$ SCFT $\Tf_{\rm UV}$ with flavor symmetry $\ff_{\rm UV}$ can be mass deformed to another SCFT $\Tf_{\rm IR}$ with flavor symmetry $\ff_{\rm IR}\subset \ff_{\rm UV}$, then the bottom theories of $\Tf_{\rm IR}$ should be one of the following: 
\begin{itemize}
    \item[(i)] One of the bottom theories of $\Tf_{\rm UV}$.
    \item[(ii)] A mass deformation of one of the bottom theories of $\Tf_{\rm UV}$.
    \item[(iii)] Connected by Higgs branch flow to one of the bottom theories of $\Tf_{\rm UV}$.
\end{itemize} 
Furthermore the mapping between the two sets of bottom theories should be compatible with the mass deformation associated to the breaking $\ff_{\rm UV}\to \ff_{\rm IR}$.
\end{fact}
\end{tcolorbox}\vspace{0.5em}

In most cases establishing whether criterion \ref{Higgs} is satisfied is rather straightforward and it only entails comparing the bottom theories and checking that they are indeed obtainable by mass deformations of the original ones. But below we will encounter examples where things work in a much subtler way and a detailed understanding of how symmetries are matched along the HB is required, in fact cases $(i)$ and $(iii)$ can only happen in very special cases when the following phenomena happen:
\begin{itemize}
    \item[($i$)] There is a simple component, $\tilde{\ff}^\natural$, of $\ff^\natural$ which does not act on a given bottom theory, $\widetilde{\Tf}$. In this case the mass deformation associated to $\tilde{\ff}^\natural$ is neither acting on the bottom theory $\widetilde{\Tf}$ nor lifting a symplectic stratum. If such a mass deformation does lead to another SCFT, we expect to find $\widetilde{\Tf}$ as one of the bottom theories after the deformation.
    
    \item[($iii$)] One of the resulting simple factors of $\ff_{\rm IR}$, call it $\tilde{\ff}_{\rm IR}$, which survives a given mass deformation corresponds to a Higgs direction of one of the original bottom theories. We then expect that the theory supported on the stratum corresponding to $\tilde{\ff}_{\rm IR}$ is neither one of the original bottom theories nor its mass deformation, rather one which is connected by an HB flow to the original theory. We will see explicit examples where this happens and discuss them in detail in section \ref{checks}.
\end{itemize}

We will now take a detour in the marvelous world of five dimensions and will come back to the analysis of the \emph{4d} moduli space in section \ref{checks}.

\section{Brief summary of the observations on the 4d  compactification of 5d  SCFTs}
\label{sec:sum}

In section \ref{sec:GL} we have previously discussed the general features of the compactification of $5d$ SCFTs to $4d$. Here we shall summarize many technical observations about the compactification of specific $5d$ SCFTs. This will serve as both a summery of some of the current knowledge on this subject as well as an illustration of the more general results and approaches that were discussed in section \ref{sec:GL}.

The most well studied $5d$ SCFTs are the ones that can be engineered in string theory using brane webs\footnote{It is also possible to realize $5d$ SCFTs by the compactification of M-theory on a Calabi-Yau 3-fold, the so-called geometric engineering. This approach have undergone extensive study recently, see for instance \cite{Jefferson:2018irk,Bhardwaj:2018yhy,Bhardwaj:2018vuu,Closset:2018bjz,Apruzzi:2019vpe,Apruzzi:2019opn,Apruzzi:2019enx,Bhardwaj:2019jtr,Closset:2019juk,Bhardwaj:2020gyu,Morrison:2020ool,Albertini:2020mdx,Apruzzi:2021vcu}, and it would be interesting if it can also be applied to tackle some of the problems mentioned in this paper.} \cite{Aharony:1997bh,Aharony:1997ju}. As such for the most part, we shall concentrate on non-twisted compactifications of $5d$ SCFTs that can be engineered using brane webs in flat space (that is not in the presence of orientifold planes). The generalization of some of these results to twisted compactifications and to brane webs in the presence of orientifold planes is also known, and we may comment on this later.

The behavior of these $5d$ SCFTs under compactification seems to depend on their brane web representation. Specifically, the $5d$ SCFTs are observed to reduce to $4d$ SCFTs or IR free theories depending on the form of their brane web representation. First, we need to discuss what we mean by the form of the brane web representation. The issue here is that given a brane web, there are many different ways of generating different looking webs, that nevertheless describe the same $5d$ SCFT. Specifically, we can perform a global $SL(2,\Z)$ transformation, and also are free to move $7$-branes around\footnote{While moving $7$-branes around should result in the same interacting piece, sometimes the resulting configurations differ by a number of free hypermultiplets \cite{Zafrir:2016jpu}. We shall for the most part ignore this issue.}. As such we shall take the definition that a brane configuration is deemed to have a specific form if it can be massaged to that form via some of these motions. Alternatively, it is deemed not of this form if it is not possible to massage it in that form via some of these motions. We note that it may be difficult to determined the existence or lack of existence of such motions. Before elaborating on this, it is convenient to begin by discussing the rank $1$ case, which can be analyzed rather straightforwardly. This will allow us to set the stage for what to expect from compactifications of $5d$ SCFTs, and build upon when we discuss other cases.

\subsection{Recap of rank-1 5d SCFTs}

Most known rank $1$ $5d$ SCFTs can be realized by standard brane webs systems\footnote{The only known exception to this is the rank $1$ $5d$ SCFT recently discovered in \cite{Bhardwaj:2019jtr}. This has a brane web realization involving an O$7^+$ plane \cite{Kim:2020hhh}. We will not consider this case here.}. These can be generated by the compactification of the rank $1$ E-string theory on a circle \cite{Ganor:1996pc}. Specifically, we consider compactifying the $6d$ SCFT on a circle and taking the $R\rightarrow 0$, without any other scaling limits. This is expected to give a $5d$ SCFT of rank $1$ and with an $\ef_8$ global symmetry. This $5d$ SCFT is the $N_f=7$ case of the so called $E_{N_f+1}$ theories discovered in \cite{Seiberg:1996bd}. These are $5d$ SCFTs of rank $1$ and can be conveniently described as the UV completions of the $5d$ $\mathcal{N}=1$ gauge theory $SU(2)+N_fF$. Their global symmetry is $\ef_{N_f+1}$. From the $E_8$ SCFT one can flow to the SCFTs corresponding to lower $N_f$ via mass deformations until reaching the $E_2$ SCFT. At that stage one can flow to two theories, the $E_1$ and $\tilde{E}_1$ SCFTs. The $\tilde{E}_1$ SCFT can be further deformed to another $5d$ SCFT called the $E_0$ SCFT. This exhausts the rank $1$ SCFTs reachable in this manner.

We can next consider the reductions of these SCFTs to $4d$. Here we can analyze this explicitly by relying on the results of \cite{Ganor:1996pc}, which determined the structure of the Coulomb branch of the rank $1$ E-string theory compactified on a torus of finite area. Let's summarize these results. 

Upon compactification, the structure of the Coulomb branch is as follows. At the origin we have the singularity associated with the rank $1$ MN $E_8$ SCFT, which is thus the theory we get from the torus reduction of the rank $1$ E-string theory. Additionally there are two other $I_1$ type singularities, where a free hyper becomes massless, associated with the KK modes becoming massless. While both of these are type $I_1$ singularities, and in both a free hyper becomes massless, they differ by their monodromies. Specifically, if one is chosen so that the free hyper has unit electric charge, then at the other it will have unit magnetic charge. 

Here, as we consider reduction of the rank $1$ $5d$ $E_8$ SCFT on a circle, we need to take one of these $I_1$ singularities to infinity. The resulting Coulomb branch, describing the rank $1$ $5d$ $E_8$ SCFT on a circle, contains a singularity of type $II^*$ at the origin and one of type $I_1$ at a distance proportional to $R^{-1}$. We shall take this singularity to be the one where the free hyper has unit magnetic charge. This allows us to analyze the compactification and determine the resulting theory.

First, we consider reductions without scaling limits. If we reduce the rank $1$ $5d$ $E_8$ SCFT then we get the rank $1$ $4d$ MN $E_8$ theory, which is an SCFT. We can analyze the other cases by turning on mass deformations. This again follows the analysis in \cite{Ganor:1996pc}. For instance the $5d$ $E_8$ SCFT can flow to the $5d$ $E_7$ SCFT by turning on a mass deformation preserving the $\ef_7$ subgroup of $\ef_8$. On the Coulomb branch of the compactified theory, this corresponds to the deformation splitting the $II^*$ singularity to a $III^*$ singularity and an $I_1$ singularity. The limit where we get the $5d$ $E_7$ SCFT corresponds to taking this $I_1$ part to infinity, while the $R\rightarrow 0$ limit corresponds to taking the other $I_1$ singularity, associated with the KK modes, to infinity. We end up with only the $III^*$ singularity. This implies that the compactification of the rank $1$ $E_7$ SCFT on a circle leads to the rank $1$ MN $4d$ $E_7$ SCFT.

We can continue in this vein and flow from $E_7$ to $E_6$ and then to $E_5 = \sof(10)$. Each step requires separating an $I_1$ singularity. The result is that the reduction of the $5d$ rank $1$ $E_6$ SCFT is the $4d$ rank $1$ MN $E_6$ SCFT, while that of the $5d$ rank $1$ $E_5$ SCFT is the $4d$ IR free gauge theory $SU(2)+5F$. In this case we end up with an IR free gauge theory, that is not a $4d$ SCFT.

The fact that we can flow from the $E_8$ $5d$ SCFT to $E_7$, $E_6$ and then $E_5$ implies that we can have similar flows between the $4d$ SCFTs, that is flow from the MN $E_8$ theory to the $E_7$ one, following by the $E_6$ MN theory, and from there to the IR free $SU(2)+5F$ gauge theory. This matches the flow pattern observed using $4d$ methods, see for instance \cite{Argyres:2015gha}. In other words, here the two deformations given by the compactification and the flavor mass deformation commute. This is readily observed in the Coulomb branch of the compactified theory as both are associated with taking an $I_1$ singularity to infinity and the order in which these are taken does not affect the final theory. We shall later see evidence that this appears to work quite well also for many rank $2$ theories when the mass deformation starts from $4d$ SCFTs. This confirms with our expectations as the resulting $4d$ theories are all either SCFTs or IR free gauge theories.

Consider now the case of the compactification of the $5d$ $E_4$ SCFT. For this we consider the Coulomb branch of the $5d$ $E_5$ SCFT, and look for the deformation preserving the $\suf(5)$ subgroup of $\sof(10)$. The correct deformation is the one taking the $I^*_1$ singularity at the origin of the Coulomb branch (corresponding to the $SU(2)$ with five flavors) and separating it to two $I_1$ singularities and an $I_5$ singularity. This can be understood directly by looking at the gauge theory close to the origin, which we recall is $SU(2)+5F$. To preserve the $\suf(5)$ subgroup of $\sof(10)$, we need to give all the flavors the same mass. We then have a pure $SU(2)$ gauge theory at the origin, which gives the two $I_1$ singularities where we have the massless monopole and dyon. However, there is also a point on the Coulomb branch where the $5$ flavors become massless, and we get a $U(1)+5F$ gauge theory. This is the point associated with the $I_5$ singularity.

We next want to take the mass deformation such that we get a non-trivial Higgs branch supporting an $\suf(5)$ global symmetry. As previously, this implies that we need to take the $I_1$ singularity that was stripped from the $\sof(10)$ theory to infinity. However, in this case we have two of them, leaving us with the $I_5$ singularity, the additional $I_1$ singularity that was separated from it and the $I_1$ singularity associated with the KK modes. As we shall soon argue, these should be taken to have a free magnetic and a free dyonic hypermultiplet. Taking these two $I_1$ singularities to infinity as well leads to the $U(1)+5F$ $4d$ gauge theory. This theory indeed has the correct Higgs branch to be the result of the direct compactification of the $E_4$ SCFT.

Before moving on to the next case, a few comments are in order. First in this case, unlike the other cases, the singularity actually splits and moves away from the origin of the Coulomb branch. This means that in the reduction we may need to shift the origin of the Coulomb branch. This is not a significant issue. As we are interested in the compactification of $5d$ SCFTs, it is reasonable to fix the origin of the Coulomb branch to be at the point where the full Higgs branch can be accessed. This indeed correspond to the $I_5$ point. Alternatively, this can be described as compactifying with a scaling limit to the Coulomb branch vev. Nevertheless, as we have previously argued, if the Higgs branch contains disconnected components, then this may not be possible.

A second issue we wish to point out is the failure of commutativity of some of the mass deformations in this case. Specifically, the $5d$ $E_4$ SCFT has a mass deformation leading to the $SU(2)+3F$ gauge theory. From the previous discussion, we may expect for the resulting $4d$ theory to inherit this mass deformation. However, it is apparent that one cannot flow from the $U(1)+5F$ theory to the $SU(2)+3F$ gauge theory. We can understand the way this flow works out by considering the Coulomb branch of the $5d$ $E_4$ SCFT that we have previously motivated. Specifically, to get the $SU(2)+3F$ gauge theory we need to turn on a mass deformation breaking the $\suf(5)$ global symmetry to $\uf(1)\times \suf(4)$. In the Coulomb branch this corresponds to breaking the $I_5$ singularity to an $I_1$ and $I_4$ singularities. When performed on the $U(1)+5F$ theory this leads to $U(1)+4F$ theory living at the $I_4$ singularity, and the $U(1)+1F$ theory living at the $I_1$ singularity.

\begin{figure}[t]
\centering
\begin{adjustbox}{center,max width=1.1\textwidth}
\begin{tikzpicture}
[
auto,
good/.style={rectangle,rounded corners,fill=green!50,inner sep=2pt},
bad/.style={rectangle,rounded corners,fill=red!15,inner sep=2pt},
ugly/.style={rectangle,rounded corners,fill=blue!10,inner sep=2pt},
Green/.style={->,>=stealth[round],shorten >=1pt,line width=.4mm,ForestGreen},
Yellow/.style={->,>=stealth[round],shorten >=1pt,line width=.4mm,Goldenrod}
]
\begin{scope}[yshift=-0cm]
\node at (6,0) {{\large {\fontfamily{qcs}\selectfont\textsc{5d/4$d$  compactification commutativity}}}};
\end{scope}
%E8 compactification:
\begin{scope}[yshift=-2cm,xshift=-2cm] 
\fill[color=gray!10, rounded corners] (-1.5,1.5) rectangle (18,-7.3);

\node (top) at (-.7,.7) {};
\node (radius) at (-1.1,.2) {\Fgreen{\tiny{$1/R$}}};
\node (coupl) at (-.2,1) {\Fgreen{\tiny{$m_0=1/g^2$}}};
\node (down) at (0.5,.7) {};
\node (right) at (-.7,-.5) {};
\node (E8) at (0,0) [,align=center] {$\bred{E_8}$};
\node (IIsI1) at (0,-2) [,align=center] {$\{\bred{II^*},I_1^{S^1}\}$};
\node (SU27FI1) at (3,0) [,align=center] {$\{SU(2)+7F,I_1^{m_0}\}$};
\node (I3sI12) at (3,-2) [,align=center] {$\{I_3^*,I_1^{S^1},I_1^{m_0}\}$};
\draw[Green] (top) to (down) {};
\draw[Green] (top) to (right) {};
\draw[Yellow] (E8) to (IIsI1) {};
\draw[Yellow] (E8) to (SU27FI1) {};
\draw[Yellow] (SU27FI1) to (I3sI12) {};
\draw[Yellow] (IIsI1) to (I3sI12) {};
\end{scope}

%E7 compactification:
\begin{scope}[yshift=-2cm,xshift=4.5cm] 
\node (top) at (-.7,.7) {};
\node (radius) at (-1.1,.2) {\Fgreen{\tiny{$1/R$}}};
\node (coupl) at (-.2,1) {\Fgreen{\tiny{$m_0=1/g^2$}}};
\node (down) at (0.5,.7) {};
\node (right) at (-.7,-.5) {};
\node (E8) at (0,0) [,align=center] {$\bred{E_7}$};
\node (IIsI1) at (0,-2) [,align=center] {$\{\bred{III^*},I_1^{S^1}\}$};
\node (SU27FI1) at (3,0) [,align=center] {$\{SU(2)+6F,I_1^{m_0}\}$};
\node (I3sI12) at (3,-2) [,align=center] {$\{I_2^*,I_1^{S^1},I_1^{m_0}\}$};
\draw[Green] (top) to (down) {};
\draw[Green] (top) to (right) {};
\draw[Yellow] (E8) to (IIsI1) {};
\draw[Yellow] (E8) to (SU27FI1) {};
\draw[Yellow] (SU27FI1) to (I3sI12) {};
\draw[Yellow] (IIsI1) to (I3sI12) {};
\end{scope}

%E6 compactification:
\begin{scope}[yshift=-2cm,xshift=11cm] 
\node (top) at (-.7,.7) {};
\node (radius) at (-1.1,.2) {\Fgreen{\tiny{$1/R$}}};
\node (coupl) at (-.2,1) {\Fgreen{\tiny{$m_0=1/g^2$}}};
\node (down) at (0.5,.7) {};
\node (right) at (-.7,-.5) {};
\node (E8) at (0,0) [,align=center] {$\bred{E_6}$};
\node (IIsI1) at (0,-2) [,align=center] {$\{\bred{IV^*},I_1^{S^1}\}$};
\node (SU27FI1) at (3,0) [,align=center] {$\{SU(2)+5F,I_1^{m_0}\}$};
\node (I3sI12) at (3,-2) [,align=center] {$\{I_1^*,I_1^{S^1},I_1^{m_0}\}$};
\draw[Green] (top) to (down) {};
\draw[Green] (top) to (right) {};
\draw[Yellow] (E8) to (IIsI1) {};
\draw[Yellow] (E8) to (SU27FI1) {};
\draw[Yellow] (SU27FI1) to (I3sI12) {};
\draw[Yellow] (IIsI1) to (I3sI12) {};
\end{scope}

%E5 compactification:
\begin{scope}[yshift=-6.5cm,xshift=-2cm] 
\node (top) at (-.7,.7) {};
\node (radius) at (-1.1,.2) {\Fgreen{\tiny{$1/R$}}};
\node (coupl) at (-.2,1) {\Fgreen{\tiny{$m_0=1/g^2$}}};
\node (down) at (0.5,.7) {};
\node (right) at (-.7,-.5) {};
\node (E8) at (0,0) [,align=center] {$\bred{E_5}$};
\node (IIsI1) at (0,-2) [,align=center] {$\{I_1^*,I_1^{S^1}\}$};
\node (SU27FI1) at (3,0) [,align=center] {$\{SU(2)+4F,I_1^{m_0}\}$};
\node (I3sI12) at (3,-2) [,align=center] {$\{\bred{I_0^*},I_1^{S^1},I_1^{m_0}\}$};
\draw[Green] (top) to (down) {};
\draw[Green] (top) to (right) {};
\draw[Yellow] (E8) to (IIsI1) {};
\draw[Yellow] (E8) to (SU27FI1) {};
\draw[Yellow] (SU27FI1) to (I3sI12) {};
\draw[Yellow] (IIsI1) to (I3sI12) {};
\end{scope}

%E4 compactification:
\begin{scope}[yshift=-6.5cm,xshift=4.5cm] 
\node (top) at (-.7,.7) {};
\node (radius) at (-1.1,.2) {\Fgreen{\tiny{$1/R$}}};
\node (coupl) at (-.2,1) {\Fgreen{\tiny{$m_0=1/g^2$}}};
\node (down) at (0.5,.7) {};
\node (right) at (-.7,-.5) {};
\node (E8) at (0,0) [,align=center] {$\bred{E_4}$};
\node (IIsI1) at (0,-2) [,align=center] {$\{I_5,(I_1^{S^1})^2\}$};
\node (SU27FI1) at (3,0) [,align=center] {$\{SU(2)+3F,I_1^{m_0}\}$};
\node (I3sI12) at (3,-2) [,align=center] {$\{I_4,(I_1^{S^1})^2,I_1^{m_0}\}$};
\draw[Green] (top) to (down) {};
\draw[Green] (top) to (right) {};
\draw[Yellow] (E8) to (IIsI1) {};
\draw[Yellow] (E8) to (SU27FI1) {};
\draw[Yellow] (SU27FI1) to (I3sI12) {};
\draw[Yellow] (IIsI1) to (I3sI12) {};
\end{scope}

%E3 compactification:
\begin{scope}[yshift=-6.5cm,xshift=11cm] 
\node (top) at (-.7,.7) {};
\node (radius) at (-1.1,.2) {\Fgreen{\tiny{$1/R$}}};
\node (coupl) at (-.2,1) {\Fgreen{\tiny{$m_0=1/g^2$}}};
\node (down) at (0.5,.7) {};
\node (right) at (-.7,-.5) {};
\node (E8) at (0,0) [,align=center] {$\bred{E_3}$};
\node (IIsI1) at (0,-2) [,align=center] {$\{I_3,I_2,I_1^{S^1}\}$};
\node (SU27FI1) at (3,0) [,align=center] {$\{SU(2)+2F,I_1^{m_0}\}$};
\node (I3sI12) at (3,-2) [,align=center] {$\{I_2^2,I_1^{S^1},I_1^{m_0}\}$};
\draw[Green] (top) to (down) {};
\draw[Green] (top) to (right) {};
\draw[Yellow] (E8) to (IIsI1) {};
\draw[Yellow] (E8) to (SU27FI1) {};
\draw[Yellow] (SU27FI1) to (I3sI12) {};
\draw[Yellow] (IIsI1) to (I3sI12) {};
\end{scope}
\end{tikzpicture}
\caption{Graphical depiction of the behavior of the \emph{5d} rank-1 $E_n$ theories under both circle compactification (vertical direction) and mass deformation (horizontal direction) associated to \emph{5d} gauge coupling. Superconformal theories are labeled in red.}
\label{MapEnf}
\end{adjustbox}
\end{figure}
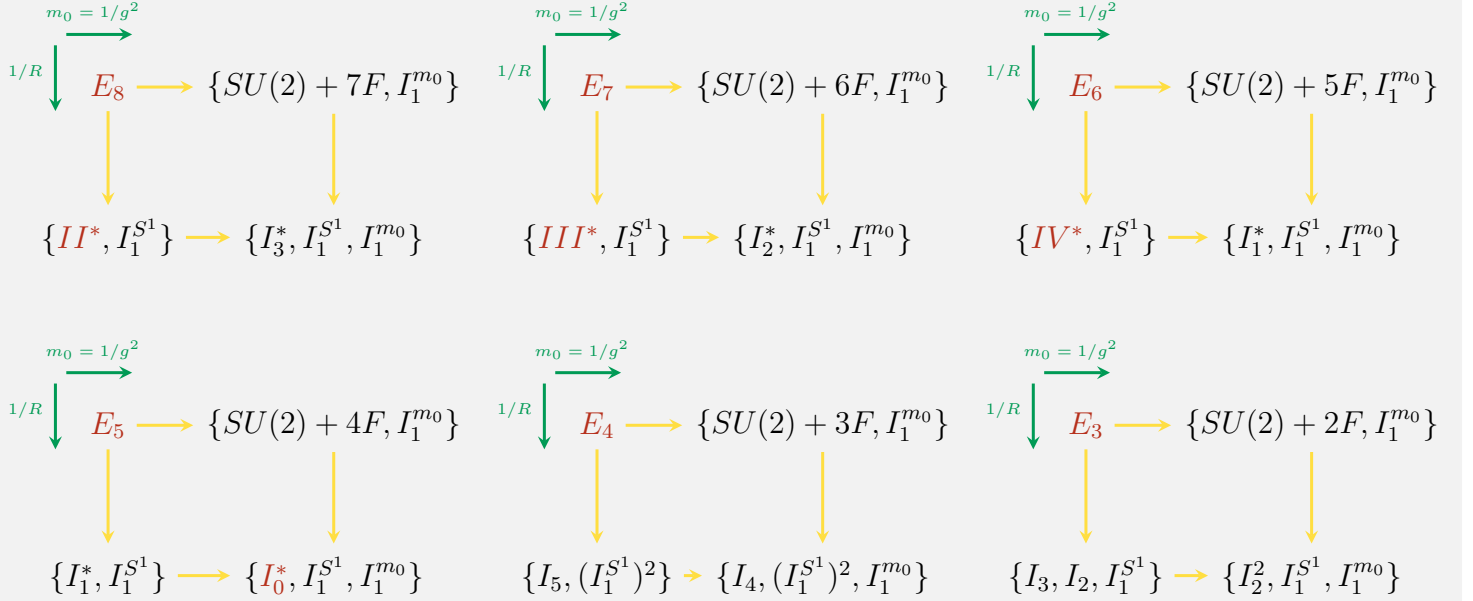

The Coulomb branch of the $SU(2)+3F$ theory has precisely two singularities of the same type, $I_1$ and $I_4$. However, while these are mutually local in the $U(1)+5F$ theory, they are not so for the $SU(2)+3F$ theory. However, recall that in the case with a finite circle we had two additional $I_1$ singularities which were not mutually local with the $I_5$ one. Therefore, we can describe the flow to the $SU(2)+3F$ theory as follows. We deform the $I_5$ singularity, breaking it to an $I_1$ and $I_4$ singularities, and then take the $I_1$ singularity to infinity. This is done while keeping the distance between the $I_4$ singularity and one of the other $I_1$ singularities, that is not mutually local with it, finite. This limit then indeed gives the structure of the Coulomb branch of the $SU(2)+3F$ theory. The issue then is that in this flow we need the additional $I_1$ singularities that were sent to infinity in the reduction to the $U(1)+5F$ theory. This again confirms with our expectations, and can be attributed to the fact that $SU(2)+3F$ theory is asymptotically free. Specifically, the distance between the $I_1$ and $I_4$ singularities is precisely related to the strong coupling scale of the $SU(2)+3F$ theory. 

An even more exotic behavior takes place in the case of the $5d$ $E_3$ SCFT. To see this, consider the mass deformation sending the $E_4$ SCFT to the $E_3$ SCFT. On the Coulomb branch this amounts to separating the $I_5$ singularity to an $I_3$, an $I_2$ and an $I_1$ singularity, and then sending the $I_1$ singularity to infinity. This leaves us with the $I_3$, $I_2$ and the two $I_1$ singularities we had before. When we take the limit of zero radius we expect to zoom in on one of them, determined by how we scale the Coulomb branch parameter as we perform the reduction, while sending the others to infinity. However, in this way we can get the theory corresponding to the $I_3$ singularity, \emph{i.e.} the $U(1)+3F$ gauge theory, or to the $I_2$ singularity, \emph{i.e.} the $U(1)+2F$. The first case manifests the $\suf(3)$ part of $E_3$, and the second manifests the $\suf(2)$ part of $E_3$, but there is no theory manifesting both. In other words, we can naturally get two different theories by reducing the $E_3$ SCFT on a circle, depending on how we scale the Coulomb branch parameters, but each will manifest a different part of the flavor symmetry, the other part acting trivially on it. This is starkly different from the previous cases where the theory manifests the full global symmetry and it is a realization of a phenomenon we introduced in our general discussion of section \ref{sec:GL}.

%To better understand this, it is convenient to look at the Higgs branch of the theory. This is as first the global symmetry acts on the Higgs branch. Second, the Higgs branch should be invariant under dimensional reduction. As such we would expect the 4$d$  theory to inherit the Higgs branch of the 5\emph{d}SCFT, implying that the global symmetry of the 5\emph{d}SCFT must act non-trivially on it. However, there is a loophole in this argument that is exploited in the case of the $E_3$ SCFT to lead to the result we mentioned.

This phenomenon comes about because the Higgs branch of the $E_3$ SCFT is made of two distinct cones, one supporting the 1-instanton moduli space of $\suf(3)$, and the other supporting the 1-instanton moduli space of $\suf(2)$. We note that the $\suf(3)$ part of the global symmetry acts only on one of them and similarly for the $\suf(2)$ part. These two branches touch one another and the Coulomb branch at the SCFT point, as dictated by superconformal symmetry. As we noted previously, while the structure of the Higgs branch should be invariant under quantum corrections, the property that the origin of the two branches intersect at a single point is not invariant. A well known example of this is the case of the 4$d$  $SU(2)+2F$ gauge theory, which classically has a Higgs branch consisting of two branches, intersecting at a point, but which are separated by quantum corrections\cite{Seiberg:1994aj}. This is not coincidental as its 5\emph{d} version is the theory that has the $E_3$ SCFT as its UV completion, and these two branches are embedded in those of the 5\emph{d} SCFT.

Consider the compactification of the $E_3$ SCFT on a circle, in light of what was said so far. We noted that conformal symmetry dictates that the two branches intersect at the origin of the Coulomb branch. However, that property is not protected from corrections. Once we compactify, we break conformal invariance, and there is no longer any reason for it to hold. As such we expect it to change and that the points of intersection with the Coulomb branch of the two branches to be separated by a distance proportional to the inverse radius. Therefore, in the radius goes to zero limit the distance between them becomes infinite and we can only stay with one of them. This leads to the picture we seen previously.  

We can continue in this vein and consider other cases, but at this point it is apparent that these will give IR free theories so we will not continue further. Instead we shall use what observed here to motivate some general criteria for when a 5\emph{d} SCFT reduces to a 4$d$  SCFT, that were observed experimentally.

We noted that of the rank 1 theories in the $E_{N_f+1}$ family, only the cases of $N_f=5,6$ and $7$ reduce to 4$d$  SCFTs. We can then ask what does these theories have in common? One interesting property that they share is the shape of their brane web representation. All these theories can be represented by a brane web that is made from a collection of $5$-brane junctions of $(1,0)$, $(0,1)$ and $(1,1)$ 5-branes \cite{Benini:2009gi}. These end on their respective $7$-branes, where we allow for multiple $5$-branes to end on the same $7$-branes. In fact this class of $5d$ SCFTs is known for quite some time to reduce to 4$d$  class S theories, which are $\mathcal{N}=2$ SCFTs.

It appears that this class of 5\emph{d} SCFTs is the only one that reduces to 4$d$  SCFTs if we restrict our attention to only non-twisted reductions of 5\emph{d} SCFTs that can be engineered via standard brane webs without orientifold planes. This is indeed consistent with what we observed previously for the rank 1 cases. This is mostly an experimental observation, and we shall elucidate some of the evidence in favor of it. However, we do note that it is quite possible for counterexample to exist that simply lie outside the cases worked out so far.

\subsection{Higher rank 5d SCFTs - brane web untwisted}

The behavior of rank 1 theories, combined with properties of RG flows for theories with 8 supercharges that we mentioned previously, allows us to identify the general behavior of a larger class of theories: 5\emph{d} SCFTs which can be Higgsed to one of the $E_{N_f+1}$ rank 1 theories. In particular we wish to identify those which reduce to 4$d$ SCFTs. As we mentioned, the Higgs branch is invariant under dimensional reduction\footnote{Here we are assuming that the reduction is done so that the Higgs branch intersects the 4$d$  Coulomb branch at the origin.}, so these putative 4$d$  SCFTs should also have a Higgs branch leading to 4$d$ reductions of the rank 1 $E_{N_f+1}$ theories the 5$d$ SCFTs reduce to\footnote{Here we assume that Higgs branch flows and dimensional reduction commute. This is supported by the string theory picture using branes, at least for the type of dimensional reduction we consider here.}. As we already mentioned, a special property of Higgs branch flows is that they preserve the $\mathcal{N}=2$ $\uf(1)_R$ R-symmetry. As such the resulting 4$d$  theory must also be an SCFT. However, we have noted that this is only true if the 5$d$ rank 1 SCFT is of type $E_6$, $E_7$ or $E_8$ and we thus conclude that 5\emph{d} SCFTs that can be Higgsed to a rank 1 $E_{N_f+1}$ SCFT for $N_f<5$ should not reduce to 4$d$  SCFTs. An interesting property of the family of 5\emph{d} SCFTs we previously mentioned, that is the one made from the junction of $(1,0)$, $(0,1)$ and $(1,1)$ $5$-branes, is that the Higgs branch keeps you within this family. %As such these are indeed the natural cases obeying this requirement.

This also points out about potential counterexamples. Specifically, cases that cannot be Higgsed to any rank 1 $5d$ SCFT are one such case. There are indeed known 5\emph{d} SCFTs that either don't have an Higgs branch, or where the Higgs branch ``stops'' with theories of higher rank never reaching any of the rank-1 theories. Another potential option is that there might be cases Higgsable only to rank 1 $E_{N_f+1}$ SCFTs for the mentioned values of $N_f$, but whose brane web is not of the form discussed, though we are not aware of any candidate for this.

We have noted that for a particular class of 5\emph{d} SCFTs, one describable by a collection of 5-brane junctions, the 4$d$  reduction is known. There are several other cases of families of 5\emph{d} SCFTs where there is a proposal for the expected 4$d$  theory passing some tests. These tests usually involve the matching of the structure of the Higgs branch and the dimension of the Coulomb branch, which are invariant under dimensional reduction. Next, we shall look at some of these cases and see how these illustrate our previous discussion.

The first case involves 5\emph{d} SCFTs that can be engineered as the low-energy theories on the intersection of NS and D5-branes. Here we again take the 5-branes to end on their respective 7-branes, and allow multiple 5-branes to end on the same 7-brane, as long as there are at least two 7-branes on each of the 4 sides\footnote{If this is not true, that is all the 5-branes end on the same 7-brane on at least one side, then one can show that this configuration can be deformed to a case of the previously discussed family.}. In this case the proposal for the 4$d$  theories resulting from the compactification of these theories appeared in \cite{Ohmori:2015pia}. These are made from a gauging of two class S theories. The exact form is not important to us here, but we do note that the gauging is such that the theory is IR free, thus these theories are not SCFTs.

It is interesting to note that the $E_5$ SCFT belongs to this class, and furthermore all theories in this class can be Higgsed down to the rank 1 $E_5$ SCFT giving a consistent picture with what previously observed. Finally, we note that in certain cases these theories may be Higgsed also to $E_{N_f+1}$ SCFTs with $N_f>4$. This does not contradict our discussion, as the $\uf(1)_R$ symmetry of the resulting 4$d$  SCFT can be viewed as an accidental enhancement.  

The results thus far derived, can be generalized also to brane systems that can be generically Higgsed to the rank $1$ $E_4$ SCFT, and potentially also to other $E_{N_f+1}$ SCFTs with $N_f>3$. These are described by the combination of the two previous systems, that is the intersection of a system of NS$5$-branes, D$5$-branes and $5$-brane junctions of NS$5$-branes, D$5$-branes and $(1,1)$ $5$-branes. Like the previous cases, we allow multiple 5-branes to end on the same 7-brane, but some restrictions are needed so that the theory cannot be deformed to one of the previously studied cases. We can use the discussion in \cite{Ferlito:2017xdq} to form a reasonable conjecture for the $4d$ theory expected to result from the compactification of theories in this class. These are in general IR free gauge theories so are not SCFTs, again as excepted from our general argument.

Finally, we note that these classes are generally related by mass deformations, that is the theories in the first class can first be deformed to other theories in the same class, and eventually to theories in the second class. Theories in the second class can be mass deformed to theories in the third class. In the rank $1$ case, this is the well known flow pattern: $E_8\rightarrow E_7\rightarrow E_6 \rightarrow E_5\rightarrow E_4 \rightarrow ...$, with the first three theories belonging to the first class, and the fourth and fifth theories belonging to the second and third classes, respectively. This flow pattern seems to hold generically, for theories in this classes.

We can try to continue and consider theories that can be Higgsed to the $E_3$ SCFTs, and potentially also cases with $N_f>2$. However, the Higgs branch of theories in this class usually contains multiple branches, so following our previous discussion we expect to get multiple interesting $4d$ theories depending on how we tune the Coulomb branch parameters. This makes the study of reductions of these theories more complicated.

\subsection{Remarks about other classes of 5d SCFTs}

So far we concentrated on the case of untwisted compactifications of 5\emph{d} SCFTs that can be engineered via brane webs without any orientifold planes. While, this is probably the most well understood case, there are many other cases. We shall for the most part concentrate on 5\emph{d} SCFTs with a brane description, that can facilitate in the study of this problem. It would be interesting to try to extend this study also to cases with no known brane description, for instance by relaying on constructions involving geometric engineering. 

\subsubsection*{Cases involving brane webs with orientifold planes}

The first case we consider are 5\emph{d} SCFTs that have a brane description involving orientifold planes. To preserve supersymmetry the orientifold planes that we can use are either an O$7$ or an O$5$ planes. Constructions involving the former were studied in \cite{Bergman:2015dpa}, while those involving the latter were studied in \cite{Brunner:1997gk,Zafrir:2015ftn,Hayashi:2015vhy}. The behavior depends on the specific orientifold used, and we shall next briefly discuss each case in turn.

Under quantum corrections, the O$7^-$ plane decomposes to two $7$-branes\cite{Sen:1996vd}. This allows us to map configurations with O$7^-$ planes to those without them, that we previously looked at. As such we do not get any new cases here.

This is not true for configurations involving an O$7^+$ plane, which does lead to a different class of systems. Unfortunately, not much is known about the 4$d$  reductions of theories in this class. As an example of theories, we have the 5\emph{d} SCFTs that UV complete the 5\emph{d} gauge theories $SO(N)+nV$, where to have a 5\emph{d} fixed point we need to have that $n<N-2$ \cite{Bergman:2015dpa}. These can usually be Higgsed to one another until one reaches the 5\emph{d} SCFTs that UV complete the 5\emph{d} gauge theories $SO(N-n)$ \footnote{Here we are using the Higgs branch direction that are not lifted by the gauge theory deformation. It is possible that the 5\emph{d} SCFT possesses other interesting Higgs branch flows that are lifted when we turn on the mass deformation.}. For $n=N-3$, these theories can be Higgsable down to a rank 1 case, which is expected to be the $E_1$ SCFT. As such our previous discussion would suggest that these do not lead to 4$d$  SCFTs. Cases with smaller $n$ can be reached by mass deformations from the $n=N-3$ case, suggesting these reduce to 4$d$  theories that can be generated from mass deformations of 4$d$  theories that are not SCFTs, though that is not enough to rule these theories out as SCFTs. More complicated systems can usually be Higgsed to this class of theories. Thus, there are some indications that theories in this class do not give interesting 4$d$  SCFTs though this analysis is by no means complete.

Finally, we have the case of configurations involving O$5$-planes. Unfortunately, this case has some issues making the discussion more complicated. First the behavior of the system here depends on the asymptotic charges of the O$5$-planes at the two sides, specifically, whether these have an O$5^-$ or O$5^+$-planes\footnote{These configurations generally involve an O$5$-plane intersected by NS$5$ charge carrying $5$-brane. The latter causes the type of the O$5$-plane to change from $+$ to $-$ and vice versa\cite{Hanany:2000fq}, which leads to the option for the asymptotic O$5$-plane charges to be different on the two sides.}. Of these, the cases where we have an O$5^+$ on one of the sides are relatively unstudied. There are, however, some known results for the case where we have an O$5^-$ plane on both sides. Interestingly these configurations also contain systems that provide an alternative realization of the rank 1 $E_{N_f+1}$ SCFTs. As such, much of what we said about the previous cases apply also to here. Specifically, there are known brane web configurations that are thought to reduce to 4$d$  SCFTs\cite{Zafrir:2016jpu} (see also \cite{Ohmori:2015pia}), here of D type class S, and also cases thought to reduce to specific 4$d$  IR free theories\cite{Ohmori:2016ahw}, that are constructed from an IR free gauging of two class S theories. Unfortunately, the configurations are more involved so we will not repeat them here. However, this is far from a complete answer. For instance, the brane configuration with the O$5$-plane describing the rank 1 $E_6$ SCFT is not of the type considered in \cite{Zafrir:2016jpu}, suggesting that there might be other interesting cases. 

\subsubsection*{Twisted compactifications}

We can consider also twisted compactifications of various systems. Here there is the added complications that one must first figure out the discrete symmetries of 5\emph{d} SCFTs that commute with supersymmetry. There is not much known about this, with the exception of one case that is relatively well studied. This was first considered in \cite{Zafrir:2016wkk}, which we next review.

This case involves 5\emph{d} SCFTs that can be engineered via a standard brane web configuration, that is without orientifold planes. We can then consider such brane configurations that are invariant under the combination of an $SL(2,\mathbb{Z})$ transformation and a rotation in the plane of the web. Specifically, the $SL(2,\mathbb{Z})$ transformation is taken to be in a $\mathbb{Z}_k$ discrete subgroup, and the rotation then is with angle $\frac{2\pi}{k}$. The former implies that $k=2,3,4$ or $6$. The claim is that 5\emph{d} SCFTs whose brane realization possess this symmetry have a $\mathbb{Z}_k$ discrete symmetry that commutes with supersymmetry. Furthermore, as the symmetry is manifested in the brane construction, we can use the brane realization to understand how this symmetry acts on the global symmetry and moduli space of the 5\emph{d} SCFT. This in turn can provide us with useful information about the resulting 4$d$  theory.

We can then raise the question of when do these types of twisted compactifications give 4$d$  SCFTs. This can be tackled in the same way as previously, specifically, by studying low-rank cases, and figuring out which cases reduce to 4$d$  SCFTs. The logic then is that cases that can only be Higgsed to cases that reduce to 4$d$  SCFTs can be SCFTs themselves. While, this does not prove that these reduce to 4$d$  SCFTs, it is a strong indication that they do. Additionally, the brane construction can be used to infer various dualities between the gaugings of these theories. In some cases, one side can be related to known theories, which again strongly supports the claim that these are 4$d$  SCFTs. Even if neither side can be identified with known theories, the presence of such dualities suggests that these are 4$d$  SCFTs.

Like the previous cases, this analysis tends to lead to families of theories, sharing a specific form of their brane webs representation, that should reduce to 4$d$  SCFTs. They can be conveniently organized in flow families, where there is a top theory, and the other theories in the family can be generated via mass deformations. 

We next give a brief review of the type of configurations expected to reduce to 4$d$  SCFTs. This is based on the study of rank 1 cases done in \cite{Giacomelli:2020gee}. As such, it is not necessarily complete, and we shall comment about some open questions at the end. One issue is the case of $k=6$, which does not appear for rank 1, and we shall comment about it at the end.

First, let us consider the case of $k=2,3,4$. We can organize the theories into families starting from a top $5d$ SCFT, possessing the required discrete symmetry, from which the other cases in the family can be generated via mass deformations that preserve the discrete symmetry. This top $5d$ SCFT can be generated from the compactification of a $6d$ $(1,0)$ SCFT, and the $4d$ we get from its twisted compactification can also be realized by the compactification of the $6d$ $(1,0)$ SCFT on a torus with two almost commuting holonomies\cite{Ohmori:2018ona}. We refer the reader to the reference for the exact form of the relation, and for the form of the brane web representation of the top $5d$ SCFT for these cases.

Additional theories can be generated via mass deformations from the top $5d$ SCFT. The analysis of the rank $1$ case suggests that these behave similarly to the $E_{N_f+1}$ family with the $E_8$ playing the role of the top $5d$ SCFT. That is we can perform up to two mass deformations before we stop getting interesting theories.

The case of $k=2$ has been studied extensively in \cite{Zafrir:2016wkk}, where the properties of some of the theories in this class have been worked out. We refer the reader to the reference for the explicit form of the brane webs for the cases expected to reduce to $4d$ SCFTs. The case of $k=3$ as also been partially studied in \cite{Zafrir:2016wkk,Ohmori:2018ona}, though not to the same extent as the $k=2$ case. The case of $k=4$ remains relatively unstudied, apart from the cases of the top $5d$ SCFTs. 

This leaves the case of $k=6$. Here there is no rank $1$ case, and the first case happens at rank $2$. This case seems to reduce to the $\mathcal{N}=4$ $G_2$ $4d$ SCFT. As such in this case it seems that only the top theory gives interesting cases, though this requires further checking.

\subsubsection*{Other cases}

Finally, there are a few cases that do not fit nicely into the cases that we previously discussed. Most of these involve twisted compactifications of $5d$ SCFTs whose manifestation in the brane description is not apparent. We shall next describe some of these cases.

We previously discussed the case of twisted compactifications of $5d$ SCFTs and noted their relation to the compactification of $6d$ $(1,0)$ SCFT on a torus with two almost commuting holonomies. However, there is a family of $6d$ $(1,0)$ SCFTs that can be compactified on a torus with two almost commuting holonomies to get $4d$ SCFTs, but don't have a $5d$ realization of the type we described. The issue is that these cases would naively give a case with $k=5$, but $\mathbb{Z}_5$ is not a subgroup of $SL(2,\mathbb{Z})$, and so it cannot be manifested in the way we described. However, the arguments leading to the relation between the two compactifications should still hold. We thus expect the $4d$ SCFTs to also be given by the $\mathbb{Z}_5$ twisted compactification of the $5d$ SCFTs, but here it is not immediately apparent how the discrete symmetry is manifested in the brane system.

Additionally, during the course of this work, we found several cases of known $4d$ SCFTs whose properties can be nicely explained if they originate from the twisted compactification of $5d$ SCFTs. However, as in the $\Z_5$ case just discussed, it is not clear how the discrete symmetry involved in the twist is realized in the brane system. 

%In 5\emph{d}the gauge coupling is irrelevant and, relatedly, all gauge theories IR-free. Thus there is a clear sense in which $1/g^2$ is a relevant operator initiating the flow connecting the SCFT, in this case the $E_n$ rank-1, in the UV and the gauge theory description, in this case $SU(2)+(n-1)F$, in the IR. In 4$d$  the situation is obviously more complex as $\cN=2$ gauge theories can be anything from IR-free, to superconformal to asymptotically free. In the latter case, dimensional transmutation makes it such that the gauge theory description ceases to be meaningful in the IR and the RG-flow from the UV fixed point to the IR description has to account for this.

\section{Compactification from 5d and mass deformations}
\label{sec:def}

In this section we discuss the results of specific compactifications of 5\emph{d} SCFTs leading to 4$d$  $\mathcal{N}=2$ SCFTs of rank 2.

\subsection{Direct compactifications of 5d SCFTs with simple brane webs}

We shall start with the case of untwisted compactifications of 5\emph{d} SCFTs that have a simple brane web construction. From our previous discussion, these are the 5\emph{d} SCFTs whose brane webs are of the form of a collection of $(1,0)$, $(0,1)$ and $(1,1)$ $5$-brane junction. 

These types of compactifications account for the entire $\ef_8-\sof(20)$ series, in the notation of \cite{Martone:2021ixp}. This works as follows. Four dimensional Rank-2 $E_{N_f+1}$ Minahan-Nemeschansky (MN) theories \cite{Banks:1996nj,Sen:1996vd,Minahan:1996fg,Minahan:1996cj} can be obtained by compactifying the 5\emph{d} SCFTs UV completing $USp(4)+1\AS+N_f F$ and $SU(3)_{\frac{11-N_f}2}+(N_f+1)F$ gauge theories. In 5\emph{d} the $SU(2)^0 \times SU(2)- 5F$, where the superscript indicate the value of the $\theta$ angle which is physical in the absence of flavors, at strong coupling goes to an SCFT with flavor symmetry $\ef_8$. By computing its Higgs branch via the magnetic quiver, it is straightforward to infer that the compactification of this 5\emph{d} theory to 4$d$ should give the $[\ef_8]_{20}$ theory. There is an intricate inner relation between $USp(4)$, $SU(3)$ and $SO(4)=SU(2) \times SU(2)$ gauge theories in 5$d$. Specifically, in addition to the $USp(4)$ and $SU(3)$ descriptions, the 5\emph{d} rank 2 MN $E_8$ theory also has the description as the UV completion of the $1F- SU(2) \times SU(2) -5F$ gauge theory. Integrating the flavor on the left, if done so that the theta angle of the left $SU(2)$ after the flavor is integrated out is $\pi$, is dual to integrating out a fundamental flavor from the $USp(4)$ description. This then leads to the 5\emph{d} rank 2 MN $E_7$ SCFT. Finally, if you integrate the flavor on the left, such that the theta angle of the left $SU(2)$ after the flavor is integrated out is 0, you get precisely the 5\emph{d} SCFT with $\ef_8$ global symmetry which we just discussed, that reduces to the 4$d$  SCFT $[\ef_8]_{20}$. We therefore claim that the two 4$d$  SCFT are related by mass deformations.

%\footnote{To be precise, the compactification of the 5\emph{d}SCFT gives in 4$d$  an extra free hypermultiplet contribution which doesn't affect in any way the substance of our discussion.}

Let's now perform a similar analysis with the $\sof(20)_{16}$, $\sof(16)_{12}\times\suf(2)_8$ and $\sof(14)_{10}\times \uf(1)$ in 4$d$  by compactifying the 5\emph{d} SCFTs which UV complete $SU(3)_{\frac12}+9F$, $SU(3)_{1}+8F$ and $SU(3)_{\frac32}+7F$ respectively and which are all connected by decoupling a $F$ with a positive real mass. This immediately shows that there is a mass deformation between the $E_8$ MN theory and the $\sof(14)_{10}\times \uf(1)$ which corresponds on the gauge theory description to decoupling one flavor with a negative real mass. 
Finally the $\suf(10)_{10}$ and $\suf(2)_6\times \suf(8)_8$ can be obtained from the compactification of the 5\emph{d} SCFTs which UV complete an $SU(3)_{0}+8F$ and a $SU(3)_{\frac12}+7F$ respectively. This last observation suggests that these two four dimensional theories ought to be related by a mass deformation, which from the gauge theory perspective in 5$d$, is realized as a positive real mass for one of the $F$. Similarly a negative real mass from the 5\emph{d} gauge theory perspective shows that there exists a mass deformation connecting the $\sof(20)_{16}$ with $\suf(10)_{10}$ as well as the $\sof(16)_{12}\times \suf(2)_8$ with the $\suf(2)_6\times \suf(8)_8$. 

An immediate check that these compactifications work out the way we just described, is matching the Higgs branches of the original 5\emph{d} SCFTs with those of the resulting 4$d$  theories. Since we are here considering the case where no twist is introduced, the Higgs branch is unaffected by the compactification and the check is straightforward. The Higgs branches the 5\emph{d} theories are described in detail in \cite{vanBeest:2020civ} while those of the 4$d$  theories are collected in \cite{Martone:2021ixp}. All in all, the mass deformations which we have just derived are reproduced in the top left corner of figure \ref{MapR2}.

Notice that the CB of the 4$d$  SCFTs which we have just discussed, have a known integrable system description which is another perspective allowing to study the intricate network of mass deformations. Curiously our analysis points out the existence of more deformations than those seen from the integrable system perspective, see for example \cite{kawakami2018fourdimensional}. It would be interesting to understand whether our predicted flows, can also be seen directly as deformations of the corresponding integrable systems.

\subsection{Direct compactifications of 5d SCFTs with brane webs involving orientifolds}

Here we discuss untwisted compactifications of 5\emph{d} SCFTs that have a brane web construction involving orientifold planes. Following our previous discussion, we shall here take the more well understood cases were the orientifold is of type O$5^-$ at both ends, and further restrict to the cases discussed in \cite{Zafrir:2016jpu}, for which there is evidence that they reduce to $4d$ SCFTs. As we noted, there might be other brane configurations that reduce to $4d$ SCFTs, but for simplicity we shall concentrate on the more well understood cases.

The $5d$ SCFTs that we get from these constructions are generically the UV completions of $USp(2N)$ groups with fundamental matter and $SO(N)$ groups with vector and spinor matter. For the case of rank $2$ these mostly give cases that were covered previously, and are related to D type class S realizations of $4d$ SCFTs that are also realizable from A type class S. There are two exceptional cases, though, corresponding to $USp(4)$ with two antisymmetric hypermultiplets and fundamental matter, and $G_2$ with fundamental matter, which we can get by Higgsing down an $SO(7)$ gauge theory with spinor matter \cite{Hayashi:2018bkd}. These are in fact dual to one another, that is the $5d$ SCFTs that UV completes $G_2+(N_f+2)F$ and $USp(4)^{\pi}+2AS+N_fF$ are the same\footnote{This holds for $3\geq N_f \geq 0$. The superscript denotes the $\theta$ angle when $N_f=3$.} \cite{Jefferson:2018irk}. Of these, the $N_f=3$ case has a brane description of the form considered in \cite{Zafrir:2016jpu}. It reduces to the $\spf(12)_{11}$ $4d$ SCFT.

The case of $N_f=2$ is of the form that is expected to reduce to an IR free theory from \cite{Ohmori:2018ona}, and as such this case exhaust the new $4d$ SCFTs that we can get from this class of theories. We do note, however, that we can in principle flow from the $5d$ SCFT associated to the $N_f=3$ case to the one associated with $N_f=2$, which should give an IR free theory, and from their to the gauge theory description, but now in $4d$. This implies that $\spf(12)_{11}$ $4d$ SCFT can be mass deformed to the $\spf(4)_5 \times \sof(4)_8$ and $\spf(8)_7$ SCFTs, though the deformation goes through IR free theories.  

\subsection{Twisted compactifications of 5d SCFTs with simple brane webs}

Here we discussed the rank $2$ $4d$ SCFTs that can be realized by the twisted compactifications of 5\emph{d} SCFTs with a description in terms of a brane web without orientifold planes. We shall further concentrate only on the cases where the symmetry is manifested as in \cite{Zafrir:2016wkk}. We shall discuss other cases in the next subsection. We shall separate our discussion based on the order of the twist.

\subsubsection{$\mathbb{Z}_2$ twist}

We start with the case where the twist is a $\mathbb{Z}_2$ symmetry. This accounts for the flow lines emanating from the $\suf(2)_9{\times}\suf(6)_{16}$, $\suf(2)_{26}{\times}\spf(8)_{13}$, $\spf(14)_9$ $4d$ SCFTs, and for most of the entries in the $\spf(12)_8$ and $[\ff_4]_{12}{\times }\suf(2)^2_7$ families.

Specifically, the $\suf(2)_{26}{\times}\spf(8)_{13}$ and $[\ff_4]_{12}{\times }\suf(2)^2_7$ flow lines are associated with the rank $2$ S and T theories, respectively, which were introduced in \cite{Giacomelli:2020gee}. In terms of compactifications of $5d$ SCFTs, the $\suf(2)_{26}{\times}\spf(8)_{13}$ $4d$ SCFT is given from the compactification of the $5d$ SCFT UV completing the $5d$ gauge theory $SU(5)_0+2AS+6F$, where the compactification is done with a twist in a $\mathbb{Z}_2$ global symmetry. The latter acts on the gauge theory as charge conjugation. Similarly, the $\suf(2)_{18} {\times}\suf(2)_{16} {\times} \spf(4)_{9}$ and $\suf(2)_{7} {\times} \suf(2)_{14} {\times} \uf(1)$ $4d$ SCFTs are given by the compactification of the $5d$ SCFTs UV completing the $5d$ gauge theories $SU(5)_0+2AS+4F$ and $SU(5)_0+2AS+2F$, respectively. In these cases as well the compactification is done with a twist in a $\mathbb{Z}_2$ global symmetry, which acts on the gauge theory as charge conjugation. The mass deformations between them are then naturally explained as $\mathbb{Z}_2$ preserving mass deformations between the corresponding SCFTs. 

The $[\ff_4]_{12}{\times }\suf(2)^2_7$ $4d$ SCFT is given from the compactification of the $5d$ SCFT UV completing the $5d$ gauge theory $SU(4)_0+2AS+6F$. Again the compactification is done with a twist in a $\mathbb{Z}_2$ global symmetry, which acts on the gauge theory as charge conjugation. Similarly, the $\sof(7)_{8}{\times} \suf(2)^2_{5}$ and $\suf(3)_{6} {\times} \suf(2)^2_{4}$ $4d$ SCFTs are given by the compactification of the $5d$ SCFTs UV completing the $5d$ gauge theories $SU(4)_0+2AS+4F$ and $SU(4)_0+2AS+2F$, respectively, with the compactification done with a $\mathbb{Z}_2$ twist. Additionally, the $SU(4)_0+2AS+6F$ $5d$ gauge theory has a dual description in terms of an $1F+SU(2)\times [SU(2)+2F]\times SU(2)+1F$ $5d$ gauge theory, where the $\mathbb{Z}_2$ symmetry acts as a reflection of the quiver. Similarly, the $SU(4)_0+2AS+4F$ $5d$ gauge theory has a dual description in terms of an $SU(2)^{\pi}\times [SU(2)+2F]\times SU(2)^{\pi}$ $5d$ gauge theory. However, the $5d$ gauge theory $SU(2)^{0}\times [SU(2)+2F]\times SU(2)^{0}$ has no $SU(4)$ dual, and compactifying it with a twist in its quiver reflection discrete symmetry gives the $[\ff_4]_{10}{\times} \uf(1)$ $4d$ SCFT. The relation between the various gauge theory descriptions of the $5d$ SCFTs suggests relations between the corresponding $4d$ SCFTs given by mass deformations.

The $\spf(12)_8$ and $\spf(14)_9$ are generated from the compactification of the $5d$ SCFTs UV completing the $5d$ gauge theories $SU(4)_0+10F$ and $SU(5)_0+12F$, respectively. As in the previous cases, the compactification is done with a twist in a $\mathbb{Z}_2$ discrete symmetry that acts on the gauge theories as charge conjugation. Other $4d$ SCFTs in these families are generated from similar twisted compactifications of $SU(4)$ and $SU(5)$ gauge theories with smaller number of fundamental flavors. Specifically, the $\spf(8)_6 \times \suf(2)_8$ and $\spf(6)_5{\times}\uf(1)$ $4d$ SCFTs are given by the twisted compactification of the $5d$ SCFTs associated with the gauge theories $SU(4)_0+8F$ and $SU(4)_0+6F$, respectively. Similarly, the $\spf(10)_7{\times} \suf(2)_8$ and $\spf(8)_6 \times \uf(1)$ $4d$ SCFTs are given by the twisted compactification of the $5d$ SCFTs associated with the gauge theories $SU(5)_0+10F$ and $SU(5)_0+8F$, respectively. Once again, theories in this family are related by mass deformations, which manifest similar relations between the $5d$ SCFTs. Additionally, some theories in this family are expected to be related to theories in the previous families, where the relation is seen as integrating the antisymmetric hypermultiplets in the $5d$ gauge theories that the corresponding $5d$ SCFTs UV complete.

Finally, we have the $\suf(2)_9{\times}\suf(6)_{16}$ family. The $\suf(2)_9{\times}\suf(6)_{16}$ theory can be generated by the compactification of the $5d$ SCFT UV completing the $5d$ gauge theory $4F+SU(3)_{\frac{1}{2}} \times SU(3)_{-\frac{1}{2}}+4F$. The compactification is done with a $\mathbb{Z}_2$ twist that acts on the gauge theory as a combination of charge conjugation and quiver reflection. The other cases have similar constructions, but with other $5d$ SCFTs related to the previous one by mass deformations. Specifically, for the $4d$ SCFT $\suf(3)_{10} {\times} \suf(3)_{10}{\times}\uf(1)$ the parent $5d$ SCFT is the one UV completing the gauge theory $3F+SU(3)_{0} \times SU(3)_{0}+3F$. For the $4d$ SCFT $\suf(4)_{12} {\times} \suf(2)_{7}{\times} \uf(1)$ the parent $5d$ SCFT is the one UV completing the gauge theory $3F+SU(3)_{1} \times SU(3)_{-1}+3F$. For the $4d$ SCFT $\suf(3)_{10} {\times} \suf(2)_{6}{\times} \uf(1)$ the parent $5d$ SCFT is the one UV completing the gauge theory $2F+SU(3)_{\frac{3}{2}} \times SU(3)_{-\frac{3}{2}}+2F$. For the $4d$ SCFT $\suf(2)_{8}{\times} \suf(2)_{8}{\times} \uf(1)^2$ the parent $5d$ SCFT is the one UV completing the gauge theory $2F+SU(3)_{\frac{1}{2}} \times SU(3)_{-\frac{1}{2}}+2F$. Like in the previous cases, the mass deformation relations between the $5d$ SCFTs imply similar relations between the associated $4d$ SCFTs.  

\subsubsection{$\mathbb{Z}_3$ twist}

Next, we consider the case where the twist is in a $\mathbb{Z}_3$ discrete symmetry. This accounts for the flow lines emanating from the $\suf(3)_{26} {\times} \uf(1)$, $[\gf_2]_{8}{\times} \suf(2)_{14}$ and $\suf(5)_{16}$ families.

Specifically, the $\suf(3)_{26} {\times} \uf(1)$ and $[\gf_2]_{8}{\times} \suf(2)_{14}$ flow lines are associated with the rank $2$ S and T theories, respectively, which were introduced in \cite{Giacomelli:2020gee}. These can be represented by $\mathbb{Z}_3$ twisted compactifications of various $5d$ SCFTs, where we refer the reader to \cite{Giacomelli:2020gee} for more details.

Finally, the $\suf(5)_{16}$ family can also be realized by the $\mathbb{Z}_3$ twisted compactification of certain $5d$ SCFTs. Specifically, the $\suf(5)_{16}$ SCFT can be generated by the compactification of the $5d$ $T_5$ SCFT, which is the $5d$ version of the famous $4d$ $T_N$ theories. Here the compactification is done with a $\mathbb{Z}_3$ twist that acts on the $5d$ SCFT by the cyclic permutation of its three $SU(5)$ global symmetry groups. The other cases are given by the twisted compactification of the $5d$ SCFTs generated from the $5d$ $T_5$ theory by $\mathbb{Z}_3$ preserving mass deformations.

\begin{figure}
\center
\includegraphics[width=1.08\textwidth]{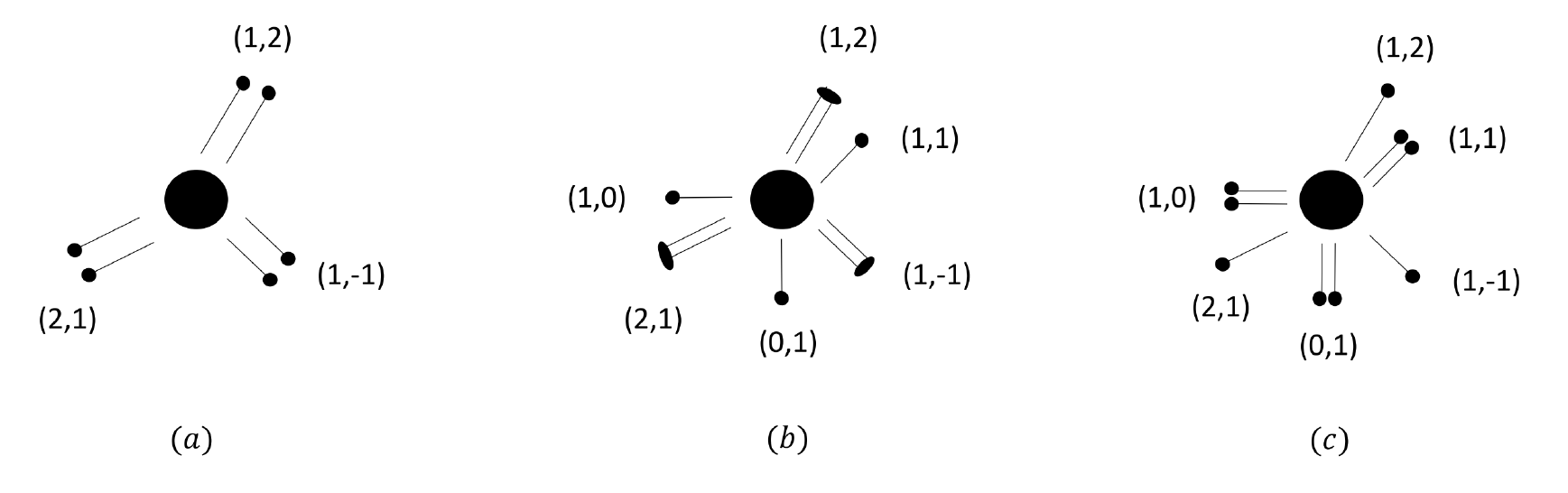} 
\caption{The brane web representations for some of the $\mathbb{Z}_3$ symmetric $5d$ SCFTs. (a) shows the case of the $5d$ SCFT whose compactifications gives theory 60, that is the $\mathcal{N}=4$ $SU(3)$ theory. (b) shows the case of the $5d$ SCFT whose compactifications gives theory 63, which is one of the $\mathcal{N}=3$ theories. (c) shows the case of the $5d$ SCFT whose compactifications gives theory 46. In all cases we are only interested in the external legs and not in how these are connected to one another, which is instead represented by a black blob. The parenthesis gives the $7$-brane charges of the corresponding $7$-branes, or alternatively, the $5$-brane charges of the $5$-branes ending on them.}
\label{WebDiag}
\end{figure}

The $5d$ SCFTs associated with the final $4d$ SCFTs in each of these three families don't appear to inherit the Lagrangian description used by the previous members of the family. Specifically, the $\mathbb{Z}_3$ symmetric mass deformation used to get these from the previous theories don't seem to be manifest in the gauge theory description of these theories. Said mass deformation can be seen from the brane web representation of the $5d$ SCFTs, which can be used to get a brane representation of the $5d$ SCFTs. For completeness we have included these brane realizations, for the three SCFTs in question, in figure \ref{WebDiag}. It is possible that mass deformations leading to Lagrangian theories can be found for some of these $5d$ SCFTs though we shall not pursue this.

\subsubsection{$\mathbb{Z}_4$ twist}

Next, we consider the case where the twist is in a $\mathbb{Z}_4$ discrete symmetry. This accounts for the flow lines emanating from the $\suf(2)_{16} \times \uf(1)$, $\suf(2)_{6}{\times} \suf(2)_{8}$ families. These are associated with the rank $2$ S and T theories, respectively, which were introduced in \cite{Giacomelli:2020gee}. These can be represented by $\mathbb{Z}_4$ twisted compactifications of various $5d$ SCFTs, where we refer the reader to \cite{Giacomelli:2020gee} for more details.

\subsubsection{$\mathbb{Z}_6$ twist}

There is only one case, $\suf(2)_{14}$, that can be generated by the $\mathbb{Z}_6$ twisted compactification of a $5d$ SCFT, see \cite{Giacomelli:2020gee} for more details. However, this case is just the $\mathcal{N}=4$ $G_2$ SCFT, which we can also construct from other methods.

\subsection{Predictions for new discrete symmetries}

Here we consider several cases that do not fit nicely into the cases that we previously discussed. Most of these involve twisted compactifications of $5d$ SCFTs where the manifestation of the discrete symmetry is not apparent in the brane description. Nevertheless we are led to conjecture that such a symmetry exists by looking at the predictions that follow from assuming the twisted compactification exists. Below we explain all these constraints in detail.

\subsubsection{$\mathbb{Z}_2$ symmetry - $SU(4)_0+1 AS+N_f F$}

We noted previously that there are $4d$ SCFTs associated with the twisted compactifications of the $5d$ SCFTs UV completing the $5d$ gauge theories $SU(4)_0+2\AS+N^{(1)}_fF$ and $SU(4)_0+N^{(2)}_fF$, for some ranges of $N^{(1)}_f, N^{(2)}_f$. However, there was no $4d$ SCFT associated with the twisted compactifications of the $5d$ SCFTs UV completing the $5d$ gauge theories $SU(4)_0+1\AS+N_fF$. This is because the brane web associated with this SCFT does not seem to manifest a $\mathbb{Z}_2$ symmetry in the same manner as discussed in \cite{Zafrir:2016wkk}. Nevertheless, there are $4d$ SCFTs whose properties are remarkably consistent with being the result of the compactification of these $5d$ SCFTs with a $\mathbb{Z}_2$ twist in a global symmetry that acts on the gauge theory as charge conjugation. We shall next describe these cases.

There are two $4d$ SCFTs, the ones with global symmetry $\suf(2)_5 {\times} \spf(6)_6{\times} \uf(1)$ and $\spf(4)_7 {\times} \spf(8)_8$, whose properties fit with being the result of the twisted compactifications of the $5d$ SCFTs UV completing the $5d$ gauge theories $SU(4)_0+1\AS+N_fF$. We begin with discussing the case of $\suf(2)_5 {\times} \spf(6)_6{\times} \uf(1)$, which appears consistent with being the result of the twisted compactifications of the $5d$ SCFTs UV completing the $5d$ gauge theory $SU(4)_0+1\AS+6F$.

\subsubsection*{$N_f=6$}

This $5d$ SCFT is expected to have an $\suf(2){\times} \suf(6){\times} \uf(1)^3$ global symmetry, with the $\uf(1)^3$ associated with the fundamental baryonic, antisymmetric baryonic and topological symmetries. If we compactify it with a twist, acting on the gauge theory as charge conjugation, then we expect to get a $4d$ theory with $\suf(2) {\times} \spf(6){\times}\uf(1)$ global symmetry, with the $\uf(1)$ coming from the topological symmetry. This matches the properties of the $4d$ SCFT.

Additionally, various properties of the Higgs branch match between the two theories. The $\suf(2)_5 {\times} \spf(6)_6{\times} \uf(1)$ $4d$ SCFT has been realized in the class S construction of \cite{Chacaltana:2012ch}, from which we can infer various properties associated with its Higgs branch (see also the discussion on this theory in \cite{Martone:2021ixp}). As the Higgs branch is invariant under dimensional reduction, these are particularly convenient for our purposes. Specifically, there are two interesting Higgs branch flows associated with this theory. The first, corresponding to Higgsing through its $\spf(6)$ global symmetry factor, leads to the $4d$ rank 1 SCFT with $\suf(2)_8\times \spf(6)_6$ global symmetry. The second, corresponding to Higgsing through its $\suf(2)$ global symmetry factor, leads to the rank 2 Lagrangian SCFT with gauge group $SU(3)$ and six fundamental hypermultiplets. As we shall next illustrate, both of these are naturally reproduced by the proposed $5d$ picture. 

Consider the $5d$ SCFT UV completing the $SU(4)_0+1\AS+6F$ gauge theory. Since the processes of Higgs branch flows and mass deformation compatible with them commute, and as the global symmetry is fully manifest in the gauge theory, we can analyze the Higgs branch flows using the gauge theory description. In this description, the Higgsing through the $\spf(6)$ factor is described by given a vev to the mesons, which are the moment map operators of the $\suf(6)$ global symmetry, which becomes the $\spf(6)$ factor after the reduction. This leads to the $SU(3)_0+6F$ gauge theory, and as such we expect the $5d$ SCFT to be Higgsed to the $5d$ SCFT UV completing the $SU(3)_0+6F$ gauge theory. At the $4d$ level then, this should lead to the twisted reduction of the $5d$ SCFT UV completing this gauge theory, which is known to be precisely the rank 1 $\suf(2)_8\times \spf(6)_6$ SCFT \cite{Zafrir:2016wkk}.

We can analyze the flow through the $\suf(2)$ factor in a similar manner. Now, we are giving a vev to the quadratic invariant made from the antisymmetric hyper. This breaks $SU(4)$ to $USp(4)$, and we expect to end up with the $5d$ SCFT UV completing the $5d$ gauge theory $USp(4)+6F$. Note that charge conjugation acts trivially on this theory. To be more precise, the action of charge conjugation in this case on the flavor factor is an inner automorphism rather than an outer automorphism. This implies that the twisted reduction we considered becomes a standard reduction with a flavor holonomy, which becomes a mass deformation in $4d$. The specific mass deformation is the one that sits in a $\uf(1)$ Cartan of the $\sof(12)$ flavor symmetry, such that its commutant is $\suf(6)$. The $4d$ compactification of the $5d$ SCFT associated with the $5d$ gauge theory $USp(4)+6F$ can be worked out from the results of \cite{Ohmori:2015pia}. It turns out to be the IR free theory described by an $SU(2)+1F$ gauging of the rank 1 MN $E_7$ SCFT. From our previous discussion, the result of the Higgs branch flow associated with the $SU(2)$ factor for the $4d$ theory we get from the twisted compactification of the considered $5d$ SCFT should be the mass deformation of this theory that initiates the breaking of $\sof(12) \to \uf(1)\times \suf(6)$. This is the mass deformation sending the rank 1 MN $E_7$ SCFT to the rank 1 MN $E_6$ SCFT, so we end up with the $4d$ SCFT described by the $SU(2)+1F$ gauging of the rank 1 MN $E_6$ SCFT. The latter is equivalent to the $4d$ Lagrangian SCFT $SU(3)+6F$, as expected.

Finally, we can consider the spectrum of Higgs branch chiral ring generators of the $4d$ SCFT. These can be extracted from the class S description using the results of \cite{Gaiotto:2012uq,Gaiotto:2012xa}. We find that in addition to the moment map operators, there are several other Higgs branch chiral ring generators. Specifically, we have one in the $(\textbf{2},\textbf{14})$ of $\suf(2)\times \spf(6)$ and the $\textbf{4}$ of $\suf(2)_R$, and another one in the $(\textbf{1},\textbf{6})$ of $\suf(2)\times \spf(6)$, charge $\pm 1$ of the $\uf(1)$ and in the $\textbf{4}$ of $\suf(2)_R$. Finally, there is a Higgs branch chiral ring generator in the $(\textbf{2},\textbf{14'})$ of $\suf(2)\times \spf(6)$, charge $\pm 1$ of the $\uf(1)$ and in the $\textbf{5}$ of $\suf(2)_R$. There might be other ones at higher R-charges, but these would suffice for our purposes. We can match these against our expectations from the $5d$ SCFT. These in turn can be read from the gauge theory, where they are represented as BPS multiplets build from the hypermultiplets, potentially in the presence of instantons. Here the instanton number zero sector accounts for the Higgs branch present in the gauge theory, while the sectors with non-trivial instanton number fill the additional operators that are only present in the Higgs branch of the $5d$ SCFT. Looking at the $SU(4)_0+1\AS+6F$ gauge theory, we see that in addition to the moment map operators, we also have an operator made from two fundamental and one antisymmetric\footnote{The only other independent invariant are the baryons. However, these are in the $\textbf{5}$ $\suf(2)_R$, which is the same representation as the square of the moment map operators. In the $5d$ SCFT these two can be distinguished by their $\uf(1)$ baryon charge, but after the twisting, the baryonic symmetry is broken and these become indistinguishable. As such we will not consider them here.}. This gives a Higgs branch generator in the $(\textbf{2},\textbf{15})$ of $\suf(2)\times \suf(6)$ and in the $\textbf{4}$ of $\suf(2)_R$. When compactified with a charge conjugation twist, this should descend to one of the Higgs branch chiral ring generators we mentioned.

The other two should come from the sector with instanton charge $1$, due to their charges under the $\uf(1)$. We can use instanton counting techniques to determine these. Specifically, we can use the simplified instanton counting methods introduced in \cite{Tachikawa:2015mha}. These were further developed to determine contributions of instanton to Higgs branch chiral ring operators in \cite{Zafrir:2015uaa}, and applied to the case of $SU$ groups with antisymmetric hypermultiplets in \cite{Yonekura:2015ksa} (see also \cite{Zafrir:2015rga}). Using these techniques, we indeed find that the 1-instanton sector gives two Higgs branch chiral ring operators, one in the $(\textbf{1},\textbf{6})\oplus(\textbf{1},\bar{\textbf{6}})$ of $\suf(2)\times \suf(6)$ and in the $\textbf{4}$ of $\suf(2)_R$, and one in the $(\textbf{2},\textbf{20})$ of $\suf(2)\times \suf(6)$ and in the $\textbf{5}$ of $\suf(2)_R$. When compactified with a charge conjugation twist, these should descend to the two other Higgs branch chiral ring generators we mentioned.

\subsubsection*{$N_f=8$}

We next consider the case of the $\spf(4)_7 {\times} \spf(8)_8$ $4d$ SCFT. We conjecture that this theory is the result of the $\mathbb{Z}_2$ twisted compactification of the $5d$ SCFT UV completing the $SU(4)_0+1\AS+8F$ gauge theory, where the $\mathbb{Z}_2$ acts on the gauge theory as charge conjugation. Our evidence for this claims is similar to that of the previous case. Specifically, this $5d$ SCFT has an $\suf(4) \times \suf(8)$ global symmetry \cite{Yonekura:2015ksa}, with the $\suf(4)$ factor being broken by the gauge theory deformation. As such, when compactified with a twist we expect to get a rank $2$ theory with $\spf(4) \times \spf(8)$ global symmetry. 

We can also consider its Higgs branch flows. There are again two of them associated with each of its global symmetry factors, which can again be inferred from its class S realizations\cite{Chacaltana:2011ze,Chacaltana:2013oka} (see also the discussion of this theory in \cite{Martone:2021ixp}). The one associated with the $\spf(8)$ factor leads to the rank $1$ $\spf(10)$ SCFT (or $\cS^{(1)}_{E_6,2}$), while the one associated with the $\spf(4)$ factor leads to the $R_{(0,4)}$ $4d$ SCFT (or $\suf(2)_6{\times}\suf(8)_8$). These match the $5d$ expectations. Specifically, Higgsing through the $\spf(8)$ corresponds to Higgsing the $5d$ SCFT to the one UV completing the $5d$ gauge theory $SU(3)_0+8F$, whose twisted compactification yield the rank $1$ $\spf(10)$ SCFT \cite{Zafrir:2016wkk}. Alternatively, Higgsing through the $\spf(4)$ corresponds to Higgsing the $5d$ SCFT to the one UV completing the $5d$ gauge theory $USp(4)+8F$. The latter is acted trivially by the twisting so we expect it to just be an holonomy in the $\sof(16)$ global symmetry preserving its $\uf(1)\times \suf(8)$ subgroup. As previously mentioned, this $5d$ SCFT reduces to the $\suf(2)_8 \times \sof(16)_{12}$ $4d$ SCFT, and the mass deformation expected due to the holonomy should cause it to flow to the $R_{(0,4)}$ $4d$ SCFT.

Finally, we can again consider the spectrum of Higgs branch chiral ring generators of the $4d$ SCFT. These can similarly be extracted from the class S description using the results of \cite{Gaiotto:2012uq,Gaiotto:2012xa,Lemos:2012ph}. We find that in addition to the moment map operators, there are two other Higgs branch chiral ring generators. Specifically, we have one in the $(\textbf{4},\textbf{28})$ of $\spf(4)\times \spf(8)$ and the $\textbf{4}$ of $\suf(2)_R$, and one in the $(\textbf{5},\textbf{42})$ of $\spf(4)\times \spf(8)$ and in the $\textbf{5}$ of $\suf(2)_R$. There might be other ones at higher R-charges, but these would suffice for our purposes. We can match these against our expectations from the $5d$ SCFT. Here there is the complication that the full symmetry is not manifest in the gauge theory so to understand the structure of the multiplets requires both perturbative and non-perturbative states. Fortunately, for our present purposes knowledge of the perturbative and 1-instanton sector suffice, which we can determine using our previous methods. Besides the moment map operators associated with the enhanced symmetry\footnote{These actually require 2-instanton contributions, in addition to the perturbative and 1-instanton contribution. However, there is ample evidence for this enhancement from both field theoretic and string theoretic considerations, see for instance \cite{Yonekura:2015ksa,Bergman:2015dpa,Bhardwaj:2020avz}.}, we also find two other Higgs branch chiral ring operators. One is in the $(\textbf{4},\textbf{28})+c.c$ of $\suf(4)\times \suf(8)$ and in the $\textbf{4}$ of $\suf(2)_R$. It is made from the perturbative gauge invariant made from the antisymmetric hyper and two fundametal hyper, and a 1-instanton state in the $\textbf{28}$ of $\suf(8)$ and with baryon charge $2$, plus the complex conjugate. Another one is in the $(\textbf{6},\textbf{70})$ of $\suf(4)\times \suf(8)$ and in the $\textbf{5}$ of $\suf(2)_R$. It is made from the perturbative gauge invariant associated with the $\suf(4)$ baryons, and a 1-instanton state in the $(\textbf{2},\textbf{70})$ of the $\suf(2)\times \suf(8)$ symmetry visible in the gauge theory. When compactified with a charge conjugation twist, these should descend to the two Higgs branch chiral ring generators we mentioned.

\begin{center}
\rule[1mm]{2cm}{.4pt}\hspace{1cm}$\circ$\hspace{1cm} \rule[1mm]{2cm}{.4pt}
\end{center}

Finally, let's investigate the implications of these claims in terms of the mass deformations of the corresponding $4d$ SCFTs. First, we should be able to flow from the $\spf(4)_7 {\times} \spf(8)_8$ SCFT to the $\suf(2)_5 {\times} \spf(6)_6 {\times} \uf(1)$ SCFT, corresponding to integrating out fundamental flavors. Additionally, we can consider mass deformations given by integrating out antisymmetric hypermultiplets. This leads to mass deformations between the $\spf(4)_7 {\times} \spf(8)_8$ SCFT and the $\suf(2)_8 \times \spf(8)_6$ SCFT, and the $\suf(2)_5 {\times} \spf(6)_6 {\times} \uf(1)$ SCFT and the $\uf(1) {\times} \spf(6)_5$ SCFT. Similarly, it suggests a mass deformation between the $[\ff_4]_{12}{\times} \suf(2)^2_7$ SCFT and the $\suf(2)_5 {\times} \spf(6)_6 {\times} \uf(1)$ SCFT. These conclusion will also be checked below in a purely 4 dimensional language.

Finally, you can mass deform the $SU(4)_0+1\AS+6F$ theory to the 5\emph{d} gauge theory $SU(4)_0+1\AS+4F$, and one can ask if the compactification of the associated 5\emph{d} SCFT leads to a 4$d$  SCFT or not. There is the possibility that this is so, leading to a 4$d$  SCFT with $\uf(1) \times \suf(2) \times\spf(4)$ global symmetry. It should also be Higgsable to the rank-1 $\uf(1) \times \spf(4)_4$ ($\cS^{(1)}_{A_2,2}$) SCFT.

\subsubsection{$\mathbb{Z}_3$ - $SO(8)+2\V+2\S+2\C$}

Finally, we consider the $\suf(2)_8 \times \spf(4)_{14}$ $4d$ SCFT. As we shall next describe, some of its properties are consistent with being the $\Z_3$ twisted reduction of a $5d$ SCFT.

Consider the 5\emph{d} SCFT with an $\suf(2) \times \spf(4)^3$ global symmetry which UV completes the 5\emph{d} gauge theory $SO(8)+2\V+2\S+2\C$. The gauge theory has a $\Z_3$ symmetry acting on the gauge theory as the outer automorphism transformation. We can then consider the compactification of the 5\emph{d} SCFT with a twist in this symmetry. This will give a rank 2 theory with $\suf(2) \times \spf(4)$ global symmetry which can be readily identified with the 4$d$  $\suf(2)_{8}\times\spf(4)_{14}$. An important consistency check is that if you Higgs this 5\emph{d} SCFT by breaking the $SU(2)$ group, you can show that you get the 5\emph{d} $T_4$ SCFT and thus this identification predicts a Higgsing between the 4$d$  rank-2 theory to the rank-1 $\cS^{(1)}_{D_4,3}$ SCFT which indeed is precisely matched by the $4d$ data.

\subsection{Theories with no known 5d realization}

We finally end with several cases for which the realization in terms of the compactification of $5d$ SCFTs is lacking. The first case we consider is the $SU(2)_{14}$ SCFT with Coulomb branch operators of dimensions $\frac{12}{5}$ and $6$. As we previously mentioned, this theory can be engineered by the compactification of a certain $6d$ SCFT with a non-trivial flavor background. These types of compactifications are usually related to twisted compactifications of $5d$ SCFTs. However, in this case the twist is expected to be in a $\Z_5$ group, and as such is not of the type that was previously studied. This makes us expect that this $4d$ SCFT can be realized by the $\Z_5$ twisted compactification of a $5d$ SCFT, though the lack of a good understanding of its action makes it difficult to provide precise statements.

There are several cases of $4d$ SCFTs for which we don't currently have a candidate realization as the compactification of a $5d$ SCFT. One such example is the Lagrangian SCFT consisting of a $USp(4)$ gauge group with a half-hyper in its representation of dimension sixteen. The analogous $5d$ gauge theory does not obey the criteria of \cite{Jefferson:2017ahm}, and so is thought not to have an SCFT UV completion. It is possible that this theory be realized by the compactification of the few known rank $2$ $5d$ SCFTs with no Higgs branch and global symmetry, but the lack of both makes it hard to provide enough supportive evidence for this statement.

There is also a strongly coupled $4d$ SCFT for which we do not know of a $5d$ parent, being the $\suf(2)_5 \times \spf(8)_7$ $4d$ SCFT.

\section{Consistency of the 4d  story} \label{checks}

It is possible to check the consistencies of the RG-trajectories that we derived from the 5\emph{d} analysis directly in 4$d$  by checking that the moduli space of vacua of all theories in the same series are related as described in section \ref{sec:4dsetup}. Given that the analysis is quite intricate we will explicitly depict both the CB and HB stratification, see figures \ref{fig:CBE8r2}-\ref{fig:CBthree}, and comment on the most subtle points. The mass deformations of the $\cS$ and $\cT$ theories were already analyzed in \cite{Giacomelli:2020jel,Giacomelli:2020gee}, thus we will not discuss this further and skip the analysis of the $\spf(8)-\suf(2)^2$, $\gf_2$, $\suf(3)$ and $\suf(2)$ series. We will also not discuss the set of isolated theories in table \ref{tab:r23} since it is currently unknown to the authors whether these theories belong to a larger series of theories connected by RG-flows.

\subsection{Setting things up}

In performing the study of the mass deformations from the CB perspective we will heavily rely on the analysis of the mass deformations of the rank-1 theories which has been thoroughly studied in \cite{Argyres:2015ffa,Argyres:2015gha,Argyres:2016xua,Argyres:2016xmc}. Below we will use the following notation:
\beq\label{NotCB}
\Tf_0\quad \to\quad \{\Tf_1,...,\Tf_n\}
\eeq
$\{\Tf_1,...,\Tf_n\}$ is the aforementioned \emph{deformation pattern} and this indicates that a given rank-1 theory $\Tf_0$ can be mass deformed to the $n$ theories on the RHS. More precisely, $\Tf_0$ has a one dimensional CB and the SCFT ``lives'' at the origin. Turning on the mass deformation which induces the flow in \eqref{NotCB} splits the singularity associated to $\Tf_0$ in a series of $n$ singularities associated to $\Tf_1$ to $\Tf_n$. As described in sec \ref{sec:4dsetup}, a necessary condition for two rank-2 theories to be connected by mass deformation is that the CB stratification of one is a deformation of the other. Key in establishing this relation is the decoration of the CB stratification which is the set of the rank-1 theories supported on the irreducible complex co-dimension one strata. And a necessary condition for two theories to be related to one another is that the decoration of one should be a deformation of the other in the sense of section \ref{sec:4dsetup}. To not confuse the decoration with the deformation pattern, recall that we will use angle bracket for the former, see \eqref{deco}, and curly brackets for the latter as in \eqref{NotCB}. Going through two explicit examples, one in which the two theories are connected by mass deformation, and one in which the two theories aren't, might help clarifying this criterion.

Consider first the flow from the $[\ef_8]_{24}{\times}\suf(2)_{13}\to[\ef_7]_{16}{\times}\suf(2)_{9}$. The former has decoration $\langle \cT^{(1)}_{E_8,1},\blue{\cS^{(1)}_{\varnothing,2}} \rangle$ while the latter $\langle \cT^{(1)}_{E_7,1},\blue{\cS^{(1)}_{\varnothing,2}} \rangle$, see figure \ref{fig:CBE8r2}. Since the following deformation pattern exists:
\beq
\cT^{(1)}_{E_8,1}\to\{\cT^{(1)}_{E_7,1},[I_1,\varnothing]\}
\eeq
it follows that the CB stratification of $[\ef_7]_{16}{\times}\suf(2)_{9}$ is indeed a deformation of the CB stratification of $[\ef_8]_{24}{\times}\suf(2)_{13}$. Conversely in figure \ref{MapR2} we claim that no mass deformation exists among $[\ef_8]_{24}{\times}\suf(2)_{13}$ and $\sof(20)_{16}$. And we can check that indeed neither CB stratification is the deformation of the other.

From figure \ref{fig:CBE8r2} we can readily read off the CB stratification of $\sof(20)_{16}$ which has decoration $\langle[I_1,\varnothing],[I_6^*,\sof(20)]\rangle$. Now we have two possibilities:
\begin{itemize}
    \item The CB stratification of $[\ef_8]_{24}{\times}\suf(2)_{13}$ is a deformation of that of $\sof(20)_{16}$. This option can be excluded since the $[I_6^*,\sof(20)]$ is an IR-free theory while $\cT^{(1)}_{E_8,1}$ is an SCFT at a fixed value of the holomorphic coupling away from the weakly coupled cusp.
    \item  The CB stratification of $\sof(20)_{16}$ is a deformation of that of $[\ef_8]_{24}{\times}\suf(2)_{13}$. There certainly exists a deformation pattern which connects $\blue{\cS^{(1)}_{\varnothing,2}}$ and $[I_1,\varnothing]$ and it is certainly possible to flow from a strongly coupled SCFT to an IR-free theory. But the \emph{dangerously irrelevant conjecture} \cite{Argyres:2015ffa} implies that the rank of the flavor symmetry strictly decreases under mass deformation and thus no deformation pattern of $\cT^{(1)}_{E_8,1}$ could ever contain $[I_6^*,\sof(20)]$.
\end{itemize}
We then conclude that neither options work and indeed there cannot be any mass deformation connecting the two theories.

As far as the HB goes, the main criterion to check for the two moduli spaces to be compatible of being related by mass deformation is what we called criterion \ref{Higgs}. It is worth recalling that the HB is a stratified symplectic variety and by bottom symplectic strata we refer to those strata which only include the zero dimensional stratum (the suprconformal vacuum) and similarly the bottom theories are the SCFTs supported on the corresponding symplectic leaves. In other words the bottom strata are the lowest strata in the Hasse diagrams in figures \ref{fig:HBE8r2}-\ref{fig:HBSp12}. In most cases it is fairly straightforward, after taking into account how the mass deformations at the superconformal vacuum act on the HB, to check that the bottom theories before and after a mass deformation are indeed appropriately related. But there are cases where things work out in a rather surprising way, we will discuss these in detail. 

To perform the HB analysis, we need an in-depth understanding of how the flavor symmetry is matched along the HB. This study was already performed in \cite{Martone:2021ixp} but for convenience is summarized in appendix \ref{app:Higgs}, the main results are reported in the, rather dense, tables \ref{Higgs:one}-\ref{Higgs:three}. Let's consider an example to highlight the type of subtleties we might encounter. Consider the HB of the $\sof(20)_{16}$ and the one of $\suf(2)_8{\times}\sof(16)_{12}$. Following figure \ref{MapR2}, we expect these two theories to be related by a mass deformation and their corresponding HBs to satisfy criterion \ref{Higgs}.

From figure \ref{fig:HBE8r2}, the bottom theory on the HB of the $\sof(20)_{16}$ is $\cT^{(1)}_{E_8,1}$. To make progress we need to understand how the mass deformations of the $\sof(20)$ maps to those of the $\ef_8$ of the bottom theory, for that we turn to table \ref{Higgs:one}. As reported in column nine of the $\sof(20)_{16}$ row, the bottom symplectic leaf is associated to the $\sof(20)\to\sof(16)\times\suf(2)$ spontaneous symmetry breaking, thus in this case $\ff^\natural\equiv\sof(16)\times\suf(2)$. Since $\ff^\natural$ is a rank nine group it cannot be embedded into $\ef_8$. This apparent contradiction can be resolved by carefully accounting for the extra goldstone bosons associated to the spontaneous symmetry breaking, which are reported in column ten of the table \ref{Higgs:one}, and matching the levels accordingly. Doing so we discover that the $\suf(2)$ only acts on the goldstones while $\sof(16)$ embeds into $\ef_8$ as one of its maximal subgroups which also nicely reproduce the correct flavor symmetry level, as it can be readily checked using the information in column twelve of the table as well as the group theory fact that the fundamental of $\sof(16)$ has Dynkin index $T_2({\bf16})=2$\footnote{Recall that we use a normalization for the Dynkin index for which the fundamental representation of $\suf(n)$ has $T_2({\bf n})=1$.}. Now that we have understood how things work for $\sof(20)_{16}$ let's compare with the HB structure of $\suf(2)_8{\times}\sof(16)_{12}$.

Again we can rely on \ref{fig:HBE8r2} to quickly read off its HB structure. Since the flavor symmetry is semi-simple, not surprisingly the HB has two irreducible components carrying each an action of the two simple factors. The spontaneous breaking of the $\sof(16)$ higgses the SCFT to $\cT^{(1)}_{E_7,1}$ which it is indeed a mass deformation of the bottom theory of the $\sof(20)_{16}$. But the $\suf(2)$ breaking seem instead to give a contradictory answer since the bottom theory there is again $\cT^{(1)}_{E_8,1}$! How can it be possible? The resolution is given by the detailed mapping of mass deformations along the HB which we just performed. We in fact showed that only mass deformations associated with $\sof(16)\subset \sof(20)$ map to mass deformations of the $\cT^{(1)}_{E_8,1}$ while the commutant of $\sof(16)$ in $\sof(20)$ simply does not act on that theory. Thus the result we find is not only compatible but the only one possible. We conclude that indeed the HB of these two theories are deformable into one another. 

Let's now turn into a detailed analysis of each series in turn.

\subsection{$\ef_8-\sof(20)$ series}

\begin{figure}
\begin{tikzpicture}[decoration={markings,
mark=at position .5 with {\arrow{>}}},
Marrow/.style={->,>=stealth[round],shorten >=1pt,line width=.4mm,coolblack},
IRarrow/.style={->,>=stealth[round],shorten >=1pt,line width=.4mm,deepcarmine,dashed}]
\begin{scope}[scale=1.4] %E8 rank2 and SO(20)%
\node[bbc,scale=.5] (E8r2_top) at (0,0) {};       %%%%%%%%  E8r2   %%%%%%%%

\draw [draw=black] (-.9,.1) rectangle (.9,.4) {};
\filldraw [fill=white, draw=black] (-.9,.1) rectangle (.9,.4) {};
\node[scale=.8] (title1) at (0,.25) {\small{$[\ef_8]_{24}{\times}\suf(2)_{13}$}};

\draw [draw=black] (-.9,-1.5) rectangle (.9,.1) {};
\node[bbc,scale=.5] (E8r2_down) at (0,-1.4) {};
\node[scale=.8] (E8r2_1a) at (-.6,-.7) {$\blue{\cS^{(1)}_{\varnothing,2}}$};
\node[scale=.8] (E8r2_1b) at (.6,-.7) {$\cT^{(1)}_{E_8,1}$};
\draw[red] (E8r2_top) -- (E8r2_1a) {};
\draw[red] (E8r2_top) -- (E8r2_1b) {};
\draw[red] (E8r2_down) -- (E8r2_1a) {};
\draw[red] (E8r2_down) -- (E8r2_1b) {};
\node[bbc,scale=.5] (SO20_top) at (4.5,0) {};       %%%%%%%%  SO20   %%%%%%%%
\draw [draw=black] (3.4,.1) rectangle (5.7,.4) {};
\filldraw [fill=white, draw=black] (3.4,.1) rectangle (5.7,.4) {};
\node[scale=.8] (title1) at (4.5,.25) {\small{$\sof(20)_{16}$}};

\draw [draw=black] (3.4,-1.5) rectangle (5.7,.1) {};
\node[bbc,scale=.5] (SO20_down) at (4.5,-1.4) {};
\node[scale=.8] (SO20_1a) at (3.9,-.7) {$[I_1,\varnothing]$};
\node[scale=.8] (SO20_1b) at (5.1,-.7) {$[I_6^*,\sof(20)]$};
\draw[red] (SO20_top) -- (SO20_1a) {};
\draw[red] (SO20_top) -- (SO20_1b) {};
\draw[red] (SO20_down) -- (SO20_1a) {};
\draw[red] (SO20_down) -- (SO20_1b) {};
\end{scope}

\begin{scope}[scale=1.4,yshift=-2cm,xshift=0cm] %E820, E7 rank2 SO16 and SU(10)%
\node[bbc,scale=.5] (E820_top) at (-2.3,0) {};       %%%%%%%%  E820   %%%%%%%%
\draw [draw=black] (-1.35,-1.5) rectangle (-3.25,.1) {};
\node[bbc,scale=.5] (E820_down) at (-2.3,-1.4) {};
\node[scale=.8] (E820_1a) at (-1.7,-.7) {$[I_1,\varnothing]$};
\node[scale=.8] (E820_1b) at (-2.9,-.7) {$\cT^{(1)}_{E_8,1}$};
\draw[red] (E820_top) -- (E820_1a) {};
\draw[red] (E820_top) -- (E820_1b) {};
\draw[red] (E820_down) -- (E820_1a) {};
\draw[red] (E820_down) -- (E820_1b) {};

%%%%% The E7r2 is below to make things work out properly

\node[bbc,scale=.5] (SO16_top) at (3,0) {};       %%%%%%%%  SO16   %%%%%%%%
\draw [draw=black] (1.5,.1) rectangle (4.55,.4) {};
\filldraw [fill=white, draw=black] (1.5,.1) rectangle (4.55,.4) {};
\node[scale=.8] (title1) at (3,.25) {\small{$\suf(2)_8{\times}\sof(16)_{12}$}};

\draw [draw=black] (1.5,-1.5) rectangle (4.55,.1) {};
\node[bbc,scale=.5] (SO16_down) at (3,-1.4) {};
\node[scale=.8] (SO16_1a) at (2.1,-.7) {$[I_1,\varnothing]$};
\node[scale=.8] (SO16_1b) at (3,-.7) {$[I_2,\suf(2)]\ \ $};
\node[scale=.8] (SO16_1c) at (3.9,-.7) {\ \ $[I^*_4,\sof(16)]$};
\draw[red] (SO16_top) -- (SO16_1a) {};
\draw[red] (SO16_top) -- (SO16_1b) {};
\draw[red] (SO16_top) -- (SO16_1c) {};
\draw[red] (SO16_down) -- (SO16_1a) {};
\draw[red] (SO16_down) -- (SO16_1b) {};
\draw[red] (SO16_down) -- (SO16_1c) {};
\node[bbc,scale=.5] (SU10_top) at (5.9,0) {};       %%%%%%%%  SU10   %%%%%%%%
\draw [draw=black] (4.7,-1.5) rectangle (7.05,.1) {};
\node[bbc,scale=.5] (SU10_down) at (5.9,-1.4) {};
\node[scale=.8] (SU10_1a) at (6.5,-.7) {$[I_1,\varnothing]$};
\node[scale=.8] (SU10_1b) at (5.3,-.7) {$[I_{10},\suf(10)]$};
\draw[red] (SU10_top) -- (SU10_1a) {};
\draw[red] (SU10_top) -- (SU10_1b) {};
\draw[red] (SU10_down) -- (SU10_1a) {};
\draw[red] (SU10_down) -- (SU10_1b) {};

\draw[Marrow] (SO20_1b) to (SU10_1b) {};
\draw[Marrow] (SO20_1b) to (SO16_1b) {};
\draw[Marrow] (SO20_1b) to (SO16_1c) {};

\draw[Marrow] (E8r2_1a) to (E820_1a) {};

%%%%%% Titles for the tables %%%%%%%%

\draw [draw=black] (-1.35,.1) rectangle (-3.25,.4) {};
\filldraw [fill=white, draw=black] (-1.35,.1) rectangle (-3.25,.4) {};
\node[scale=.8] (title1) at (-2.3,.25) {\small{$[\ef_8]_{20}$}};

\node[scale=.8] (SO14_1a) at (1.6,-2.7) {$[I_1,\varnothing]$};
\draw[Marrow] (E8r2_1a) to (SO14_1a) {};
\filldraw [fill=white, draw=black] (-.9,-1.5) rectangle (.9,.1) {};
\draw [draw=black] (-.9,.1) rectangle (.9,.4) {};
\node[bbc,scale=.5] (E7r2_top) at (0,0) {};       %%%%%%%%  E7r2   %%%%%%%%
\draw [draw=black] (-.9,-1.5) rectangle (.9,.1) {};
\node[bbc,scale=.5] (E7r2_down) at (0,-1.4) {};
\node[scale=.8] (E7r2_1a) at (-.6,-.7) {$\cT^{(1)}_{E_7,1}$};
\node[scale=.8] (E7r2_1b) at (.6,-.7) {$\blue{\cS^{(1)}_{\varnothing,2}}$};
\draw[red] (E7r2_top) -- (E7r2_1a) {};
\draw[red] (E7r2_top) -- (E7r2_1b) {};
\draw[red] (E7r2_down) -- (E7r2_1a) {};
\draw[red] (E7r2_down) -- (E7r2_1b) {};
\draw[Marrow] (E8r2_1b) to (E7r2_1a) {};
\filldraw [fill=white, draw=black] (-.9,.1) rectangle (.9,.4) {};
\node[scale=.8] (title1) at (0,.25) {\small{$[\ef_7]_{16}{\times}\suf(2)_9$}};

\draw [draw=black] (1.5,.1) rectangle (4.55,.4) {};
\filldraw [fill=white, draw=black] (1.5,.1) rectangle (4.55,.4) {};
\node[scale=.8] (title1) at (3,.25) {\small{$\suf(2)_8{\times}\sof(16)_{12}$}};

\draw [draw=black] (4.7,.1) rectangle (7.05,.4) {};
\filldraw [fill=white, draw=black] (4.7,.1) rectangle (7.05,.4) {};
\node[scale=.8] (title1) at (5.9,.25) {\small{$\suf(10)_{10}$}};

\end{scope}

\begin{scope}[scale=1.4,yshift=-4cm,xshift=0cm] %E6 rank2, SO14 and SU(8)%

%%%%%% the CB stratification of the E6r2 will be below to make the figure more readable %%%%%%

\node[bbc,scale=.5] (SO14_top) at (2.2,0) {};       %%%%%%%%  SO14   %%%%%%%%
\draw [draw=black] (1,-1.5) rectangle (3.4,.1) {};
\node[bbc,scale=.5] (SO14_down) at (2.2,-1.4) {};
\node[scale=.8] (SO14_1b) at (2.8,-.7) {$[I^*_3,\sof(14)]$};
\draw[red] (SO14_top) -- (SO14_1a) {};
\draw[red] (SO14_top) -- (SO14_1b) {};
\draw[red] (SO14_down) -- (SO14_1a) {};
\draw[red] (SO14_down) -- (SO14_1b) {};
\node[bbc,scale=.5] (SU8_top) at (6.1,0) {};       %%%%%%%%   SU8   %%%%%%%%

\draw [draw=black] (4.6,-1.5) rectangle (7.55,.1) {};
\node[bbc,scale=.5] (SU8_down) at (6.1,-1.4) {};
\node[scale=.8] (SU8_1a) at (5.15,-.7) {$[I_8,\suf(8)]$};
\node[scale=.8] (SU8_1b) at (6.1,-.7) {$[I_2,\suf(2)]$};
\node[scale=.8] (SU8_1c) at (7.05,-.7) {$[I_1,\varnothing]$};
\draw[red] (SU8_top) -- (SU8_1a);
\draw[red] (SU8_top) -- (SU8_1b);
\draw[red] (SU8_top) -- (SU8_1c);
\draw[red] (SU8_down) -- (SU8_1a);
\draw[red] (SU8_down) -- (SU8_1b);
\draw[red] (SU8_down) -- (SU8_1c);

\draw[Marrow] (E8r2_1b) to (SO14_1b);

\draw[Marrow] (SO16_1c) to (SU8_1a);
\draw[Marrow] (SO16_1c) to (SO14_1b);

\draw[Marrow] (SU10_1b) to (SU8_1b);
\draw[Marrow] (SU10_1b) to (SU8_1a);

%%%%%% Titles for the tables %%%%%%%%

\node[scale=.8] (SO12_1a) at (-3.3,-2.7) {$[I_1,\varnothing]$};
\node[scale=.8] (SO12_1b) at (-2.4,-2.7) {$[I^*_2,\sof(12)]$};
\node[scale=.8] (SO12_1c) at (-1.5,-2.7) {$[I_1,\varnothing]$};
\draw[IRarrow] (E7r2_1a) to (SO12_1a);
\draw[IRarrow] (E7r2_1a) to (SO12_1b);
\draw[IRarrow] (E7r2_1b) to (SO12_1c);
\filldraw [fill=white, draw=black] (-1.7,-1.5) rectangle (0.1,.1);
\node[bbc,scale=.5] (E6r2_top) at (-.8,0) {};       %%%%%%%%  E6r2   %%%%%%%%
\draw [draw=black] (-1.7,-1.5) rectangle (0.1,.1);
\node[bbc,scale=.5] (E6r2_down) at (-.8,-1.4) {};
\node[scale=.8] (E6r2_1a) at (-1.4,-.7) {$\blue{\cS^{(1)}_{\varnothing,2}}$};
\node[scale=.8] (E6r2_1b) at (-.2,-.7) {$\cT^{(1)}_{E_6,1}$};
\draw[red] (E6r2_top) -- (E6r2_1a);
\draw[red] (E6r2_top) -- (E6r2_1b);
\draw[red] (E6r2_down) -- (E6r2_1a);
\draw[red] (E6r2_down) -- (E6r2_1b);

%%%%%% Title for E6 x SU(2) %%%%%%%%

\draw[Marrow] (E7r2_1a) to (E6r2_1b);
\draw [draw=black] (-1.7,.1) rectangle (0.1,.4);
\filldraw [fill=white, draw=black] (-1.7,.1) rectangle (0.1,.4);
\node[scale=.8] (title1) at (-.8,.25) {\small{$[\ef_6]_{12}{\times}\suf(2)_7$}};

\draw [draw=black] (1,.1) rectangle (3.4,.4);
\filldraw [fill=white, draw=black] (1,.1) rectangle (3.4,.4);
\node[scale=.8] (title1) at (2.2,.25) {\small{$\sof(14)_{10}{\times}\uf(1)$}};

\draw [draw=black] (4.6,.1) rectangle (7.55,.4);
\filldraw [fill=white, draw=black] (4.6,.1) rectangle (7.55,.4);
\node[scale=.8] (title1) at (6.1,.25) {\small{$\suf(2)_6{\times}\suf(8)_8$}};

\end{scope}

\begin{scope}[scale=1.4,yshift=-6cm,xshift=0cm] %Lagrangian level, SO(12), SO(8), U(6), SU(2)^5%
\node[bbc,scale=.5] (SO12_top) at (-2.4,0) {};       %%%%%%%%  SO12   %%%%%%%%
\draw [draw=black] (-3.65,-1.5) rectangle (-1.15,.1);
\node[bbc,scale=.5] (SO12_down) at (-2.4,-1.4) {};
\draw[red] (SO12_top) -- (SO12_1a);
\draw[red] (SO12_top) -- (SO12_1b);
\draw[red] (SO12_top) -- (SO12_1c);
\draw[red] (SO12_down) -- (SO12_1a);
\draw[red] (SO12_down) -- (SO12_1b);
\draw[red] (SO12_down) -- (SO12_1c);
\node[bbc,scale=.5] (D4r2_top) at (0,0) {};       %%%%%%%%  D4r2   %%%%%%%%
\draw [draw=black] (-.9,-1.5) rectangle (.9,.1);
\node[bbc,scale=.5] (D4r2_down) at (0,-1.4) {};
\node[scale=.8] (D4r2_1a) at (-.6,-.7) {$\blue{\cS^{(1)}_{\varnothing,2}}$};
\node[scale=.8] (D4r2_1b) at (.6,-.7) {$\cT^{(1)}_{D_4,1}$};
\draw[red] (D4r2_top) -- (D4r2_1a);
\draw[red] (D4r2_top) -- (D4r2_1b);
\draw[red] (D4r2_down) -- (D4r2_1a);
\draw[red] (D4r2_down) -- (D4r2_1b);
\node[bbc,scale=.5] (SU6_top) at (2.2,0) {};       %%%%%%%%  SU6   %%%%%%%%
\draw [draw=black] (.95,-1.5) rectangle (3.6,.1);
\node[bbc,scale=.5] (SU6_down) at (2.2,-1.4) {};
\node[scale=.8] (SU6_1a) at (1.3,-.7) {$[I_1,\varnothing]$};
\node[scale=.8] (SU6_1b) at (2.2,-.7) {$[I_1,\varnothing]$};
\node[scale=.8] (SU6_1c) at (3.1,-.7) {$[I_6,\suf(6)]$};
\draw[red] (SU6_top) -- (SU6_1a);
\draw[red] (SU6_top) -- (SU6_1b);
\draw[red] (SU6_top) -- (SU6_1c);
\draw[red] (SU6_down) -- (SU6_1a);
\draw[red] (SU6_down) -- (SU6_1b);
\draw[red] (SU6_down) -- (SU6_1c);
\node[bbc,scale=.5] (SU25_top) at (6.1,0) {};        %%%%%%%%  SU25   %%%%%%%%
\draw [draw=black] (3.7,-1.5) rectangle (8.5,.1);
\node[bbc,scale=.5] (SU25_down) at (6.1,-1.4) {};
\node[scale=.8] (SU25_1a) at (4.2,-.7) {$[I_2,\suf(2)]$};
\node[scale=.8] (SU25_1b) at (5.15,-.7) {$[I_2,\suf(2)]$};
\node[scale=.8] (SU25_1c) at (6.1,-.7) {$[I_2,\suf(2)]$};
\node[scale=.8] (SU25_1d) at (7.05,-.7) {$[I_2,\suf(2)]$};
\node[scale=.8] (SU25_1e) at (8,-.7) {$[I_2,\suf(2)]$};
\draw[red] (SU25_top) -- (SU25_1a);
\draw[red] (SU25_top) -- (SU25_1b);
\draw[red] (SU25_top) -- (SU25_1c);
\draw[red] (SU25_top) -- (SU25_1d);
\draw[red] (SU25_top) -- (SU25_1e);
\draw[red] (SU25_down) -- (SU25_1a);
\draw[red] (SU25_down) -- (SU25_1b);
\draw[red] (SU25_down) -- (SU25_1c);
\draw[red] (SU25_down) -- (SU25_1d);
\draw[red] (SU25_down) -- (SU25_1e);

\draw[IRarrow] (E820_1b) to (SO12_1a);
\draw[IRarrow] (E820_1b) to (SO12_1b);

\draw[IRarrow] (E6r2_1b) to (D4r2_1b);

\draw[IRarrow] (SO14_1b) to (SO12_1b);
\draw[IRarrow] (SO14_1b) to (SO12_1c);
\draw[IRarrow] (SO14_1b) to (SU6_1b);
\draw[IRarrow] (SO14_1b) to (SU6_1c);

\draw[IRarrow] (SU8_1a) to (SU25_1a);
\draw[IRarrow] (SU8_1a) to (SU25_1b);
\draw[IRarrow] (SU8_1a) to (SU25_1c);
\draw[IRarrow] (SU8_1a) to (SU25_1d);
\draw[IRarrow] (SU8_1b) to (SU25_1e);

%%%%%% Titles for the tables %%%%%%%%

\draw [draw=black] (-3.65,.1) rectangle (-1.15,.4);
\filldraw [fill=yellow!40, draw=black] (-3.65,.1) rectangle (-1.15,.4);
\node[scale=.8] (title1) at (-2.4,.25) {\small{$USp(4)+6F$}};

\draw [draw=black] (-.9,.1) rectangle (.9,.4);
\filldraw [fill=yellow!40, draw=black] (-.9,.1) rectangle (.9,.4);
\node[scale=.8] (title1) at (0,.25) {\small{$USp(4)+4F+V$}};

\draw [draw=black] (.95,.1) rectangle (3.6,.4);
\filldraw [fill=yellow!40, draw=black] (.95,.1) rectangle (3.6,.4);
\node[scale=.8] (title1) at (2.2,.25) {\small{$SU(3)+6F$}};

\draw [draw=black] (3.7,.1) rectangle (8.5,.4);
\filldraw [fill=yellow!40, draw=black] (3.7,.1) rectangle (8.5,.4);
\node[scale=.8] (title1) at (6.1,.25) {\small{$2F-SU(2)-SU(2)-2F$}};

\end{scope}

\end{tikzpicture}
{\caption{\label{fig:CBE8r2} Coulomb branch analysis of the mass deformations among the entries of the $\ef_8-\sof(20)$ series.}}
\end{figure}

The analysis of this set of theories is reported explicitly in figure \ref{fig:CBE8r2} and \ref{fig:HBE8r2}. The CB analysis is quite straightforward and relies on the following deformation patterns which can be derived from an analysis of the SW curve describing the corresponding rank-1 theory:
\begin{itemize}
    \item $\qquad\cT_{E_8,1}^{(1)}\quad \to\quad
    \left\{\begin{array}{l}
    \{[I_3^*,\sof(14)],[I_1,\varnothing]\}\\
    \{\cT^{(1)}_{E_7,1},[I_1,\varnothing]\}
    \end{array}\right.$.
    
    \item $\qquad\cT_{E_7,1}^{(1)}\quad \to\quad
    \left\{\begin{array}{l}
    \{[I_2^*,\sof(12)],[I_1,\varnothing]\}\\
    \{\cT^{(1)}_{E_6,1},[I_1,\varnothing]\}
    \end{array}\right.$.
    
    \item $\qquad\cT_{E_6,1}^{(1)}\quad \to\quad
    \left\{\begin{array}{l}
    \{[I_6,\suf(6)],[I_1,\varnothing]^2\}\\
    \{\cT^{(1)}_{D_4,1},[I_1,\varnothing]^2\}
    \end{array}\right.$.
    
    \item $\qquad\blue{\cS_{\varnothing,2}^{(1)}}\quad \to\quad \{[I_1,\varnothing]^2,[I_4,\varnothing]\}$.
    
    \item $\qquad [I_n^*,\sof(2(n+4))]\quad \to\quad
    \left\{\begin{array}{l}
    \{[I_{n-1}^*,\sof(2(n+3)],[I_1,\varnothing]\}\\
    \{[I_{n-2}^*,\sof(2(n+2)],[I_2,\suf(2)]\}\\
    \{[I_{n+4},\suf(n+4)],[I_1,\varnothing]^2\}\\
    \{\cT^{(1)}_{D_4,1},[I_1,\varnothing]^2\}
    \end{array}\right.$.
    
    \item $\qquad [I_n,\suf(n)]\quad \to\quad \{[I_{n_1},\suf(n_1)],...,[I_{n_m},\suf(n_m)]\}$ with $\sum_{i=1}^m n_i=n$.
\end{itemize}

\begin{figure}
\begin{adjustbox}{center,max width=.75\textwidth}
\begin{tikzpicture}[decoration={markings,
mark=at position .5 with {\arrow{>}}},
Marrow/.style={->,>=stealth[round],shorten >=1pt,line width=.4mm,coolblack},
IRarrow/.style={->,>=stealth[round],shorten >=1pt,line width=.4mm,deepcarmine,dashed}]

\begin{scope}[scale=1.4,yshift=4cm].   %%%%% E8r2 and SO20 %%%%%%

\draw [draw=black] (-1.7,-3.2) rectangle (1.5,.3);
\draw [draw=black] (-1.7,.3) rectangle (1.5,.6);
\node[scale=.8] (title1) at (0,.45) {\small{$[\ef_8]_{24}{\times}\suf(2)_{13}$}};

\node[scale=.5] (E8r2_top) at (0,0) {};
\node[scale=.8] (E8r2_dim) at (0,.1) {$\H^{59}$};
\node[scale=.8] (E8r2_s1) at (-.2,-.5) {$\ef_8$};
\node[scale=.8] (E8r2_ts1) at (0,-1) {$\cT^{(1)}_{E_8,1}$};
\node[scale=.8] (E8r2_t2a) at (-.7,-1.4) {$\ef_8$};
\node[scale=.8] (E8r2_t2b) at (.7,-1.4) {$\af_1$};
\node[scale=.8] (E8r2_ts2a) at (-1,-2) {$\cT^{(1)}_{E_8,1}{\times}\cT^{(1)}_{E_8,1}$};
\node[scale=.8] (E8r2_ts2b) at (1,-2) {$\cT^{(1)}_{E_8,1}{\times}\H$};
\node[scale=.8] (E8r2_s3a) at (-.7,-2.7) {$\af_1\times \af_1$};
\node[scale=.8] (E8r2_s3b) at (.7,-2.7) {$\ef_8$};
\node[bbc,scale=.5] (E8r2_down) at (0,-3) {};
\draw[blue] (E8r2_top) -- (E8r2_ts1);
\draw[blue] (E8r2_ts1) -- (E8r2_ts2a);
\draw[blue] (E8r2_ts1) -- (E8r2_ts2b);
\draw[blue] (E8r2_ts2a) -- (E8r2_down);
\draw[blue] (E8r2_ts2b) -- (E8r2_down);
\node[] (E8b1) at (-1,-3.2) {};
\node[] (E8b2) at (-.3,-3.2) {};
\node[] (E8b3) at (.5,-3.2) {};

\draw [draw=black] (3.3,-2.7) rectangle (4.7,-0.25);  %%%%%%%%  SO20   %%%%%%%%
\draw [draw=black] (3.3,-0.25) rectangle (4.7,0.05);
\node[scale=.8] (title1) at (4,-0.1) {\small{$\sof(20)_{16}$}};

\node[scale=.8] (SO20_top) at (4,-.5) {};       
\node[scale=.8] (SO20_dim) at (4,-.4) {$\H^{46}$};
\node[scale=.8] (SO20_s1) at (3.8,-1) {$\ef_8$};
\node[scale=.8] (SO20_ts1) at (4,-1.5) {$\cT^{(1)}_{E_8,1}$};
\node[scale=.8] (SO20_s2) at (3.8,-2) {$\df_{10}$};
\node[bbc,scale=.5] (SO20_down) at (4,-2.5) {};
\draw[blue] (SO20_top) -- (SO20_ts1);
\draw[blue] (SO20_ts1) -- (SO20_down);
\node (SO20b1) at (3.5,-2.8) {};
\node (SO20b2) at (4.5,-2.8) {};

\end{scope}

\begin{scope}[scale=1.4,yshift=-4.5cm,xshift=-4cm] %E820, E7 rank2 SO16 and SU(10)%

\draw[draw=black] (-.7,1.3) rectangle (.7,3.75);    %%%%%%%%  E820   %%%%%%%%
\draw[draw=black] (-.7,3.75) rectangle (.7,4.05);
\node[scale=.8] (title1) at (0,3.9) {\small{$[\ef_8]_{20}$}};

\node[] (E820t) at (0,4.2) {};
\node[scale=.8] (E820_top) at (0,3.5) {};      
\node[scale=.8] (E820_dim) at (0,3.6) {$\H^{46}$};
\node[scale=.8] (E820_s1) at (-.2,3) {$\ef_7$};
\node[scale=.8] (E820_ts1) at (0,2.5) {$\cT^{(1)}_{E_7,1}$};
\node[scale=.8] (E820_s2) at (-.2,2) {$\ef_8$};
\node[bbc,scale=.5] (E820_down) at (0,1.5) {};
\draw[blue] (E820_top) -- (E820_ts1);
\draw[blue] (E820_ts1) -- (E820_down);
\node[] (E820b) at (0,1.3) {};

\node[] (E7r2t) at (3.8,4.5) {};
\draw [draw=black] (1.3,.8) rectangle (4.5,4.3);   %%%%%% E7r2 %%%%%%%%
\draw [draw=black] (1.3,4.3) rectangle (4.5,4.6);
\node[scale=.8] (title1) at (3,4.45) {\small{$[\ef_7]_{16}{\times}\suf(2)_9$}};

\node[scale=.5] (E7r2_top) at (3,4) {};    
\node[scale=.8] (E7r2_dim) at (3,4.1) {$\H^{35}$};
\node[scale=.8] (E7r2_s1) at (2.8,3.5) {$\ef_7$};
\node[scale=.8] (E7r2_ts1) at (3,3) {$\cT^{(1)}_{E_7,1}$};
\node[scale=.8] (E7r2_t2a) at (2.3,2.6) {$\ef_7$};
\node[scale=.8] (E7r2_t2b) at (3.7,2.6) {$\ef_7$};
\node[scale=.8] (E7r2_ts2a) at (2,2) {$\cT^{(1)}_{E_7,1}{\times}\cT^{(1)}_{E_7,1}$};
\node[scale=.8] (E7r2_ts2b) at (4,2) {$\cT^{(1)}_{E_7,1}{\times}\H$};
\node[scale=.8] (E7r2_s3a) at (2.3,1.3) {$\af_1\times \af_1$};
\node[scale=.8] (E7r2_s3b) at (3.7,1.3) {$\ef_7$};
\node[bbc,scale=.5] (E7r2_down) at (3,1) {};
\draw[blue] (E7r2_top) -- (E7r2_ts1);
\draw[blue] (E7r2_ts1) -- (E7r2_ts2a);
\draw[blue] (E7r2_ts1) -- (E7r2_ts2b);
\draw[blue] (E7r2_ts2a) -- (E7r2_down);
\draw[blue] (E7r2_ts2b) -- (E7r2_down);
\node[] (E7r2b1) at (1.5,.8) {};
\node[] (E7r2b2) at (2.2,.8) {};

\node[] (SO16t) at (7,4.2) {};
\draw [draw=black] (5.7,1.3) rectangle (8.3,3.75);    %%%%%% SO16 %%%%%%%%
\draw [draw=black] (5.7,3.75) rectangle (8.3,4.05);
\node[scale=.8] (title1) at (7,3.9) {\small{$\sof(16)_{12}{\times}\suf(2)_8$}};

\node[scale=.5] (SO16_top) at (7,3.5) {};   
\node[scale=.8] (SO16_dim) at (7,3.6) {$\H^{30}$};
\node[scale=.8] (SO16_t1a) at (6.3,3.1) {$\ef_8$};
\node[scale=.8] (SO16_t1b) at (7.7,3.1) {$\ef_7$};
\node[scale=.8] (SO16_ts1a) at (6,2.5) {$\cT^{(1)}_{E_8,1}$};
\node[scale=.8] (SO16_ts1b) at (8,2.5) {$\cT^{(1)}_{E_7,1}$};
\node[scale=.8] (SO16_s2a) at (6.3,1.8) {$\af_1$};
\node[scale=.8] (SO16_s2b) at (7.7,1.8) {$\df_8$};
\node[bbc,scale=.5] (SO16_down) at (7,1.5) {};
\draw[blue] (SO16_top) -- (SO16_ts1a);
\draw[blue] (SO16_top) -- (SO16_ts1b);
\draw[blue] (SO16_ts1a) -- (SO16_down);
\draw[blue] (SO16_ts1b) -- (SO16_down);
\node[] (SO16b1) at (6.5,1.3) {};
\node[] (SO16b2) at (7.5,1.3) {};

\node (SU10t) at (11,4.2) {};
\draw [draw=black] (10.3,1.3) rectangle (11.7,3.75);    %%%%%%%%  SU10   %%%%%%%%
\draw [draw=black] (10.3,3.75) rectangle (11.7,4.05);
\node[scale=.8] (title1) at (11,3.9) {\small{$\suf(10)_{10}$}};

\node[scale=.8] (SU10_top) at (11,3.5) {};       
\node[scale=.8] (SU10_dim) at (11,3.6) {$\H^{26}$};
\node[scale=.8] (SU10_s1) at (10.8,3) {$\ef_7$};
\node[scale=.8] (SU10_ts1) at (11,2.5) {$\cT^{(1)}_{E_7,1}$};
\node[scale=.8] (SU10_s2) at (10.8,2) {$\af_9$};
\node[bbc,scale=.5] (SU10_down) at (11,1.5) {};
\draw[blue] (SU10_top) -- (SU10_ts1);
\draw[blue] (SU10_ts1) -- (SU10_down);
\node[] (SU10b) at (11,1.3) {};

\end{scope}

\begin{scope}[scale=1.4,yshift=-5cm,xshift=-2cm] %E6 rank2, SO14 and SU(8)%

\node (E6r2t) at (0,.6) {};
\draw [draw=black] (-1.7,-3.2) rectangle (1.5,.3);   %%%%%% E6r2 %%%%%%%%

\node (SO12t) at (-1.9,-4.3) {};
\draw[IRarrow] (E7r2b1) to (SO12t);

\filldraw [fill=white, draw=black] (-1.7,-3.2) rectangle (1.5,.3);
\node[scale=.5] (E6r2_top) at (0,0) {};    
\node[scale=.8] (E6r2_dim) at (0,.1) {$\H^{23}$};
\node[scale=.8] (E6r2_s1) at (-.2,-.5) {$\ef_6$};
\node[scale=.8] (E6r2_ts1) at (0,-1) {$\cT^{(1)}_{E_6,1}$};
\node[scale=.8] (E6r2_t2a) at (-.7,-1.4) {$\ef_6$};
\node[scale=.8] (E6r2_t2b) at (.7,-1.4) {$\ef_6$};
\node[scale=.8] (E6r2_ts2a) at (-1,-2) {$\cT^{(1)}_{E_6,1}{\times}\cT^{(1)}_{E_6,1}$};
\node[scale=.8] (E6r2_ts2b) at (1,-2) {$\cT^{(1)}_{E_6,1}{\times}\H$};
\node[scale=.8] (E6r2_s3a) at (-.7,-2.7) {$\af_1\times \af_1$};
\node[scale=.8] (E6r2_s3b) at (.7,-2.7) {$\ef_6$};
\node[bbc,scale=.5] (E6r2_down) at (0,-3) {};
\draw[blue] (E6r2_top) -- (E6r2_ts1);
\draw[blue] (E6r2_ts1) -- (E6r2_ts2a);
\draw[blue] (E6r2_ts1) -- (E6r2_ts2b);
\draw[blue] (E6r2_ts2a) -- (E6r2_down);
\draw[blue] (E6r2_ts2b) -- (E6r2_down);
\node[] (E6r2b1) at (1,-3.2) {};
\node[] (E6r2b2) at (2,-3.2) {};
\draw [draw=black] (-1.7,.3) rectangle (1.5,.6);
\filldraw [fill=white, draw=black] (-1.7,.3) rectangle (1.5,.6);
\node[scale=.8] (title1) at (0,.45) {\small{$[\ef_6]_{12}{\times}\suf(2)_7$}};

\node[] (SO14t) at (4,.2) {};
\draw[draw=black] (3.2,-2.7) rectangle (4.8,-.25);    %%%%%%%%  SO14   %%%%%%%%
\draw[draw=black] (3.2,-.25) rectangle (4.8,.05);
\node[scale=.8] (title1) at (4,-.1) {\small{$\sof(14)_{10}{\times}\uf(1)$}};

\node[scale=.8] (SO14_top) at (4,-.5) {};      
\node[scale=.8] (SO14_dim) at (4,-0.4) {$\H^{22}$};
\node[scale=.8] (SO14_s1) at (3.8,-1) {$\ef_6$};
\node[scale=.8] (SO14_ts1) at (4,-1.5) {$\cT^{(1)}_{E_6,1}$};
\node[scale=.8] (SO14_s2) at (3.8,-2) {$\df_7$};
\node[bbc,scale=.5] (SO14_down) at (4,-2.5) {};
\draw[blue] (SO14_top) -- (SO14_ts1);
\draw[blue] (SO14_ts1) -- (SO14_down);
\node[] (SO14b1) at (3.5,-2.7) {};
\node[] (SO14b2) at (4.5,-2.7) {};

\node[] (SU8t) at (7,.2) {};
\draw [draw=black] (5.7,-2.7) rectangle (8.3,-.25);    %%%%%% SU8 %%%%%%%%
\draw [draw=black] (5.7,-.25) rectangle (8.3,.05);
\node[scale=.8] (title1) at (7,-.1) {\small{$\suf(2)_6{\times}\suf(8)_8$}};

\node[scale=.5] (SU8_top) at (7,-.5) {};    
\node[scale=.8] (SU8_dim) at (7,-.4) {$\H^{18}$};
\node[scale=.8] (SU8_t1a) at (6.3,-.9) {$\ef_7$};
\node[scale=.8] (SU8_t1b) at (7.7,-.9) {$\ef_6$};
\node[scale=.8] (SU8_ts1a) at (6,-1.5) {$\cT^{(1)}_{E_7,1}$};
\node[scale=.8] (SU8_ts1b) at (8,-1.5) {$\cT^{(1)}_{E_6,1}$};
\node[scale=.8] (SU8_s2a) at (6.3,-2.2) {$\af_1$};
\node[scale=.8] (SU8_s2b) at (7.7,-2.2) {$\af_7$};
\node[bbc,scale=.5] (SU8_down) at (7,-2.5) {};
\draw[blue] (SU8_top) -- (SU8_ts1a);
\draw[blue] (SU8_top) -- (SU8_ts1b);
\draw[blue] (SU8_ts1a) -- (SU8_down);
\draw[blue] (SU8_ts1b) -- (SU8_down);
\node[] (SU8b1) at (6.5,-2.7) {};
\node[] (SU8b2) at (7.5,-2.7) {};

\end{scope}

\begin{scope}[scale=1.4,yshift=-13.5cm,xshift=-4cm] %Lagrangian level, SO12, D4r2, U6, SU25%

\node (SO12t) at (0,4.2) {};
\draw [draw=black] (-.8,1.3) rectangle (.8,3.75);    %%%%%%%%  SO12   %%%%%%%%
\draw [draw=black] (-.8,3.75) rectangle (.8,4.05);
\filldraw [fill=yellow!40, draw=black] (-.8,3.75) rectangle (.8,4.05);
\node[scale=.8] (title1) at (0,3.9) {\small{$USp(4)+6F$}};

\node[scale=.8] (SO12_top) at (0,3.5) {};      
\node[scale=.8] (SO12_dim) at (0,3.6) {$\H^{14}$};
\node[scale=.8] (SO12_s1) at (-.2,3) {$\df_4$};
\node[scale=.8] (SO12_ts1) at (0,2.5) {$\cT^{(1)}_{D_4,1}$};
\node[scale=.8] (SO12_s2) at (-.2,2) {$\df_6$};
\node[scale=.8] (SO12_down) at (0,1.5) {};
\node[bbc,scale=.5] (SO12_down) at (0,1.5) {};
\draw[blue] (SO12_top) -- (SO12_ts1);
\draw[blue] (SO12_ts1) -- (SO12_down);

\node (D4r2t) at (3.5,4.65) {};
\draw [draw=black] (1.8,.8) rectangle (5,4.3);    %%%%%% D4r2 %%%%%%%%
\draw [draw=black] (1.8,4.3) rectangle (5,4.6);
\filldraw [fill=yellow!40, draw=black] (1.8,4.3) rectangle (5,4.6);
\node[scale=.8] (title1) at (3.5,4.45) {\small{$USp(4)+4F+V$}};

\node[scale=.5] (D4r2_top) at (3.5,4) {};   
\node[scale=.8] (D4r2_dim) at (3.5,4.1) {$\H^{11}$};
\node[scale=.8] (D4r2_s1) at (3.3,3.5) {$\df_4$};
\node[scale=.8] (D4r2_ts1) at (3.5,3) {$\cT^{(1)}_{D_4,1}$};
\node[scale=.8] (D4r2_t2a) at (2.8,2.6) {$\df_4$};
\node[scale=.8] (D4r2_t2b) at (4.2,2.6) {$\df_4$};
\node[scale=.8] (D4r2_ts2a) at (2.5,2) {$\cT^{(1)}_{D_4,1}{\times}\cT^{(1)}_{D_4,1}$};
\node[scale=.8] (D4r2_ts2b) at (4.5,2) {$\cT^{(1)}_{D_4,1}{\times}\H$};
\node[scale=.8] (D4r2_s3a) at (2.8,1.3) {$\af_1\times \af_1$};
\node[scale=.8] (D4r2_s3b) at (4.2,1.3) {$\df_4$};
\node[bbc,scale=.5] (D4r2_down) at (3.5,1) {};
\draw[blue] (D4r2_top) -- (D4r2_ts1);
\draw[blue] (D4r2_ts1) -- (D4r2_ts2a);
\draw[blue] (D4r2_ts1) -- (D4r2_ts2b);
\draw[blue] (D4r2_ts2a) -- (D4r2_down);
\draw[blue] (D4r2_ts2b) -- (D4r2_down);

\node (SU6t) at (7,4.2) {};
\draw [draw=black] (6.2,1.3) rectangle (7.8,3.75);    %%%%%%%%  SU6   %%%%%%%%
\draw [draw=black] (6.2,3.75) rectangle (7.8,4.05);
\filldraw [fill=yellow!40, draw=black] (6.2,3.75) rectangle (7.8,4.05);
\node[scale=.8] (title1) at (7,3.9) {\small{$SU(3)+6F$}};

\node[scale=.8] (SU6_top) at (7,3.5) {};       
\node[scale=.8] (SU6_dim) at (7,3.6) {$\H^{10}$};
\node[scale=.8] (SU6_s1) at (6.8,3) {$\df_4$};
\node[scale=.8] (SU6_ts1) at (7,2.5) {$\cT^{(1)}_{D_4,1}$};
\node[scale=.8] (SU6_s2) at (6.8,2) {$\af_5$};
\node[bbc,scale=.5] (SU6_down) at (7,1.5) {};
\draw[blue] (SU6_top) -- (SU6_ts1);
\draw[blue] (SU6_ts1) -- (SU6_down);

\node (SU25t) at (11,4.2) {};
\draw [draw=black] (9.6,1.3) rectangle (12.4,3.75);    %%%%%%%%  SU25   %%%%%%%%
\draw [draw=black] (9.6,3.75) rectangle (12.4,4.05);
\filldraw [fill=yellow!40, draw=black] (9.6,3.75) rectangle (12.4,4.05);
\node[scale=.8] (title1) at (11,3.9) {\small{$2F-SU(2)-SU(2)-2F$}};

\node[scale=.8] (SU25_top) at (11,3.5) {};      
\node[scale=.8] (SU25_dim) at (11,3.6) {$\H^{6}$};
\node[scale=.8] (SU25_s1) at (10.8,3) {$\df_4$};
\node[scale=.8] (SU25_ts1) at (11,2.5) {$\cT^{(1)}_{D_4,1}$};
\node[scale=.8] (SU25_s2) at (10.8,2) {$\af_1$};
\node[bbc,scale=.5] (SU25_down) at (11,1.5) {};
\draw[blue] (SU25_top) -- (SU25_ts1);
\draw[blue] (SU25_ts1) -- (SU25_down);
\end{scope}

%%%% starting from level 1 %%%%%

\draw[Marrow] (E8b1) to (E820t);
\draw[Marrow] (E8b2) to (E7r2t);
\draw[Marrow] (E8b3) to (SO14t);
\draw[Marrow] (SO20b1) to (SO16t);
\draw[Marrow] (SO20b2) to (SU10t);

%%%% starting from level 2 %%%%%

\draw[Marrow] (E7r2b2) to (E6r2t);
\draw[Marrow] (SO16b1) to (SO14t);
\draw[Marrow] (SO16b2) to (SU8t);
\draw[Marrow] (SU10b) to (SU8t);

\draw[IRarrow] (E820b) to (SO12t);

%%%% starting from level 3 %%%%%

\draw[IRarrow] (E6r2b1) to (D4r2t);
\draw[Marrow] (E6r2b2) to (SU6t);

\draw[IRarrow] (SO14b1) to (SO12t);
\draw[IRarrow] (SO14b2) to (SU6t);
\draw[IRarrow] (SU8b1) to (SU6t);
\draw[IRarrow] (SU8b2) to (SU25t);

\end{tikzpicture}
{\caption{\label{fig:HBE8r2} Higgs branch analysis of the mass deformations among the entries of the $\ef_8-\sof(20)$ series.}}
\end{adjustbox}
\end{figure}

In most cases, and by using the data from the table \ref{Higgs:one}, it is quite straightforward to check that the HBs of theories related by mass deformations in figure \ref{fig:HBE8r2} do indeed pass criterion \ref{Higgs}. We will here only discuss the most subtle deformations:
\begin{itemize}
    \item The case $\sof(20)_{16}\to\suf(2)_8{\times}\sof(16)_{12}$ has the apparent contradiction that the $\cT^{(1)}_{E_8,1}$ is left undeformed by the mass deformation, but we already discussed above that indeed the particular mass parameter which triggers the flow does not act on $\cT^{(1)}_{E_8,1}$.
    
    \item The mass deformation from $\suf(2)_8{\times}\sof(16)_{12}\to\suf(2)_6{\times}\suf(8)_8$ presents again a similar problem, with the $\cT^{(1)}_{E_7,1}$ present in the list of bottom theories in both the initial SCFT and after turning on a mass deformation. As in the previous case, the resolution lies in the fact that an $\suf(2)$ subgroup of the unbroken flavor symmetry along the $\df_8$ leaf simply does not act on the $\cT^{(1)}_{E_7,1}$. The reader can easily convince themselves with the help of the data in table \ref{Higgs:one}.
    
    \item In all the following cases: $[\ef_8]_{24}{\times}\suf(2)_{13}\to[\ef_8]_{20}$, $\suf(2)_8{\times}\sof(16)_{12}\to\sof(14){\times}\uf(1)$, $\suf(2)_6{\times}\suf(8)_8\to\uf(6)_6$ as well as $\suf(2)_6{\times}\suf(8)_8\to\suf(2)^5_4$, the initial SCFTs have an $\suf(2)$ flavor symmetry and the mass deformations which initiates the flows explicitly break this $\suf(2)$ factor. Therefore we expect that the SCFT ``loses'' an irreducible bottom symplectic leaf, compatibly with what we see in figure \ref{fig:HBE8r2}.
\end{itemize}

\subsection{$\spf(12)-\spf(8)-\ff_4$ series}

Let's now start performing the analysis the moduli spaces of the theories in the $\spf(12)-\spf(8)-\ff_4$ series. The CB stratifications are depicted in figure \ref{fig:CBsu12} while the HB in figure \ref{fig:HBSp12}.

The CB analysis relies on the results of mass deformations of discretely gauged rank-1 theories \cite{Argyres:2016yzz}. We report here the main ones:
\begin{itemize}
    \item $[\cT^{(1)}_{E_6,1}]_{\Z_2}\quad\to\quad\left\{
    \begin{array}{l}
    \{[\cT_{D_4,1}^{(1)}]_{\Z_2},\}\\
    \{[I_6,\suf(6)]_{\Z_2},\}
    \end{array}
    \right.$.
    
    \item $[\cT^{(1)}_{D_4,1}]_{\Z_2}\quad\to\quad\{[\cT_{A_2,1}^{(1)}]_{\Z_2},\}$.
    
    \item $[I_{2n},\suf(2n)]_{\Z_2}\quad\to\quad\{[I_{2(n-m)},\suf(2(n-m))]_{\Z_2},[I_m,\suf(m)]\}$.
    
    \item $[I_1^*,\spf(4)]\quad \to \quad\left\{
    \begin{array}{l}
    \{\blue{\cS^{(1)}_{\varnothing,2}},[I_1,\varnothing]\}\\
    \{[I_2,\suf(2)],[I_1,\varnothing],[I_4,\varnothing]\}
    \end{array}
    \right.$.
\end{itemize}

\begin{figure}
\begin{tikzpicture}[decoration={markings,
mark=at position .5 with {\arrow{>}}},
Marrow/.style={->,>=stealth[round],shorten >=1pt,line width=.4mm,coolblack},
IRarrow/.style={->,>=stealth[round],shorten >=1pt,line width=.4mm,deepcarmine,dashed}]
\begin{scope}[scale=1.4] %SP12, SP4SP8 and F4SU22%
\node[bbc,scale=.5] (Sp12_top) at (-3.1,0) {};       %%%%%%%%  Sp12   %%%%%%%%
\draw [draw=black] (-4.6,.1) rectangle (-1.4,.4);
\filldraw [fill=white, draw=black] (-4.6,.1) rectangle (-1.4,.4);
\node[scale=.8] (title1) at (-3.1,.25) {\small{$\spf(12)_8$}};
\draw [draw=black] (-4.6,-1.5) rectangle (-1.4,.1);
\node[bbc,scale=.5] (Sp12_down) at (-3.1,-1.4) {};
\node[scale=.8] (Sp12_1a) at (-4.1,-.7) {\ \ \ \ $[I_1,\varnothing]$};
\node[scale=.8] (Sp12_1b) at (-3.1,-.7) {$[I_1,\varnothing]$\ \ };
\node[scale=.8] (Sp12_1c) at (-2.1,-.7) {$[I_{12},\suf(12)]_{\Z_2}$};
\draw[red] (Sp12_top) -- (Sp12_1a);
\draw[red] (Sp12_top) -- (Sp12_1b);
\draw[red] (Sp12_top) -- (Sp12_1c);
\draw[red] (Sp12_down) -- (Sp12_1a);
\draw[red] (Sp12_down) -- (Sp12_1b);
\draw[red] (Sp12_down) -- (Sp12_1c);
\node[bbc,scale=.5] (Sp4Sp8_top) at (.2,0) {};       %%%%%%%%  Sp4Sp8   %%%%%%%%
\draw [draw=black] (-1.3,.1) rectangle (1.8,.4);
\filldraw [fill=white, draw=black] (-1.3,.1) rectangle (1.8,.4);
\node[scale=.8] (title1) at (.2,.25) {\small{$\spf(4)_7{\times}\spf(8)_8$}};
\draw [draw=black] (-1.3,-1.5) rectangle (1.8,.1);
\node[bbc,scale=.5] (Sp4Sp8_down) at (.2,-1.4) {};
\node[scale=.8] (Sp4Sp8_1a) at (-.8,-.7) {$[I_1,\varnothing]$};
\node[scale=.8] (Sp4Sp8_1b) at (0.2,-.7) {$[I_1^*,\spf(4)]$\ \ \ };
\node[scale=.8] (Sp4Sp8_1c) at (1.2,-.7) {$[I_8,\suf(8)]_{\Z_2}$};
\draw[red] (Sp4Sp8_top) -- (Sp4Sp8_1a);
\draw[red] (Sp4Sp8_top) -- (Sp4Sp8_1b);
\draw[red] (Sp4Sp8_top) -- (Sp4Sp8_1c);
\draw[red] (Sp4Sp8_down) -- (Sp4Sp8_1a);
\draw[red] (Sp4Sp8_down) -- (Sp4Sp8_1b);
\draw[red] (Sp4Sp8_down) -- (Sp4Sp8_1c);
\node[bbc,scale=.5] (F4SU22_top) at (3.1,0) {};       %%%%%%%%  F4SU22   %%%%%%%%
\draw [draw=black] (1.85,.1) rectangle (4.35,.4);
\filldraw [fill=white, draw=black] (1.85,.1) rectangle (4.35,.4);
\node[scale=.8] (title1) at (3.1,.25) {\small{$[\ff_4]_{12}{\times}\suf(2)_7^2$}};
\draw [draw=black] (1.85,-1.5) rectangle (4.35,.1);
\node[bbc,scale=.5] (F4SU22_down) at (3.1,-1.4) {};
\node[scale=.8] (F4SU22_1a) at (2.1,-.7) {$\blue{\cS^{(1)}_{\varnothing,2}}$};
\node[scale=.8] (F4SU22_1b) at (3.1,-.7) {$[\cT^{(1)}_{E_6,1}]_{\Z_2}$};
\node[scale=.8] (F4SU22_1c) at (4.1,-.7) {$\blue{\cS^{(1)}_{\varnothing,2}}$};
\draw[red] (F4SU22_top) -- (F4SU22_1a);
\draw[red] (F4SU22_top) -- (F4SU22_1b);
\draw[red] (F4SU22_top) -- (F4SU22_1c);
\draw[red] (F4SU22_down) -- (F4SU22_1a);
\draw[red] (F4SU22_down) -- (F4SU22_1b);
\draw[red] (F4SU22_down) -- (F4SU22_1c);
\end{scope}

\begin{scope}[scale=1.4,yshift=-2cm,xshift=0cm] %Sp8SU2, Sp6SU2, SO7SU22, F4U1%
\node[bbc,scale=.5] (Sp8SU2_top) at (-4,0) {};       %%%%%%%%  Sp8SU2   %%%%%%%%
\draw [draw=black] (-5.4,-1.5) rectangle (-2.4,.1);
\node[bbc,scale=.5] (Sp8SU2_down) at (-4,-1.4) {};
\node[scale=.8] (Sp8SU2_1a) at (-5,-.7) {$[I_1,\varnothing]$};
\node[scale=.8] (Sp8SU2_1b) at (-4,-.7) {$[I_8,\suf(8)]_{\Z_2}$\ \ };
\node[scale=.8] (Sp8SU2_1c) at (-3,-.7) {\ \ $[I_2,\suf(2)]$};
\draw[red] (Sp8SU2_top) -- (Sp8SU2_1a);
\draw[red] (Sp8SU2_top) -- (Sp8SU2_1b);
\draw[red] (Sp8SU2_top) -- (Sp8SU2_1c);
\draw[red] (Sp8SU2_down) -- (Sp8SU2_1a);
\draw[red] (Sp8SU2_down) -- (Sp8SU2_1b);
\draw[red] (Sp8SU2_down) -- (Sp8SU2_1c);
\node[bbc,scale=.5] (Sp6SU2_top) at (-.7,0) {};       %%%%%%%%  Sp6SU2   %%%%%%%%

\draw [draw=black] (-2.2,-1.5) rectangle (.9,.1);
\node[bbc,scale=.5] (Sp6SU2_down) at (-.7,-1.4) {};
\node[scale=.8] (Sp6SU2_1a) at (-1.7,-.7) {$\blue{\cS^{(1)}_{\varnothing,2}}$};
\node[scale=.8] (Sp6SU2_1b) at (-.7,-.7) {$[I_1,\varnothing]$};
\node[scale=.8] (Sp6SU2_1c) at (.3,-.7) {$[I_6,\suf(6)]_{\Z_2}$};
\draw[red] (Sp6SU2_top) -- (Sp6SU2_1a);
\draw[red] (Sp6SU2_top) -- (Sp6SU2_1b);
\draw[red] (Sp6SU2_top) -- (Sp6SU2_1c);
\draw[red] (Sp6SU2_down) -- (Sp6SU2_1a);
\draw[red] (Sp6SU2_down) -- (Sp6SU2_1b);
\draw[red] (Sp6SU2_down) -- (Sp6SU2_1c);

\node[bbc,scale=.5] (F4U1_top) at (2.4,0) {};       %%%%%%%%  F4U1   %%%%%%%%

\draw [draw=black] (1.35,-1.5) rectangle (3.5,.1);
\node[bbc,scale=.5] (F4U1_down) at (2.4,-1.4) {};
\node[scale=.8] (F4U1_1b) at (1.7,-.7) {$\ \ \ [\cT^{(1)}_{E_6,1}]_{\Z_2}$};
\node[scale=.8] (F4U1_1a) at (3.1,-.7) {$[I_1,\varnothing]$};
\draw[red] (F4U1_top) -- (F4U1_1a);
\draw[red] (F4U1_top) -- (F4U1_1b);
\draw[red] (F4U1_down) -- (F4U1_1a);
\draw[red] (F4U1_down) -- (F4U1_1b);

\node[bbc,scale=.5] (SO7SU22_top) at (5.15,0) {};       %%%%%%%%  SO7SU22   %%%%%%%%

\draw [draw=black] (3.9,-1.5) rectangle (6.4,.1);
\node[bbc,scale=.5] (SO7SU22_down) at (5.15,-1.4) {};
\node[scale=.8] (SO7SU22_1a) at (6.15,-.7) {$\blue{\cS^{(1)}_{\varnothing,2}}$};
\node[scale=.8] (SO7SU22_1b) at (4.15,-.7) {$\quad\ \ [\cT^{(1)}_{D_4,1}]_{\Z_2}$};
\node[scale=.8] (SO7SU22_1c) at (5.15,-.7) {$\blue{\cS^{(1)}_{\varnothing,2}}$};
\draw[red] (SO7SU22_top) -- (SO7SU22_1a);
\draw[red] (SO7SU22_top) -- (SO7SU22_1b);
\draw[red] (SO7SU22_top) -- (SO7SU22_1c);
\draw[red] (SO7SU22_down) -- (SO7SU22_1a);
\draw[red] (SO7SU22_down) -- (SO7SU22_1b);
\draw[red] (SO7SU22_down) -- (SO7SU22_1c);

%%%%% Mass deformations arrows %%%%%%

\draw[Marrow] (Sp12_1c) -- (Sp8SU2_1c);
\draw[Marrow] (Sp12_1c) -- (Sp8SU2_1b);

\draw[Marrow] (Sp4Sp8_1b) -- (Sp6SU2_1a);
\draw[Marrow] (Sp4Sp8_1c) -- (Sp6SU2_1b);
\draw[Marrow] (Sp4Sp8_1c) -- (Sp6SU2_1c);
\draw[Marrow] (Sp4Sp8_1b) -- (Sp8SU2_1c);

\draw[Marrow] (F4SU22_1b) -- (SO7SU22_1b);
\draw[Marrow] (F4SU22_1a) -- (F4U1_1a);
\draw[Marrow] (F4SU22_1c) -- (F4U1_1a);
\draw[Marrow] (F4SU22_1a) -- (Sp6SU2_1b);
\draw[Marrow] (F4SU22_1b) -- (Sp6SU2_1c);

%%%%% Titles and tables  %%%%%%%

\draw [draw=black] (-5.4,.1) rectangle (-2.4,.4);
\filldraw [fill=white, draw=black] (-5.4,.1) rectangle (-2.4,.4);
\node[scale=.8] (title1) at (-4,.25) {\small{$\spf(8)_6{\times}\suf(2)_8$}};

\draw [draw=black] (-2.2,.1) rectangle (.9,.4);
\filldraw [fill=white, draw=black] (-2.2,.1) rectangle (.9,.4);
\node[scale=.8] (title1) at (-.7,.25) {\small{$\spf(6)_6{\times}\suf(2)_5{\times}\uf(1)$}};

\draw [draw=black] (3.9,.1) rectangle (6.4,.4);
\filldraw [fill=white, draw=black] (3.9,.1) rectangle (6.4,.4);
\node[scale=.8] (title1) at (5.15,.25) {\small{$\sof(7)_8{\times}\suf(2)_5^2$}};

\draw [draw=black] (1.35,.1) rectangle (3.5,.4);
\filldraw [fill=white, draw=black] (1.35,.1) rectangle (3.5,.4);
\node[scale=.8] (title1) at (2.4,.25) {\small{$[\ff_4]_{10}{\times}\uf(1)$}};

\end{scope}

\begin{scope}[scale=1.4,yshift=-4cm,xshift=0cm] %Sp6 and SU3SU22%
\node[bbc,scale=.5] (Sp6_top) at (-1.4,0) {};       %%%%%%%%  Sp6   %%%%%%%%
\draw [draw=black] (-2.7,-1.5) rectangle (-.1,.1);
\node[bbc,scale=.5] (Sp6_down) at (-1.4,-1.4) {};
\node[scale=.8] (Sp6_1a) at (-.7,-.7) {$[I_1,\varnothing]$};
\node[scale=.8] (Sp6_1b) at (-2.1,-.7) {$[I_6,\suf(6)]_{\Z_2}$};
\draw[red] (Sp6_top) -- (Sp6_1a);
\draw[red] (Sp6_top) -- (Sp6_1b);
\draw[red] (Sp6_down) -- (Sp6_1a);
\draw[red] (Sp6_down) -- (Sp6_1b);
\node[bbc,scale=.5] (SU3SU22_top) at (4.5,0) {};       %%%%%%%%  SU3SU22   %%%%%%%%
\draw [draw=black] (3.05,-1.5) rectangle (5.9,.1);
\node[bbc,scale=.5] (SU3SU22_down) at (4.5,-1.4) {};
\node[scale=.8] (SU3SU22_1a) at (3.5,-.7) {$[\cT^{(1)}_{A_1,1}]_{\Z_2}$};
\node[scale=.8] (SU3SU22_1b) at (4.5,-.7) {$\blue{\cS^{(1)}_{\varnothing,2}}$};
\node[scale=.8] (SU3SU22_1c) at (5.5,-.7) {$\blue{\cS^{(1)}_{\varnothing,2}}$};
\draw[red] (SU3SU22_top) -- (SU3SU22_1a);
\draw[red] (SU3SU22_top) -- (SU3SU22_1b);
\draw[red] (SU3SU22_top) -- (SU3SU22_1c);
\draw[red] (SU3SU22_down) -- (SU3SU22_1a);
\draw[red] (SU3SU22_down) -- (SU3SU22_1b);
\draw[red] (SU3SU22_down) -- (SU3SU22_1c);

%%%% Mass deformations and arrows %%%%%

\draw[Marrow] (Sp6SU2_1a) -- (Sp6_1a);

\draw[Marrow] (Sp8SU2_1b) -- (Sp6_1b);

\draw[Marrow] (SO7SU22_1b) -- (SU3SU22_1a);

%%%%% Titles and tables for the theories %%%%%%%

\draw [draw=black] (-2.7,.1) rectangle (-.1,.4);
\filldraw [fill=white, draw=black] (-2.7,.1) rectangle (-.1,.4);
\node[scale=.8] (title1) at (-1.4,.25) {\small{$\spf(6)_5{\times}\uf(1)$}};

\draw [draw=black] (3.05,.1) rectangle (5.9,.4);
\filldraw [fill=white, draw=black] (3.05,.1) rectangle (5.9,.4);
\node[scale=.8] (title1) at (4.5,.25) {\small{$\suf(3)_6{\times}\suf(2)_4^2$}};

\end{scope}

\begin{scope}[scale=1.4,yshift=-6cm,xshift=0cm] %Lagrangian level, Sp4 and SU26%
\node[bbc,scale=.5] (Sp4_top) at (-1.4,0) {};       %%%%%%%%  Sp4   %%%%%%%%

\draw [draw=black] (-3.3,-1.5) rectangle (.5,.1);
\node[bbc,scale=.5] (Sp4_down) at (-1.4,-1.4) {};
\node[scale=.8] (Sp4_1a) at (-2.9,-.7) {$[I_1,\varnothing]$};
\node[scale=.8] (Sp4_1b) at (-2.25,-.7) {$[I_1,\varnothing]$};
\node[scale=.8] (Sp4_1c) at (-1.6,-.7) {$[I_1,\varnothing]$};
\node[scale=.8] (Sp4_1d) at (-.95,-.7) {$[I_1,\varnothing]$};
\node[scale=.8] (Sp4_1e) at (-.05,-.7) {$[I_4,\suf(4)]_{\Z_2}$};
\draw[red] (Sp4_top) -- (Sp4_1a);
\draw[red] (Sp4_top) -- (Sp4_1b);
\draw[red] (Sp4_top) -- (Sp4_1c);
\draw[red] (Sp4_top) -- (Sp4_1d);
\draw[red] (Sp4_top) -- (Sp4_1e);
\draw[red] (Sp4_down) -- (Sp4_1a);
\draw[red] (Sp4_down) -- (Sp4_1b);
\draw[red] (Sp4_down) -- (Sp4_1c);
\draw[red] (Sp4_down) -- (Sp4_1d);
\draw[red] (Sp4_down) -- (Sp4_1e);
\node[bbc,scale=.5] (SU26_top) at (4.5,0) {};       %%%%%%%%  SU26   %%%%%%%%

\draw [draw=black] (3.5,-1.5) rectangle (5.5,.1);
\node[bbc,scale=.5] (SU26_down) at (4.5,-1.4) {};
\node[scale=.8] (SU26_1a) at (3.8,-.7) {$\blue{\cS^{(1)}_{\varnothing,2}}$};
\node[scale=.8] (SU26_1b) at (4.5,-.7) {$[I_0,\varnothing]$};
\node[scale=.8] (SU26_1c) at (5.2,-.7) {$\blue{\cS^{(1)}_{\varnothing,2}}$};
\draw[red] (SU26_top) -- (SU26_1a);
\draw[red,dotted] (SU26_top) -- (SU26_1b);
\draw[red] (SU26_top) -- (SU26_1c);
\draw[red] (SU26_down) -- (SU26_1a);
\draw[red,dotted] (SU26_down) -- (SU26_1b);
\draw[red] (SU26_down) -- (SU26_1c);

%%%% Mass deformations and arrows %%%%%

\draw[IRarrow] (Sp6_1b) -- (Sp4_1b);
\draw[IRarrow] (Sp6_1b) -- (Sp4_1c);
\draw[IRarrow] (Sp6_1b) -- (Sp4_1d);
\draw[IRarrow] (Sp6_1b) -- (Sp4_1e);

\draw[IRarrow] (SU3SU22_1a) -- (SU26_1b);

\draw[IRarrow] (F4U1_1b) -- (Sp4_1b);
\draw[IRarrow] (F4U1_1b) -- (Sp4_1c);
\draw[IRarrow] (F4U1_1b) -- (Sp4_1d);
\draw[IRarrow] (F4U1_1b) -- (Sp4_1e);

%%%%% Titles and tables for the theories %%%%%%%

\draw [draw=black] (-3.3,.1) rectangle (.5,.4);
\filldraw [fill=yellow!40, draw=black] (-3.3,.1) rectangle (.5,.4);
\node[scale=.8] (title1) at (-1.4,.25) {\small{$SU(2)-SU(2)$}};

\draw [draw=black] (3.5,.1) rectangle (5.5,.4);
\filldraw [fill=blue!10, draw=black] (3.5,.1) rectangle (5.5,.4);
\node[scale=.8] (title1) at (4.5,.25) {\small{$\cN=4\ SU(2)^2$}};

\end{scope}

\end{tikzpicture}
{\caption{\label{fig:CBsu12} Coulomb branch analysis of the mass deformations among the entries of the $\spf(12)-\spf(8)-\ff_4$ series.}}
\end{figure}

Again using the results in table \ref{Higgs:one}, most of the mass deformations from the HB perspective can be followed straightforwardly but there are some which show yet new features. These are in particular the flow from $\spf(12)_8$ and $\spf(4)_7{\times}\spf(8)_8$ into $\spf(8)_6{\times}\suf(2)_8$. Let's analyze them in detail.

\paragraph{$\spf(12)_8\to\spf(8)_6{\times}\suf(2)_8$} In the previous section, we have already encountered situations in which, after a mass deformation, a single bottom stratum splits into two. This happened when there was a two complex dimensional space of mass deformations which didn't act on the bottom theory and thus we could find a mass deformation which didn't lift the bottom stratum nor act on the bottom theory. In the case studied here we find a yet more exotic scenario when not only the stratum splits but the theory supported over one of the resulting strata is neither one of the original ones nor a mass deformation thereof, rather one which can be reached by a Higgs branch flow from the original bottom theory, \emph{i.e.} case $(iii)$ in criterion \ref{Higgs}.

Let's start from explaining the easy side of the HB Hasse diagram of the theory after mass deformation, this can be found in figure \ref{fig:HBSp12}. The $\cf_4$ stratum supports the $\cS^{(1)}_{D_4,2}$ theory which is a mass deformation of the $\cS^{(1)}_{E_6,2}$, the bottom theory of the original rank-2 theory. Since we know exactly what is the mass deformation which takes $\cS^{(1)}_{E_6,2}\to\cS^{(1)}_{D_4,2}$ \cite{Argyres:2016xmc}, by matching the $\spf(12)_8$ along the HB, we can identify exactly what mass deformation triggers the flow. In particular, by staring again at the appropriate entry in table \ref{Higgs:one}, the unbroken $\suf(2)$ factor of the $\spf(8)_6{\times}\suf(8)_8$ matches directly with the $\suf(8)_8$ factor of the $\cS^{(1)}_{D_4,2}$ which, following the deformation $\cS^{(1)}_{E_6,2}\to\cS^{(1)}_{D_4,2}$, tells us how this $\suf(2)$ embeds in the $\spf(10)$ flavor symmetry of $\cS^{(1)}_{E_6,2}$: the $\suf(2)$ is identified as the one associated to the highest root of $\spf(10)$. 

\begin{figure}
\begin{adjustbox}{center,max width=.7\textwidth}
\begin{tikzpicture}[decoration={markings,
mark=at position .5 with {\arrow{>}}},
Marrow/.style={->,>=stealth[round],shorten >=1pt,line width=.4mm,coolblack},
IRarrow/.style={->,>=stealth[round],shorten >=1pt,line width=.4mm,deepcarmine,dashed}]
\begin{scope}[scale=1.4,xshift=-2cm]   %%%%% Sp12, Sp4Sp8 and F4SU22  %%%%%%

\draw [draw=black] (-.9,.8) rectangle (.9,4.25);    %%%%%%%%  Sp12   %%%%%%%%
\draw [draw=black] (-.9,4.25) rectangle (.9,4.55); 
\node[scale=.8] (title1) at (0,4.4) {\small{$\spf(12)_8$}};

\node[scale=.8] (Sp12_top) at (0,4) {};       
\node[scale=.8] (Sp12_dim) at (0,4.1) {$\H^{22}$};
\node[scale=.8] (Sp12_s1) at (-.2,3.5) {$\ef_6$};
\node[scale=.8] (Sp12_ts1) at (0,3) {$\cT^{(1)}_{E_6,1}$};
\node[scale=.8] (Sp12_s2) at (-.2,2.5) {$\cf_{5}$};
\node[scale=.8] (Sp12_ts2) at (0,2) {$\cS^{(1)}_{E_6,2}$};
\node[scale=.8] (Sp12_s3) at (-.2,1.5) {$\cf_6$};
\node[bbc,scale=.5] (Sp12_down) at (0,1) {};
\draw[blue] (Sp12_top) -- (Sp12_ts1);
\draw[blue] (Sp12_ts1) -- (Sp12_ts2);
\draw[blue] (Sp12_ts2) -- (Sp12_down);
\node[] (Sp12b) at (0,.8) {};

\draw [draw=black] (2.6,.8) rectangle (5.5,4.25);   %%%%%%%%% Sp4Sp8 %%%%%%%%%%%  
\draw [draw=black] (2.6,4.25) rectangle (5.5,4.55);
\node[scale=.8] (title1) at (4,4.4) {\small{$\spf(4)_7{\times}\spf(8)_8$}};

\node[scale=.5] (Sp4Sp8_top) at (4.5,4) {};      
\node[scale=.8] (Sp4Sp8_dim) at (4.5,4.1) {$\H^{20}$};
\node[scale=.8] (Sp4Sp8_s1a) at (4,3.6) {$\ef_6$};
\node[scale=.8] (Sp4Sp8_s1b) at (5.05,3.55) {$\ef_7$};
\node[scale=.8] (Sp4Sp8_ts1a) at (3.8,3) {$\cT^{(1)}_{E_6,1}$};
\node[scale=.8] (Sp4Sp8_ts1b) at (5.2,3) {$\cT^{(1)}_{E_7,1}$};
\node[scale=.8] (Sp4Sp8_s2a) at (3.3,2.6) {$\cf_5$};
\node[scale=.8] (Sp4Sp8_s2b) at (4.05,2.4) {$\af_7$};
\node[scale=.8] (Sp4Sp8_s2c) at (5.05,2.5) {$\af_1$};
\node[scale=.8] (Sp4Sp8_ts2bc) at (4.5,2) {$\suf(2)_6{\times}\suf(8)_8$};
\node[scale=.8] (Sp4Sp8_ts2a) at (3.1,2) {$\cS^{(1)}_{E_6,2}$};
\node[scale=.8] (Sp4Sp8_s3a) at (3.3,1.4) {$\cf_4$};
\node[scale=.8] (Sp4Sp8_s3b) at (4.3,1.4) {$\cf_2$};
\node[bbc,scale=.5] (Sp4Sp8_t) at (3.8,1) {};
\draw[blue] (Sp4Sp8_top) -- (Sp4Sp8_ts1a);
\draw[blue] (Sp4Sp8_top) -- (Sp4Sp8_ts1b);
\draw[blue] (Sp4Sp8_ts1a) -- (Sp4Sp8_ts2a);
\draw[blue] (Sp4Sp8_ts1a) -- (Sp4Sp8_ts2bc);
\draw[blue] (Sp4Sp8_ts1b) -- (Sp4Sp8_ts2bc);
\draw[blue] (Sp4Sp8_ts2a) -- (Sp4Sp8_t);
\draw[blue] (Sp4Sp8_ts2bc) -- (Sp4Sp8_t);
\node[] (Sp4Sp8b1) at (3.5,.8) {};
\node[] (Sp4Sp8b2) at (4.5,.8) {};

\draw [draw=black] (6.7,.3) rectangle (9.4,4.75);   %%%%%%%%  F4SU22   %%%%%%%%
\draw [draw=black] (6.7,4.75) rectangle (9.4,5.05);
\node[scale=.8] (title1) at (8,4.9) {\small{$[\ff_4]_{12}{\times}\suf(2)_7^2$}};

\node[scale=.5] (F4SU22_top) at (8,4.5) {};          
\node[scale=.8] (F4SU22_dim) at (8,4.6) {$\H^{24}$};
\node[scale=.8] (F4SU22_s1) at (7.8,4) {$\ef_6$};
\node[scale=.8] (F4SU22_ts1) at (8,3.5) {$\cT^{(1)}_{E_6,2}$};
\node[scale=.8] (F4SU22_s2b) at (8.5,3.1) {$\ef_6$};
\node[scale=.8] (F4SU22_s2a) at (7.5,2.75) {$\cf_5$};
\node[scale=.8] (F4SU22_ts2b) at (8.7,2.5) {$\cT^{(1)}_{E_6,1}{\times}\cT^{(1)}_{E_6,1}$};
\node[scale=.8] (F4SU22_s3) at (8.55,2) {$\af_1$};
\node[scale=.8] (F4SU22_ts2a) at (7.3,2) {$\cS^{(1)}_{E_6,2}$};
\node[scale=.8] (F4SU22_ts3) at (8.7,1.5) {$[\ef_6]_{12}{\times}\suf(2)_7$};
\node[scale=.8] (F4SU22_ts4a) at (7.45,1.25) {$\ff_4$};
\node[scale=.8] (F4SU22_ts4b) at (8.6,1) {$\af_1$};
\node[bbc,scale=.5] (F4SU22_t) at (8,.5) {};
\draw[blue] (F4SU22_top) -- (F4SU22_ts1);
\draw[blue] (F4SU22_ts1) -- (F4SU22_ts2a);
\draw[blue] (F4SU22_ts1) -- (F4SU22_ts2b);
\draw[blue] (F4SU22_ts2b) -- (F4SU22_ts3);
\draw[blue] (F4SU22_ts3) -- (F4SU22_t);
\draw[blue] (F4SU22_ts2a) -- (F4SU22_t);
\node[] (F4SU22b1) at (7.3,.3) {};
\node[] (F4SU22b2) at (8,.3) {};
\node[] (F4SU22b3) at (8.7,.3) {};

\end{scope}

\begin{scope}[scale=1.4,yshift=-5.5cm,xshift=-4.5cm] %Sp8, SP6, SO7SU22 and F4U1%

\node[] (Sp8t) at (0,4.65) {};
\draw [draw=black] (-1.1,.8) rectangle (1.1,4.25);   %%%%%% Sp8 %%%%%%%%
\draw [draw=black] (-1.1,4.25) rectangle (1.1,4.55);
\node[scale=.8] (title1) at (0,4.4) {\small{$\spf(8)_6{\times}\suf(2)_8$}};

\node[scale=.5] (Sp8_top) at (0,4) {};    
\node[scale=.8] (Sp8_dim) at (0,4.1) {$\H^{12}$};
\node[scale=.8] (Sp8_s1a) at (-.6,3.25) {$\ef_6$};
\node[scale=.8] (Sp8_s1b) at (.7,3.5) {$\df_4$};
\node[scale=.8] (Sp8_ts1a) at (.7,3) {$\cT^{(1)}_{D_4,1}$};
\node[scale=.8] (Sp8_s2) at (.9,2.5) {$\cf_3$};
\node[scale=.8] (Sp8_ts2) at (.7,2) {$\cS^{(1)}_{D_4,2}$};
\node[scale=.8] (Sp8_ts1b) at (-.7,2.5) {$\cT^{(1)}_{E_6,1}$};
\node[scale=.8] (Sp8_s3a) at (-.6,1.75) {$\af_1$};
\node[scale=.8] (Sp8_s3b) at (.6,1.5) {$\cf_4$};
\node[bbc,scale=.5] (Sp8_down) at (0,1) {};
\draw[blue] (Sp8_top) -- (Sp8_ts1a);
\draw[blue] (Sp8_top) -- (Sp8_ts1b);
\draw[blue] (Sp8_ts1a) -- (Sp8_ts2);
\draw[blue] (Sp8_ts2) -- (Sp8_down);
\draw[blue] (Sp8_ts1b) -- (Sp8_down);
\node[] (Sp8b) at (0,0.8) {};

\node[] (Sp6t) at (4,4.65) {};
\draw [draw=black] (2.8,.8) rectangle (5.2,4.3);  %%%%%% Sp6 %%%%%%%%
\draw [draw=black] (2.8,4.3) rectangle (5.2,4.6);
\node[scale=.8] (title1) at (4,4.45) {\small{$\spf(6)_6{\times}\suf(2)_5{\times}\uf(1)$}};

\node[scale=.5] (Sp6_top) at (4,4) {};    
\node[scale=.8] (Sp6_dim) at (4,4.1) {$\H^{11}$};
\node[scale=.8] (Sp6_s1) at (3.8,3.5) {$\df_4$};
\node[scale=.8] (Sp6_ts1) at (4,3) {$\cT^{(1)}_{D_4,1}$};
\node[scale=.8] (Sp6_t2a) at (3.5,2.6) {$\cf_3$};
\node[scale=.8] (Sp6_t2b) at (4.5,2.6) {$\af_5$};
\node[scale=.8] (Sp6_ts2a) at (3.2,2) {$\cS^{(1)}_{D_4,2}$};
\node[scale=.8] (Sp6_ts2b) at (4.8,2) {\hspace{-.8cm}$SU(3)\text{+}6F$};
\node[scale=.8] (Sp6_s3a) at (3.5,1.3) {$\cf_3$};
\node[scale=.8] (Sp6_s3b) at (4.5,1.3) {$\af_1$};
\node[bbc,scale=.5] (Sp6_down) at (4,1) {};
\draw[blue] (Sp6_top) -- (Sp6_ts1);
\draw[blue] (Sp6_ts1) -- (Sp6_ts2a);
\draw[blue] (Sp6_ts1) -- (Sp6_ts2b);
\draw[blue] (Sp6_ts2a) -- (Sp6_down);
\draw[blue] (Sp6_ts2b) -- (Sp6_down);
\node[] (Sp6b) at (4,.8) {};

\node[] (SO7SU22t) at (8.5,5.15) {};
\draw [draw=black] (7.2,.3) rectangle (9.9,4.75);   %%%%%%%%  SO7SU22   %%%%%%%%
\draw [draw=black] (7.2,4.75) rectangle (9.9,5.05);
\node[scale=.8] (title1) at (8.5,4.9) {\small{$\sof(7)_8{\times}\suf(2)_5^2$}};

\node[scale=.5] (SO7SU22_top) at (8.5,4.5) {};          
\node[scale=.8] (SO7SU22_dim) at (8.5,4.6) {$\H^{12}$};
\node[scale=.8] (SO7SU22_s1) at (8.3,4) {$\df_4$};
\node[scale=.8] (SO7SU22_ts1) at (8.5,3.5) {$\cT^{(1)}_{D_4,1}$};
\node[scale=.8] (SO7SU22_s2b) at (9.05,3) {$\df_4$};
\node[scale=.8] (SO7SU22_s2a) at (7.8,2.75) {$\cf_3$};
\node[scale=.8] (SO7SU22_ts2b) at (9.2,2.5) {$\cT^{(1)}_{D_4,1}{\times}\cT^{(1)}_{D_4,1}$};
\node[scale=.8] (SO7SU22_s3) at (9.05,2) {$\af_1$};
\node[scale=.8] (SO7SU22_ts2a) at (7.8,2) {$\cS^{(1)}_{D_4,2}$};
\node[scale=.8] (SO7SU22_ts3) at (9.2,1.5) {\hspace{-.6cm}$USp(4)\text{+}4F\text{+}V$};
\node[scale=.8] (SO7SU22_ts4a) at (7.95,1.25) {$\bff_3$};
\node[scale=.8] (SO7SU22_ts4b) at (9.1,1) {$\af_1$};
\node[bbc,scale=.5] (SO7SU22_t) at (8.5,.5) {};
\draw[blue] (SO7SU22_top) -- (SO7SU22_ts1);
\draw[blue] (SO7SU22_ts1) -- (SO7SU22_ts2a);
\draw[blue] (SO7SU22_ts1) -- (SO7SU22_ts2b);
\draw[blue] (SO7SU22_ts2b) -- (SO7SU22_ts3);
\draw[blue] (SO7SU22_ts3) -- (SO7SU22_t);
\draw[blue] (SO7SU22_ts2a) -- (SO7SU22_t);
\node[] (SO7SU22b1) at (7.8,.3) {};
\node[] (SO7SU22b2) at (9.2,.3) {};

\node[] (F4U1t) at (13,4.65) {};
\draw [draw=black] (12.1,.8) rectangle (13.9,4.25);    %%%%%%%%  F4U1   %%%%%%%%
\draw [draw=black] (12.1,4.25) rectangle (13.9,4.55); 
\node[scale=.8] (title1) at (13,4.4) {\small{$[\ff_4]_{10}{\times}\uf(1)$}};

\node[scale=.8] (F4U1_top) at (13,4) {};      
\node[scale=.8] (F4U1_dim) at (13,4.1) {$\H^{16}$};
\node[scale=.8] (F4U1_s1) at (12.8,3.5) {$\df_4$};
\node[scale=.8] (F4U1_ts1) at (13,3) {$\cT^{(1)}_{D_4,1}$};
\node[scale=.8] (F4U1_s2) at (12.8,2.5) {$\cf_3$};
\node[scale=.8] (F4U1_ts2) at (13,2) {$\cS^{(1)}_{D_4,2}$};
\node[scale=.8] (F4U1_s3) at (12.8,1.5) {$\ff_4$};
\node[bbc,scale=.5] (F4U1_down) at (13,1) {};
\draw[blue] (F4U1_top) -- (F4U1_ts1);
\draw[blue] (F4U1_ts1) -- (F4U1_ts2);
\draw[blue] (F4U1_ts2) -- (F4U1_down);
\node[] (F4U1b) at (13,0.8) {};

\draw[Marrow] (Sp12b) -- (Sp8t);
\draw[Marrow] (Sp4Sp8b1) -- (Sp8t);
\draw[Marrow] (Sp4Sp8b2) -- (Sp6t);
\draw[Marrow] (F4SU22b1) -- (Sp6t);
\draw[Marrow] (F4SU22b2) -- (SO7SU22t);
\draw[Marrow] (F4SU22b3) -- (F4U1t);

\end{scope}

\begin{scope}[scale=1.4,yshift=-7cm,xshift=3.3cm] %Sp6U1 and SU3SU22%

\node[] (Sp6U1t) at (-4,.65) {};
\draw [draw=black] (-4.9,-3.2) rectangle (-3.1,.25);    %%%%%%%%  Sp6U1   %%%%%%%%
\draw [draw=black] (-4.9,.25) rectangle (-3.1,.55);

\node[] (Sp4t) at (-5.3,-4.2) {};
\draw[IRarrow] (SO7SU22b1) -- (Sp4t);
\draw[IRarrow] (F4U1b) -- (Sp4t);
\filldraw [fill=white, draw=black] (-4.9,-3.2) rectangle (-3.1,.25);
\node[scale=.8] (Sp6U1_top) at (-4,0) {};       
\node[scale=.8] (Sp6U1_dim) at (-4,.1) {$\H^7$};
\node[scale=.8] (Sp6U1_s1) at (-4.2,-.5) {$\af_2$};
\node[scale=.8] (Sp6U1_ts1) at (-4,-1) {$\cT^{(1)}_{A_2,1}$};
\node[scale=.8] (Sp6U1_s2) at (-4.2,-1.5) {$\cf_2$};
\node[scale=.8] (Sp6U1_ts2) at (-4,-2) {$\cS^{(1)}_{A_2,2}$};
\node[scale=.8] (Sp6U1_s3) at (-4.2,-2.5) {$\cf_3$};
\node[bbc,scale=.5] (Sp6U1_down) at (-4,-3) {};
\draw[blue] (Sp6U1_top) -- (Sp6U1_ts1);
\draw[blue] (Sp6U1_ts1) -- (Sp6U1_ts2);
\draw[blue] (Sp6U1_ts2) -- (Sp6U1_down);
\node[] (Sp6U1b) at (-4,-3.2) {};
\filldraw [fill=white, draw=black] (-4.9,.25) rectangle (-3.1,.55);
\node[scale=.8] (title1) at (-4,.4) {\small{$\spf(6)_5{\times}\uf(1)$}};

\draw [draw=black] (.7,-3.7) rectangle (3.4,.75);   %%%%%%%%  SU3SU22   %%%%%%%%
\draw [draw=black] (.7,.75) rectangle (3.4,1.05);
\filldraw [fill=white, draw=black] (.7,.75) rectangle (3.4,1.05);
\node[scale=.8] (title1) at (2,.9) {\small{$\suf(3)_6{\times}\suf(2)_4^2$}};
\filldraw [fill=white, draw=black] (.7,-3.7) rectangle (3.4,.75);
\node[] (SU3SU22t) at (2,1.15) {};
\node[scale=.5] (SU3SU22_top) at (2,.5) {};          
\node[scale=.8] (SU3SU22_dim) at (2,.6) {$\H^6$};
\node[scale=.8] (SU3SU22_s1) at (1.8,0) {$\af_2$};
\node[scale=.8] (SU3SU22_ts1) at (2,-.5) {$\cT^{(1)}_{A_2,1}$};
\node[scale=.8] (SU3SU22_s2b) at (2.55,-1) {$\af_2$};
\node[scale=.8] (SU3SU22_s2a) at (1.3,-1.25) {$\cf_2$};
\node[scale=.8] (SU3SU22_ts2b) at (2.7,-1.5) {$\cT^{(1)}_{A_2,1}{\times}\cT^{(1)}_{A_2,1}$};
\node[scale=.8] (SU3SU22_s3) at (2.55,-2) {$\af_1$};
\node[scale=.8] (SU3SU22_ts2a) at (1.3,-2) {$\cS^{(1)}_{A_2,2}$};
\node[scale=.8] (SU3SU22_ts3) at (2.7,-2.5) {\hspace{-.2cm}$\suf(3)_6{\times}\suf(2)_4$};
\node[scale=.8] (SU3SU22_ts4a) at (1.45,-2.75) {$\af_2$};
\node[scale=.8] (SU3SU22_ts4b) at (2.6,-3) {$\af_1$};
\node[bbc,scale=.5] (SU3SU22_t) at (2,-3.5) {};
\draw[blue] (SU3SU22_top) -- (SU3SU22_ts1);
\draw[blue] (SU3SU22_ts1) -- (SU3SU22_ts2a);
\draw[blue] (SU3SU22_ts1) -- (SU3SU22_ts2b);
\draw[blue] (SU3SU22_ts2b) -- (SU3SU22_ts3);
\draw[blue] (SU3SU22_ts3) -- (SU3SU22_t);
\draw[blue] (SU3SU22_ts2a) -- (SU3SU22_t);
\node[] (SU3SU22b) at (2,-3.7) {};

\draw[Marrow] (Sp8b) -- (Sp6U1t);
\draw[Marrow] (Sp6b) -- (Sp6U1t);
\draw[Marrow] (SO7SU22b2) -- (SU3SU22t);

\end{scope}

\begin{scope}[scale=1.4,yshift=-12cm,xshift=-2cm] %Lagrangian level, Sp4 and SU26%

\node[] (Sp4t) at (0,.7) {};
\draw [draw=black] (-.8,-2.2) rectangle (.8,.3);    %%%%%%%%  Sp4   %%%%%%%%
\draw [draw=black] (-.8,.3) rectangle (.8,.6);
\filldraw [fill=yellow!40, draw=black] (-.8,.3) rectangle (.8,.6);
\node[scale=.8] (title1) at (0,.45) {\small{$SU(2)-SU(2)$}};

\node[scale=.8] (Sp4_top) at (0,0) {};       
\node[scale=.8] (Sp4_dim) at (0,.1) {$\H^3\times U(1)$};
\node[scale=.8] (Sp4_s1) at (-.2,-.5) {$\af_1$};
\node[scale=.8] (Sp4_ts1) at (0,-1) {$\blue{\cS^{(1)}_{\varnothing,2}}$};
\node[scale=.8] (Sp4_s2) at (-.2,-1.5) {$\cf_2$};
\node[bbc,scale=.5] (Sp4_down) at (0,-2) {};
\draw[blue] (Sp4_top) -- (Sp4_ts1);
\draw[blue] (Sp4_ts1) -- (Sp4_down);

\node[] (N4SU22t) at (8,.7) {};
\draw [draw=black] (7,-2.2) rectangle (9,.3);   %%%%%% SU26 %%%%%%%%
\draw [draw=black] (7,.3) rectangle (9,.6);
\filldraw [fill=blue!10, draw=black] (7,.3) rectangle (9,.6);
\node[scale=.8] (title1) at (8,.45) {\small{$\cN=4\ SU(2)^2$}};

\node[scale=.5] (N4SU22_top) at (8,0) {};    
\node[scale=.8] (N4SU22_dim) at (8,.1) {$U(1){\times}U(1)$};
\node[scale=.8] (N4SU22_s1a) at (7.45,-.35) {$\af_1$};
\node[scale=.8] (N4SU22_s1b) at (8.55,-.35) {$\af_1$};
\node[scale=.8] (N4SU22_ts1a) at (8.7,-1) {$\blue{\cS^{(1)}_{\varnothing,2}}$};
\node[scale=.8] (N4SU22_ts1b) at (7.3,-1) {$\blue{\cS^{(1)}_{\varnothing,2}}$};
\node[scale=.8] (N4SU22_s1a) at (7.5,-1.65) {$\af_1$};
\node[scale=.8] (N4SU22_s1b) at (8.5,-1.65) {$\af_1$};
\node[bbc,scale=.5] (N4SU22_down) at (8,-2) {};
\draw[blue] (N4SU22_top) -- (N4SU22_ts1a);
\draw[blue] (N4SU22_top) -- (N4SU22_ts1b);
\draw[blue] (N4SU22_ts1a) -- (N4SU22_down);
\draw[blue] (N4SU22_ts1b) -- (N4SU22_down);

\draw[IRarrow] (SU3SU22b) -- (N4SU22t);
\draw[IRarrow] (Sp6U1b) -- (Sp4t);

\end{scope}

\end{tikzpicture}
{\caption{\label{fig:HBSp12} Higgs branch analysis of the mass deformations among the entries of the $\spf(12)-\spf(8)-\ff_4$ series.}}
\end{adjustbox}
\end{figure}

This instructs us on what we ought to expect to see along the $\af_1$ stratum. Moving on the $\af_1$ stratum corresponds to turning on a vev for the $\suf(2)$ which, from what we just derived, corresponds to turning on a vev for the highest root of $\spf(10)$. But this is precisely what triggers the relevant HB flow to move on the ECB of $\spf(10)$ supporting the $\cT^{(1)}_{E_6,1}$ \cite{Apruzzi:2020pmv,CCLMW2020}. We then conclude that it is the latter theory which should be supported on the $\af_1$ stratum (or rather leaf), perfectly matching what we find in figure \ref{fig:HBSp12}. 

A few comments are in order. First, as we had already announced in a few instances, $\cT^{(1)}_{E_6,1}$ is neither equal to $\cS^{(1)}_{E_6,2}$ (doh!) nor one its mass deformations, rather it is connected by an HB flow. Second, the special phenomenon which happens here is that the mass deformation breaks a simple factor to a semi-simple one where one of the two resulting simple factors happens to coincide precisely with an Higgsing direction of the bottom theory of the original SCFT.

\paragraph{$\spf(4)_7{\times}\spf(8)_8\to\spf(8)_6{\times}\suf(2)_8$} This mass deformation is another instance where case $(iii)$ of criterion \ref{Higgs} is realized. Since we discussed the previous example in great detail and the phenomenon at play is analogous we will be somewhat more schematic.

First let's analyze the HB of $\spf(4)_7{\times}\spf(8)_8$. In particular how the flavor symmetry of the $\cS^{(1)}_{E_6,2}$ supported on the $\cf_4$ leaf is matched. Staring at table \ref{Higgs:one}, we see that on the $\cf_4$ leaf $\ff^\natural\equiv \spf(4){\times}\spf(6)$, where the $\spf(4)$ factor precisely corresponds to the $\spf(4)_7$ of the original theory which goes along with the ride on this leaf. For later, it is useful to identify how the highest root of $\spf(10)$ of the $\cS^{(1)}_{E_6,2}$, which triggers the flow to the ECB, is realized in this picture. The embedding $\spf(4){\times}\spf(6)\subset \spf(10)$ readily shows that the highest root of $\spf(10)$ is identified with the root associated with the $\suf(2)$ which commutes with the highest root of the $\spf(4)$.

Now staring at the HB Hasse diagram in figure \ref{fig:HBSp12} we again see that the mass deformation from $\spf(4)_7{\times}\spf(8)_8\to\spf(8)_6{\times}\suf(2)_8$ triggers the mass deformation of the bottom theory $\cS^{(1)}_{E_6,2}\to\cS^{(1)}_{D_4,2}$. The $\spf(8)_6{\times}\suf(2)_8$ has no ECB and since mass deformations cannot lift CB directions, see discussion in section \ref{sec:4dsetup}, we conclude that the $\cf_2$ stratum should be lifted by the mass deformation, or else the mass deformation which we are considering is identified with the highest root of the $\spf(4)$ factor in the original theory. This in turn clarifies that the $\suf(2)$ surviving the mass deformation is precisely its commutant in $\spf(4)$ and which by the discussion in the previous paragraph, corresponds to the highest root of the $\spf(10)$ of the $\cS^{(1)}_{E_6,2}$. 

We are then back to the situation we saw previously. The $\suf(2)$ which survives after the mass deformation $\spf(4)_7{\times}\spf(8)_8\to\spf(8)_6{\times}\suf(2)_8$, is identified with a Higgs direction of one of the bottom theories of the orginal theory, \emph{i.e.} $\spf(4)_7{\times}\spf(8)_8$. We then conclude that we expect the theory supported on the $\af_1$ leaf to be then one obtained by said Higgs branch flow, \emph{i.e.} $\cT^{(1)}_{E_6,1}$, which precisely matches with the result in figure \ref{fig:HBSp12}.

%%%%%%% This is only for sending out the draft %%%%%%%%%

\subsection{$\spf(14)_8$ series}

\begin{figure}
\begin{adjustbox}{center,max width=.65\textwidth}
\ffigbox{
\begin{subfloatrow}
\hspace{-5cm}\ffigbox[9cm][]{
\begin{tikzpicture}[decoration={markings,
mark=at position .5 with {\arrow{>}}},
Marrow/.style={->,>=stealth[round],shorten >=1pt,line width=.4mm,coolblack},
IRarrow/.style={->,>=stealth[round],shorten >=1pt,line width=.4mm,deepcarmine,dashed}]
\begin{scope}[scale=1.4] %%%%%%%  Sp14  %%%%%%%
\node[bbc,scale=.5] (Sp14_top) at (0,0) {};       %%%%%%%%  Sp14   %%%%%%%%
\draw [draw=black] (-1.3,.1) rectangle (1.35,.4);
\filldraw [fill=white, draw=black] (-1.3,.1) rectangle (1.35,.4);
\node[scale=.8] (title1) at (0,.25) {\small{$\spf(14)_9$}};
\draw [draw=black] (-1.3,-1.5) rectangle (1.35,.1);
\node[bbc,scale=.5] (Sp14_down) at (0,-1.4) {};
\node[scale=.8] (Sp14_1a) at (-.8,-.7) {$[I_1,\varnothing]$};
\node[scale=.8] (Sp14_1b) at (.8,-.7) {$[I_6^*,\spf(14)]$};
\draw[red] (Sp14_top) -- (Sp14_1a);
\draw[red] (Sp14_top) -- (Sp14_1b);
\draw[red] (Sp14_down) -- (Sp14_1a);
\draw[red] (Sp14_down) -- (Sp14_1b);
\end{scope}

\begin{scope}[scale=1.4,yshift=-2cm,xshift=-1.5cm] %%%%% Sp10 and SP6SU2 %%%%%%%
\node[bbc,scale=.5] (Sp10_top) at (0,0) {};        %%%%%%%%  Sp10   %%%%%%%%

\draw [draw=black] (-1.45,-1.5) rectangle (1.55,.1);
\node[bbc,scale=.5] (Sp10_down) at (0,-1.4) {};
\node[scale=.8] (Sp10_1a) at (-1,-.7) {$[I_1,\varnothing]$};
\node[scale=.8] (Sp10_1b) at (0,-.7) {$[I_2,\suf(2)]$};
\node[scale=.8] (Sp10_1c) at (1,-.7) {$[I_4^*,\spf(10)]$};
\draw[red] (Sp10_top) -- (Sp10_1a);
\draw[red] (Sp10_top) -- (Sp10_1b);
\draw[red] (Sp10_top) -- (Sp10_1c);
\draw[red] (Sp10_down) -- (Sp10_1a);
\draw[red] (Sp10_down) -- (Sp10_1b);
\draw[red] (Sp10_down) -- (Sp10_1c);
\node[bbc,scale=.5] (Sp8SU2_top) at (3.5,0) {};        %%%%%%%%  SU2Sp8   %%%%%%%%

\draw [draw=black] (2.05,-1.5) rectangle (5.05,.1);
\node[bbc,scale=.5] (Sp8SU2_down) at (3.5,-1.4) {};
\node[scale=.8] (Sp8SU2_1a) at (4.5,-.7) {$[I_1,\varnothing]$};
\node[scale=.8] (Sp8SU2_1b) at (2.5,-.7) {$\blue{\cS^{(1)}_{\varnothing,2}}$};
\node[scale=.8] (Sp8SU2_1c) at (3.5,-.7) {$[I_3^*,\spf(8)]$};
\draw[red] (Sp8SU2_top) -- (Sp8SU2_1a);
\draw[red] (Sp8SU2_top) -- (Sp8SU2_1b);
\draw[red] (Sp8SU2_top) -- (Sp8SU2_1c);
\draw[red] (Sp8SU2_down) -- (Sp8SU2_1a);
\draw[red] (Sp8SU2_down) -- (Sp8SU2_1b);
\draw[red] (Sp8SU2_down) -- (Sp8SU2_1c);

%%%% Mass deformation arrows %%%%%%

\draw[Marrow] (Sp14_1b) -- (Sp10_1b);
\draw[Marrow] (Sp14_1b) -- (Sp10_1c);

%%%% Tables with titles %%%%%%

\draw [draw=black] (2.05,.1) rectangle (5.05,.4);
\filldraw [fill=white, draw=black] (2.05,.1) rectangle (5.05,.4);
\node[scale=.8] (title1) at (3.5,.25) {\small{$\suf(2)_5{\times}\spf(8)_7$}};

\draw [draw=black] (-1.45,.1) rectangle (1.55,.4);
\filldraw [fill=white, draw=black] (-1.45,.1) rectangle (1.55,.4);
\node[scale=.8] (title1) at (0,.25) {\small{$\suf(2)_8{\times}\spf(10)_7$}};

\end{scope}

\begin{scope}[scale=1.4,yshift=-4cm,xshift=0cm] %%%%% Sp8 %%%%%%%
\node[bbc,scale=.5] (Sp8_top) at (0,0) {};        %%%%%%%%  Sp8   %%%%%%%%

\draw [draw=black] (-1.35,-1.5) rectangle (1.4,.1);
\node[bbc,scale=.5] (Sp8_down) at (0,-1.4) {};
\node[scale=.8] (Sp8_1a) at (-.7,-.7) {$[I_1,\varnothing]$};
\node[scale=.8] (Sp8_1b) at (.7,-.7)  {$[I_3^*,\spf(8)]$};
\draw[red] (Sp8_top) -- (Sp8_1a);
\draw[red] (Sp8_top) -- (Sp8_1b);
\draw[red] (Sp8_down) -- (Sp8_1a);
\draw[red] (Sp8_down) -- (Sp8_1b);

%%%% Mass deformation arrows %%%%%%

\draw[Marrow] (Sp10_1b) -- (Sp8_1a);
\draw[Marrow] (Sp10_1c) -- (Sp8_1b);

\draw[Marrow] (Sp8SU2_1b) -- (Sp8_1a);

%%%% Tables with titles %%%%%%

\draw [draw=black] (-1.35,.1) rectangle (1.4,.4);
\filldraw [fill=white, draw=black] (-1.35,.1) rectangle (1.4,.4);
\node[scale=.8] (title1) at (0,.25) {\small{$\spf(8)_6{\times}\uf(1)$}};

\end{scope}

\begin{scope}[scale=1.4,yshift=-6cm,xshift=0cm] %Lagrangian level, Sp6%
\node[bbc,scale=.5] (Sp6_top) at (0,0) {};      %%%%%%%%  Sp6   %%%%%%%%

\draw [draw=black] (-1.4,-1.5) rectangle (1.5,.1);
\node[bbc,scale=.5] (Sp6_down) at (0,-1.4) {};
\node[scale=.8] (Sp6_1a) at (-1,-.7) {$[I_1,\varnothing]$};
\node[scale=.8] (Sp6_1b) at (0,-.7) {$[I_1,\varnothing]$};
\node[scale=.8] (Sp6_1c) at (1,-.7) {$[I_2^*,\spf(6)]$};
\draw[red] (Sp6_top) -- (Sp6_1a);
\draw[red] (Sp6_top) -- (Sp6_1b);
\draw[red] (Sp6_top) -- (Sp6_1c);
\draw[red] (Sp6_down) -- (Sp6_1a);
\draw[red] (Sp6_down) -- (Sp6_1b);
\draw[red] (Sp6_down) -- (Sp6_1c);

%%%% Mass deformation arrows %%%%%%

\draw[IRarrow] (Sp8_1b) -- (Sp6_1b);
\draw[IRarrow] (Sp8_1b) -- (Sp6_1c);

\draw (4.7,9) -- (4.7,-5);

\node[] (bfigure) at (0,-7.5) {};

%%%% Tables with titles %%%%%%

\draw [draw=black] (-1.4,.1) rectangle (1.5,.4);
\filldraw [fill=yellow!40, draw=black] (-1.4,.1) rectangle (1.5,.4);
\node[scale=.8] (title1) at (0,.25) {\small{$USp(4)+3V$}};

\end{scope}
\end{tikzpicture}}{\caption{Relations among the Coulomb branches of the $\spf(14)$ series.}}
\end{subfloatrow}
\begin{subfloatrow}
\hspace{3.5cm}\ffigbox[6.5cm][]{
\begin{tikzpicture}[decoration={markings,
mark=at position .5 with {\arrow{>}}},
Marrow/.style={->,>=stealth[round],shorten >=1pt,line width=.4mm,coolblack},
IRarrow/.style={->,>=stealth[round],shorten >=1pt,line width=.4mm,deepcarmine,dashed}]

%%%%%% Starting HB flows! %%%%%%%%

\begin{scope}[scale=1.4] %%%%%%%  Sp14  %%%%%%%
\draw [draw=black] (-.7,1.25) rectangle (.7,1.55);
\node[scale=.8] (title1) at (0,1.4) {\small{$\spf(14)_9$}};
\draw [draw=black] (-.7,-3.1) rectangle (.7,1.25);
\node[scale=.5] (Sp14_p4) at (0,1) {};
\node[scale=.8] (Sp14_t0a) at (0,1.1) {$\H^{29}$};
\node[scale=.8] (Sp14_tp0) at (-.2,.5) {$\ef_6$};
\node[scale=.8] (Sp14_p3) at (0,0) {$\cT^{(1)}_{E_6,1}$};
\node[scale=.8] (Sp14_tp0) at (-.2,-.5) {$\cf_5$};
\node[scale=.8] (Sp14_p2) at (0,-1) {$\cS^{(1)}_{E_6,2}$};
\node[scale=.8] (Sp14_tp2) at (-.2,-1.5) {$\cf_6$};
\node[scale=.8] (Sp14_p1) at (0,-2) {$\spf(12)_8$};
\node[scale=.8] (Sp14_tp1) at (-.2,-2.5) {$\cf_7$};
\node[bbc,scale=.5] (Sp14_p0) at (0,-3) {};
\draw[blue] (Sp14_p0) -- (Sp14_p1);
\draw[blue] (Sp14_p1) -- (Sp14_p2);
\draw[blue] (Sp14_p2) -- (Sp14_p3);
\draw[blue] (Sp14_p3) -- (Sp14_p4);
\node[] (Sp14b) at (0,-3.2) {};

\end{scope}

\begin{scope}[scale=1.4,yshift=-5.5cm,xshift=-1.5cm] %%%%% Sp10 and SP6SU2 %%%%%%%

\node[] (Sp10t) at (-.7,1.65) {};
\node[scale=.5] (Sp10_p0a) at (-.3,1) {};  %%%%%%%%  Sp10   %%%%%%%%
\node[scale=.8] (Sp10_t0a) at (-.3,1.1) {$\H^{17}$};
\node[scale=.8] (Sp10_tp2) at (.4,.5) {$\df_4$};
\node[scale=.8] (Sp10_tp2) at (-.7,.5) {$\ef_6$};
\node[scale=.8] (Sp10_p0) at (.4,0) {$\cT^{(1)}_{D_4,1}$};
\node[scale=.8] (Sp10_tp2) at (.25,-.45) {$\cf_3$};
\node[scale=.8] (Sp10_p1b) at (-1,-.5) {$\cT^{(1)}_{E_6,1}$};
\node[scale=.8] (Sp10_p1a) at (.4,-1) {$\cS^{(1)}_{D_4,2}$};
\node[scale=.8] (Sp10_tp2) at (-1.5,-1.15) {$\cf_5$};
\node[scale=.8] (Sp10_tp2) at (-.8,-1.35) {$\af_1$};
\node[scale=.8] (Sp10_tp2) at (.2,-1.5) {$\cf_4$};
\node[scale=.8] (Sp10_p2a) at (-.3,-2) {$\suf(2)_8{\times}\spf(8)_6$};
\node[scale=.8] (Sp10_p2b) at (-1.7,-2) {$\cS^{(1)}_{E_6,2}$};
\node[scale=.8] (Sp10_tp1) at (-1.5,-2.6) {$\af_1$};
\node[scale=.8] (Sp10_tp2) at (-.5,-2.6) {$\cf_5$};
\node[bbc,scale=.5] (Sp10_p0b) at (-1,-3) {};
\draw[blue] (Sp10_p0a) -- (Sp10_p0);
\draw[blue] (Sp10_p0a) -- (Sp10_p1b);
\draw[blue] (Sp10_p0) -- (Sp10_p1a);
\draw[blue] (Sp10_p1a) -- (Sp10_p2a);
\draw[blue] (Sp10_p1b) -- (Sp10_p2a);
\draw[blue] (Sp10_p1b) -- (Sp10_p2b);
\draw[blue] (Sp10_p2a) -- (Sp10_p0b);
\draw[blue] (Sp10_p2b) -- (Sp10_p0b);
\node[] (Sp10b) at (-.7,-3.2) {};

\node[scale=.5] (Sp8SU2_p0a) at (4.2,1) {};  %%%%%%%%  SU2Sp8   %%%%%%%%
\node[scale=.8] (Sp8SU2_t0a) at (4.2,1.1) {$\H^{15}$};
\node[scale=.8] (Sp8SU2_tp2) at (4,.5) {$\df_4$};
\node[scale=.8] (Sp8SU2_p0) at (4.2,0) {$\cT^{(1)}_{D_4,1}$};
\node[scale=.8] (Sp8SU2_tp2) at (3.75,-.4) {$\af_5$};
\node[scale=.8] (Sp8SU2_tp2) at (4.7,-.5) {$\cf_3$};
\node[scale=.8] (Sp8SU2_p1b) at (3.5,-1) {$SU(3)+6F$};
\node[scale=.8] (Sp8SU2_p1a) at (4.9,-1) {$\cS^{(1)}_{D_4,2}$};
\node[scale=.8] (Sp8SU2_tp2) at (3,-1.45) {$\cf_4$};
\node[scale=.8] (Sp8SU2_tp2) at (3.7,-1.55) {$\af_1$};
\node[scale=.8] (Sp8SU2_tp2) at (4.75,-1.5) {$\cf_3$};
\node[scale=.8] (Sp8SU2_p2a) at (4.2,-2) {$\suf(2)_8{\times}\spf(8)_6$};
\node[scale=.8] (Sp8SU2_p2b) at (2.8,-2) {$G_2+4F$};
\node[scale=.8] (Sp8SU2_tp1) at (3,-2.6) {$\af_1$};
\node[scale=.8] (Sp8SU2_tp2) at (4,-2.6) {$\cf_4$};
\node[bbc,scale=.5] (Sp8SU2_p0b) at (3.5,-3) {};
\draw[blue] (Sp8SU2_p0a) -- (Sp8SU2_p0);
\draw[blue] (Sp8SU2_p0) -- (Sp8SU2_p1a);
\draw[blue] (Sp8SU2_p0) -- (Sp8SU2_p1b);
\draw[blue] (Sp8SU2_p1a) -- (Sp8SU2_p2a);
\draw[blue] (Sp8SU2_p1b) -- (Sp8SU2_p2a);
\draw[blue] (Sp8SU2_p1b) -- (Sp8SU2_p2b);
\draw[blue] (Sp8SU2_p2a) -- (Sp8SU2_p0b);
\draw[blue] (Sp8SU2_p2b) -- (Sp8SU2_p0b);
\node[] (Sp8SU2b) at (3.7,-3.2) {};

%%%% Tables with titles %%%%%%

\draw [draw=black] (2.3,1.25) rectangle (5.3,1.55);
\node[scale=.8] (title1) at (3.7,1.4) {\small{$\suf(2)_5{\times}\spf(8)_7$}};
\draw [draw=black] (2.3,-3.1) rectangle (5.3,1.25);

\draw [draw=black] (-2.15,1.25) rectangle (.8,1.55);
\node[scale=.8] (title1) at (-.7,1.4) {\small{$\suf(2)_8{\times}\spf(10)_7$}};
\draw [draw=black] (-2.15,-3.1) rectangle (.8,1.25);

\end{scope}

\begin{scope}[scale=1.4,yshift=-11cm,xshift=0cm] %%%%% Sp8 %%%%%%%

\node[] (Sp8t) at (0,1.65) {};
\draw [draw=black] (-.8,-3.1) rectangle (.8,1.25);
\node[scale=.5] (Sp8_p0a) at (0,1) {};
\node[scale=.8] (Sp8_t0a) at (0,1.1) {$\H^{11}$};
\node[scale=.8] (Sp8_p0b) at (0,-2) {$\spf(6)_5{\times}\uf(1)$};
\node[scale=.8] (Sp8_p1) at (0,0) {$\cT^{(1)}_{A_2,1}$};
\node[scale=.8] (Sp8_p2) at (0,-1) {$\cS^{(1)}_{A_2,2}$};
\node[scale=.8] (Sp8_t1c) at (.3,.5) {$\af_2$};
\node[scale=.8] (Sp8_t2c) at (.3,-0.5) {$\cf_2$};
\node[scale=.8] (Sp8_t1c) at (.3,-1.5) {$\cf_3$};
\node[scale=.8] (Sp8_t1c) at (.3,-2.5) {$\cf_4$};
\node[bbc,scale=.5] (Sp8_p0c) at (0,-3) {};
\draw[blue] (Sp8_p0a) -- (Sp8_p1);
\draw[blue] (Sp8_p1) -- (Sp8_p2);
\draw[blue] (Sp8_p2) -- (Sp8_p0b);
\draw[blue] (Sp8_p0c) -- (Sp8_p0b);
\node[] (Sp8b) at (0,-3.2) {};

%%%% Tables with titles %%%%%%

\draw [draw=black] (-.8,1.25) rectangle (.8,1.55);
\node[scale=.8] (title1) at (0,1.4) {\small{$\spf(8)_6{\times}\uf(1)$}};

\end{scope}

\begin{scope}[scale=1.4,yshift=-16.5cm,xshift=0cm] %Lagrangian level, Sp6%

\node[] (Sp6t) at (0,1.65) {};
\draw [draw=black] (-.8,-2.1) rectangle (.8,1.25);
\node[scale=.8] (Sp6_p0a) at (0,1.1) {$\H^6$};
\node[scale=.8] (Sp6_t1a) at (-0.2,.5) {$\af_1$};
\node[scale=.8] (Sp6_p1) at (-0,0) {$\blue{\cS^{(1)}_{\varnothing,2}}$};
\node[scale=.8] (Sp6_t1a) at (-0.2,-.5) {$\cf_2$};
\node[scale=.8] (Sp6_p2) at (-0,-1) {$SU(2)\text{-}SU(2)$};
\node[scale=.8] (Sp6_t1b) at (-.2,-1.5) {$\cf_3$};
\node[bbc,scale=.5] (Sp6_p0b) at (0,-2) {};
\draw[blue] (Sp6_p0a) -- (Sp6_p1);
\draw[blue] (Sp6_p1) -- (Sp6_p2);
\draw[blue] (Sp6_p2) -- (Sp6_p0b);

%%%% Mass deformation arrows %%%%%%

\draw[Marrow] (Sp14b) -- (Sp10t);
\draw[Marrow] (Sp10b) -- (Sp8t);
\draw[Marrow] (Sp8SU2b) -- (Sp8t);
\draw[IRarrow] (Sp8b) -- (Sp6t);

%%%% Tables with titles %%%%%%

\draw [draw=black] (-.8,1.25) rectangle (.8,1.55);
\filldraw [fill=yellow!40, draw=black] (-.8,1.25) rectangle (.8,1.55);
\node[scale=.8] (title1) at (0,1.4) {\small{$USp(4)+3V$}};

\end{scope}

\end{tikzpicture}}{\hspace{3cm}\caption{Higgs branches of the $\spf(14)$ series.}}
\end{subfloatrow}}
{\caption{\label{fig:CBHBSp14} Coulomb $(a)$ and Higgs branch $(b)$ analysis of the mass deformations among the entries of the $\spf(14)$ series.}}
\end{adjustbox}
\end{figure}
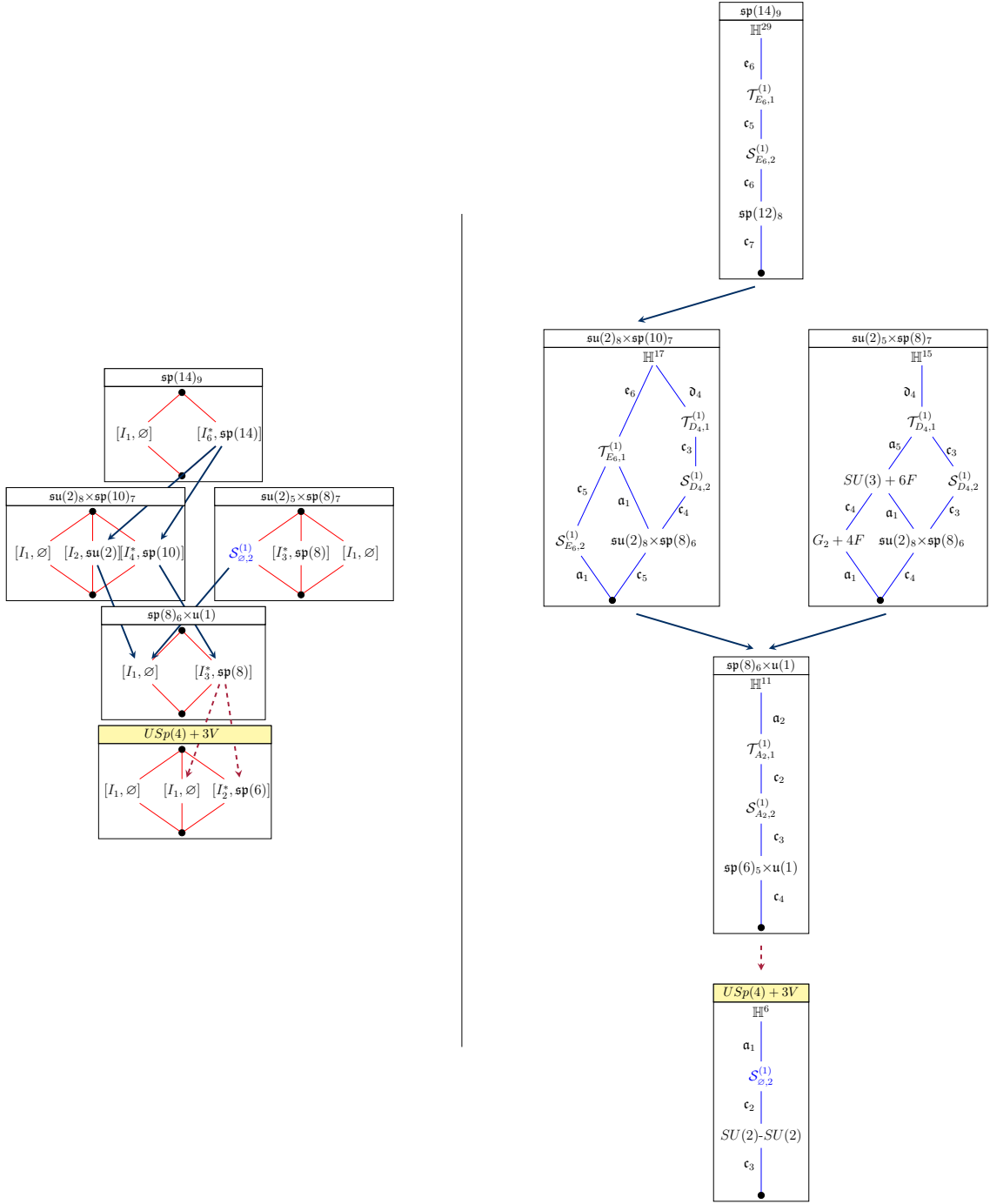

Here we will discuss the moduli space flows reported in figure \ref{fig:CBHBSp14}. As usual let's start from the CB.

The CB analysis is very straightforward and can be understood by the following general deformation pattern:
\beq\label{DefSp14}
[I_n^*,\spf(2n+2)]\quad\to\quad \left\{[I_{n-n'}^*,\spf(2(n-n')+2)],
[I_n',\suf(n')]\right\}.
\eeq
where this result can be derived by noticing that what we call $[I_n^*,\spf(2n+2)]$ is nothing but an $SU(2)$ gauge theory with $n+1$ hypers in the adjoint representation\footnote{To be technically accurate, to make this identification, we need to also rescale the normalization of the $SU(2)$ generators, see \cite[Section 4.2]{Argyres:2015ffa}.} and the above mass deformation correspond to giving an equal mass to $n'$ hypers.

Using \eqref{DefSp14}, is then obvious to see that the CBs of all the theories connected by mass deformations in figure \ref{MapR2} satisfy criterion \ref{knot}. Conversely, recalling that $\blue{\cS^{(1)}_{\varnothing,2}}\equiv[I_0^*,\spf(1)]$, the reader can convince themselves that no mass deformation connecting $\suf(2)_5{\times}\spf(8)_7$ to either $\spf(14)_9$ or $\suf(2)_8{\times}\spf(10)_7$ could exist. Let's now move to the analysis of the HBs.

Building on the analysis of previous sections we can straightforwardly observe that all HBs in figure \ref{fig:CBHBSp14} are connected as they should. Specifically:
\begin{itemize}
    \item $\spf(14)_9\to\suf(2)_8{\times}\spf(10)_7$ in this case the bottom theory of $\Tf_{\rm IR}$ are $\cS^{(1)}_{E_6,2}$ and $\suf(2){\times}\spf(8)_6$. Comparing them to the bottom theory of $\Tf_{\rm UV}$, \emph{i.e.} $\spf(12)_8$, the first satisfy criterion \ref{Higgs} $(iii)$ while the second $(i)$. The analysis to see that the matching works in such a way that indeed the $\suf(2)$ of $\Tf_{\rm IR}$ corresponds to the highest root of the $\spf(14)$ of the $\Tf_{\rm UV}$ proceeds in very much the same was as the examples in the previous section and we thus don't repeat it here.
    
    \item $\suf(2)_8{\times}\spf(10)_7\to\spf(8)_6{\times}\uf(1)$ and $\suf(2)_5{\times}\spf(8)_7\to\spf(8)_6{\times}\uf(1)$. In this case $\Tf_{\rm IR}$ has a single stratum and its bottom theory is a mass deformation of $\suf(2)_8{\times}\spf(8)_6$ which is a bottom theory for $\Tf_{\rm UV}$ in both flows. The UV $\af_1$ stratum is simply lifted by the mass deformation both for $\suf(2)_8{\times}\spf(10)_7$ and $\suf(2)_5{\times}\spf(8)_7$.
    
    \item $\spf(8)_6{\times}\uf(1)\to USp(4)+3V$. This is possibly the most straighforward case. Both $\Tf_{\rm UV}$ and $\Tf_{\rm IR}$ have a single leaf and the theories supported over it, \emph{i.e.} $\spf(6)_5{\times}\uf(1)$ and $SU(2)\text{-}SU(2)$, are one the mass deformation of the other.
    
\end{itemize}

\subsection{Remaining series}

To conclude this analysis we report the moduli space mass deformations analysis of the theories in figure \ref{fig:CBthree} which correspond to the CB structure of the $\suf(6)$ and $\suf(5)$ series. For these two series our discussion will be less detailed since our understanding of their moduli space is largely incomplete. In fact, while we can still perform non-trivial checks as far as the CB is concerned, we will not perform the HB analysis since too little is known, particularly about the bottom theories/leaves, and thus we don't have enough data to check that criterion \ref{Higgs} is indeed satisfied.

\paragraph{$\suf(6)$ series} As can be seen from figure \ref{fig:CBthree} many entries in the CB stratification of these theories are unknown thus we won't have much to say about them and will only make a few observations further constraining what these theories could be. Let's then focus first on the left side of the $\suf(6)$ flows where all strata are fully characterized.

The decoration of the $\suf(6)_{16}{\times}\suf(2)_9$ is:
\beq
\langle [I_1,\varnothing],[I_6^*,\suf(2){\times}\sof(12)]_{\Z_2}\rangle
\eeq
where the identification of the $I_6^*$ is tricky. As discussed in more detail in \cite{Martone:2021ixp}, the proposal is that it should be associated to a $\Z_2$ gauging of an $SU(2)+6F+1Adj$ and where the $\Z_2$ is the inner automorphism of $\sof(12)$ with commutant $\suf(6)$, for more details on automorphisms of Lie algebras see \cite[Theorem 8.6]{Kac} or for a more physical discussion \cite[Sec.~3.3]{Tachikawa:2011ch}. Similarly the other theories have a $[I_4^*,\suf(2){\times}\sof(8)]_{\Z_2}$ and $[I_3^*,\suf(2){\times}\sof(6)]_{\Z_2}$, which are identified with a $\Z_2$ gauged version of a $SU(2)+4F+1Adj$ and  $SU(2)+3F+1Adj$ respectively. Then naturally the allowed mass deformations among those should commute with the gauging. 

In \cite{Martone:2021ixp} the explicit action of the automorphism on the gauge theory side was not identified, we will make other compelling observations here. First we claim that the following mass deformations are allowed:
\beq\label{DefSU6}
[I_{n_{\rm in}}^*,\suf(2){\times}\sof(2n_{\rm in})]_{\Z_2}\to\left\{
[I_{n'_{\rm fin}}^*,\suf(2){\times}\sof(2n'_{\rm fin})]_{\Z_2},[I_{n''_{\rm fin}},\suf(n''_{\rm fin})]
\right\}
\eeq
where $n'_{\rm fin}+n''_{\rm fin}=n_{\rm in}$. The above mass simply correspond to give a common mass to $n''_{\rm fin}$ of the $n_{\rm in}$ fundamental flavors. Notice that there is no constraint on the fact that the number of fundamental flavors which we decouple has to be even. This suggests that the $\Z_2$ should act separately on each hyper.

\begin{figure}
\begin{adjustbox}{center,max width=.9\textwidth}
\ffigbox{
\begin{subfloatrow}
\hspace{-2cm}
\ffigbox[9cm][]{
\begin{tikzpicture}[decoration={markings,
mark=at position .5 with {\arrow{>}}},
Marrow/.style={->,>=stealth[round],shorten >=1pt,line width=.4mm,coolblack},
IRarrow/.style={->,>=stealth[round],shorten >=1pt,line width=.4mm,deepcarmine,dashed}]
\begin{scope}[scale=1.4,xshift=-3cm] %%%%%%%  SU6 SU2  %%%%%%%
\node[bbc,scale=.5] (SU6_top) at (0,0) {};       %%%%%%%%  E8r2   %%%%%%%%
\draw [draw=black] (-1.7,.1) rectangle (1.9,.4);
\filldraw [fill=white, draw=black] (-1.7,.1) rectangle (1.9,.4);
\node[scale=.8] (title1) at (0.1,.25) {\small{$\suf(6)_{16}{\times}\suf(2)_9$}};
\draw [draw=black] (-1.7,-1.5) rectangle (1.9,.1);
\node[bbc,scale=.5] (SU6_down) at (0,-1.4) {};
\node[scale=.8] (SU6_1a) at (-.8,-.7) {$[I_1,\varnothing]$};
\node[scale=.8] (SU6_1b) at (.8,-.7) {$[I_6^*,\suf(2)\times \sof(12)]_{\Z_2}$};
\draw[red] (SU6_top) -- (SU6_1a);
\draw[red] (SU6_top) -- (SU6_1b);
\draw[red] (SU6_down) -- (SU6_1a);
\draw[red] (SU6_down) -- (SU6_1b);
\node[scale=.8] (p1) at (4.3,-1) {};
\end{scope}

\begin{scope}[scale=1.4,yshift=-2cm,xshift=-3cm] %%%%% SU4SU2 and SU32 %%%%%%%
\node[bbc,scale=.5] (SU4_top) at (-1.9,0) {};        %%%%%%%%  SU4SU2   %%%%%%%%

\draw [draw=black] (-3.3,-1.5) rectangle (.45,.1);
\node[bbc,scale=.5] (SU4_down) at (-1.9,-1.4) {};
\node[scale=.8] (SU4_1a) at (-2.9,-.7) {$[I_1,\varnothing]$};
\node[scale=.8] (SU4_1b) at (-2.1,-.7) {$[I_2,\suf(2)]$};
\node[scale=.8] (SU4_1c) at (-.6,-.7) {$[I_4^*,\suf(2)\times \sof(8)]_{\Z_2}$};
\draw[red] (SU4_top) -- (SU4_1a);
\draw[red] (SU4_top) -- (SU4_1b);
\draw[red] (SU4_top) -- (SU4_1c);
\draw[red] (SU4_down) -- (SU4_1a);
\draw[red] (SU4_down) -- (SU4_1b);
\draw[red] (SU4_down) -- (SU4_1c);
\node[bbc,scale=.5] (SU32_top) at (1.6,0) {};      %%%%%%%%  SU32   %%%%%%%%

\draw [draw=black] (.5,-1.5) rectangle (2.9,.1);
\node[bbc,scale=.5] (SU32_down) at (1.6,-1.4) {};
\node[scale=.8] (SU32_1a) at (.9,-.7) {$[I_1,\varnothing]$};
\node[scale=.8] (SU32_1b) at (2.3,-.7) {$[\star {\rm w}/b=7]$};
\draw[red] (SU32_top) -- (SU32_1a);
\draw[red] (SU32_top) -- (SU32_1b);
\draw[red] (SU32_down) -- (SU32_1a);
\draw[red] (SU32_down) -- (SU32_1b);

%%%% Mass deformation arrows %%%%%%

\draw[Marrow] (SU6_1b) -- (SU4_1b);
\draw[Marrow] (SU6_1b) -- (SU4_1c);

\draw[Marrow] (SU6_1b) -- (SU32_1b);

%%%% Tables with titles %%%%%%

\draw [draw=black] (.5,.1) rectangle (2.9,.4);
\filldraw [fill=white, draw=black] (.5,.1) rectangle (2.9,.4);
\node[scale=.8] (title1) at (1.7,.25) {\small{$\suf(3)_{10}{\times}\suf(3)_{10}{\times}\uf(1)$}};

\draw [draw=black] (-3.3,.1) rectangle (.45,.4);
\filldraw [fill=white, draw=black] (-3.3,.1) rectangle (.45,.4);
\node[scale=.8] (title1) at (-1.4,.25) {\small{$\suf(4)_{12}{\times}\suf(2)_7{\times}\uf(1)$}};

\end{scope}

\begin{scope}[scale=1.4,yshift=-4cm,xshift=-3cm] %%%%% SU3SU2 and SU22 %%%%%%%
\node[bbc,scale=.5] (SU3_top) at (-2,0) {};        %%%%%%%%  SU4SU2   %%%%%%%%

\draw [draw=black] (-3.2,-1.5) rectangle (-.15,.1);
\node[bbc,scale=.5] (SU3_down) at (-2,-1.4) {};
\node[scale=.8] (SU3_1a) at (-2.8,-.7) {$[I_1,\varnothing]$};
\node[scale=.8] (SU3_1b) at (-1.2,-.7) {$[I_3^*,\suf(2)\times \sof(6)]_{\Z_2}$};
\draw[red] (SU3_top) -- (SU3_1a);
\draw[red] (SU3_top) -- (SU3_1b);
\draw[red] (SU3_down) -- (SU3_1a);
\draw[red] (SU3_down) -- (SU3_1b);
\node[bbc,scale=.5] (SU22_top) at (1.4,0) {};      %%%%%%%%  SU32   %%%%%%%%

\draw [draw=black] (.05,-1.5) rectangle (3,.1);
\node[bbc,scale=.5] (SU22_down) at (1.4,-1.4) {};
\node[scale=.8] (SU22_1a) at (0.4,-.7) {$[I_1,\varnothing]$};
\node[scale=.8] (SU22_1b) at (1.4,-.7) {$[I_2,\suf(2)]$};
\node[scale=.8] (SU22_1c) at (2.4,-.7) {$[\star {\rm w}/b=5]$};
\draw[red] (SU22_top) -- (SU22_1a);
\draw[red] (SU22_top) -- (SU22_1b);
\draw[red] (SU22_top) -- (SU22_1c);
\draw[red] (SU22_down) -- (SU22_1a);
\draw[red] (SU22_down) -- (SU22_1b);
\draw[red] (SU22_down) -- (SU22_1c);

%%%% Mass deformation arrows %%%%%%

\draw[Marrow] (SU4_1c) -- (SU3_1b);

\draw[Marrow] (SU32_1b) -- (SU22_1c);
\draw[Marrow] (SU32_1b) -- (SU22_1b);

%%%% Tables with titles %%%%%%

\draw [draw=black] (-3.2,.1) rectangle (-.15,.4);
\filldraw [fill=white, draw=black] (-3.2,.1) rectangle (-.15,.4);
\node[scale=.8] (title1) at (-1.6,.25) {\small{$\suf(3)_{10}{\times}\suf(2)_6{\times}\uf(1)$}};

\draw [draw=black] (.05,.1) rectangle (3,.4);
\filldraw [fill=white, draw=black] (.05,.1) rectangle (3,.4);
\node[scale=.8] (title1) at (1.4,.25) {\small{$\suf(2)_8{\times}\suf(2)_8{\times}\uf(1)^2$}};

\end{scope}

\begin{scope}[scale=1.4,yshift=-6cm,xshift=-3cm] %Lagrangian level, U12%
\node[bbc,scale=.5] (U12_top) at (0,0) {};      %%%%%%%%  U12   %%%%%%%%
\draw [draw=black] (-1.45,-1.5) rectangle (1.5,.1);
\node[bbc,scale=.5] (U12_down) at (0,-1.4) {};
\node[scale=.8] (U12_1a) at (-1,-.7) {$[I_1,\varnothing]$};
\node[scale=.8] (U12_1b) at (0,-.7) {$[I_1,\varnothing]$};
\node[scale=.8] (U12_1c) at (1,-.7) {$[I_6,\suf(2)]$};
\draw[red] (U12_top) -- (U12_1a);
\draw[red] (U12_top) -- (U12_1b);
\draw[red] (U12_top) -- (U12_1c);
\draw[red] (U12_down) -- (U12_1a);
\draw[red] (U12_down) -- (U12_1b);
\draw[red] (U12_down) -- (U12_1c);
\node[scale=.8] (p2) at (4.3,-.5) {};
\draw (p1) -- (p2);

%%%% Mass deformation arrows %%%%%%

\draw[IRarrow] (SU3_1b) -- (U12_1c);
\draw[IRarrow] (SU3_1b) -- (U12_1b);

\draw[IRarrow] (SU22_1b) -- (U12_1b);
\draw[IRarrow] (SU22_1c) -- (U12_1c);

%%%% Tables with titles %%%%%%

\draw [draw=black] (-1.45,.1) rectangle (1.5,.4);
\filldraw [fill=yellow!40, draw=black] (-1.45,.1) rectangle (1.5,.4);
\node[scale=.8] (title1) at (0,.25) {\small{$SU(3)+F+S$}};

\end{scope}
\end{tikzpicture}}{\caption{$\suf(6)$ series.}}
\end{subfloatrow}\hspace{1cm}
\begin{subfloatrow}
\hspace{2.5cm}\ffigbox[3.3cm][]{
\begin{tikzpicture}[decoration={markings,
mark=at position .5 with {\arrow{>}}},
Marrow/.style={->,>=stealth[round],shorten >=1pt,line width=.4mm,coolblack},
IRarrow/.style={->,>=stealth[round],shorten >=1pt,line width=.4mm,deepcarmine,dashed}]
\begin{scope}[scale=1.4,yshift=1cm] %%%%%%%  SU5  %%%%%%%
\draw [draw=black] (-1.35,.1) rectangle (1.5,.4);
\filldraw [fill=white, draw=black] (-1.35,.1) rectangle (1.5,.4);
\node[scale=.8] (title1) at (0.1,.25) {\small{$\suf(5)_{16}$}};
\draw [draw=black] (-1.35,-1.5) rectangle (1.5,.1);
\node[bbc,scale=.5] (SU5_top) at (0,0) {};       %%%%%%%%  SU5   %%%%%%%%
\node[bbc,scale=.5] (SU5_down) at (0,-1.4) {};
\node[scale=.8] (SU5_1a) at (-.8,-.7) {$[I_1,\varnothing]$};
\node[scale=.8] (SU5_1b) at (.8,-.7) {$[I_2^*,\sof(12)]_{\Z_3}$};
\draw[red] (SU5_top) -- (SU5_1a);
\draw[red] (SU5_top) -- (SU5_1b);
\draw[red] (SU5_down) -- (SU5_1a);
\draw[red] (SU5_down) -- (SU5_1b);
\end{scope}

\begin{scope}[scale=1.4,yshift=-1cm,xshift=0cm] %%%%% SU(3) %%%%%%%
\draw [draw=black] (-1.35,.1) rectangle (1.5,.4);
\draw [draw=black] (-1.35,-1.5) rectangle (1.5,.1);
\node[bbc,scale=.5] (SU3_top) at (0,0) {};        %%%%%%%%  SU3   %%%%%%%%
\node[bbc,scale=.5] (SU3_down) at (0,-1.4) {};
\node[scale=.8] (SU3_1a) at (-1,-.7) {$[I_1,\varnothing]$};
\node[scale=.8] (SU3_1b) at (0,-.7) {$[I_2,\suf(2)]$};
\node[scale=.8] (SU3_1c) at (1,-.7) {$[\cT^{(1)}_{D_4,1},]_{\Z_3}$};
\draw[red] (SU3_top) -- (SU3_1a);
\draw[red] (SU3_top) -- (SU3_1b);
\draw[red] (SU3_top) -- (SU3_1c);
\draw[red] (SU3_down) -- (SU3_1a);
\draw[red] (SU3_down) -- (SU3_1b);
\draw[red] (SU3_down) -- (SU3_1c);

%%%% Deformation arrows %%%%

\draw[Marrow] (SU5_1b) -- (SU3_1c);
\draw[Marrow] (SU5_1b) -- (SU3_1b);

%%%%% Filling the box %%%%%%

\filldraw [fill=white, draw=black] (-1.35,.1) rectangle (1.5,.4);
\node[scale=.8] (title1) at (0.1,.25) {\small{$\suf(3)_{12}{\times}\uf(1)$}};

\end{scope}

\begin{scope}[scale=1.4,yshift=-3cm,xshift=.07cm] %%%%% SU2U1 %%%%%%%
\draw [draw=black] (-1.3,.1) rectangle (1.35,.4);
\draw [draw=black] (-1.3,-1.5) rectangle (1.35,.1);
\node[bbc,scale=.5] (SU2U1_top) at (0,0) {};        %%%%%%%%  SU2U1   %%%%%%%%
\node[bbc,scale=.5] (SU2U1_down) at (0,-1.4) {};
\node[scale=.8] (SU2U1_1a) at (-.7,-.7) {$[I_1,\varnothing]$};
\node[scale=.8] (SU2U1_1b) at (.7,-.7)  {$[I_\star^*,\sof(6)]_{\Z_3}$};
\draw[red] (SU2U1_top) -- (SU2U1_1a);
\draw[red] (SU2U1_top) -- (SU2U1_1b);
\draw[red] (SU2U1_down) -- (SU2U1_1a);
\draw[red] (SU2U1_down) -- (SU2U1_1b);
\node[scale=.8] (figure_bottom) at (0,-2.5)  {};

%%%% Mass deformation arrows %%%%

\draw[Marrow] (SU3_1c) -- (SU2U1_1b);

%%%% Filling the box with the title in white %%%%%

\filldraw [fill=white, draw=black] (-1.3,.1) rectangle (1.35,.4);
\node[scale=.8] (title1) at (0,.25) {\small{$\suf(2)_{10}{\times}\uf(1)$}};

\end{scope}

\end{tikzpicture}}{\caption{$\suf(5)$ series.}}
\end{subfloatrow}}
{\caption{\label{fig:CBthree} Coulomb branch analysis of the mass deformations among the entries of the ($a$) $\suf(6)$ and ($b$) $\suf(5)$ series.}}
\end{adjustbox}
\end{figure}

Using \eqref{DefSU6}, it is straightforward to show that all CB stratification in the left side but one, satisfy criterion \ref{knot}. The exception is the last flow which ends in the Lagrangian theory $SU(3)+1F+1S$. Using the weakly coupled gauge theory description we claim that the mass deformation which allows to check criterion \ref{knot} is:
\beq
[I_3^*,\suf(2){\times}\sof(6)]_{\Z_2}\to\left\{([I_1,\varnothing],[I_2,\suf(2)])_{\Z_2},[I_6,\suf(2)]\right\}
\eeq
where this corresponds to give a common mass to two fundamental hypers and the one in the adjoint. We put the $([I_1,\varnothing],[I_2,\suf(2)])$ in parenthesis since this represents the CB singularities of what is left after decoupling the hypers, \emph{i.e.} $SU(2)+1F$, and we are agnostic on the action of the $\Z_2$ gauging on this factor. 

Let's conclude this analysis with some basic observation on the right hand side of figure \ref{fig:CBthree} $(a)$. Since all of the $[\star{\rm w/}b= \cdot]$ should be obtainable from mass deformations of IR-free $SU(2)$ theories, this strongly suggest that these have to be also IR-free and thus lagrangian. Since they have to be rank-1 they can only be either $U(1)$ or $SU(2)$ gauge theories, perhaps discretely gauged. This observation constraints effectively the realm of possibilities but we are still unable to fully characterize these strata.

\paragraph{$\suf(5)$ series} Let's conclude this analysis with the $\suf(5)$ series. To understand these mass deformations we need to only assume that the following deformation is indeed allowed:
\beq
[I_n^*,\sof(2n)]_{\mathbb{Z}_3}\to\left\{
[I_{n'}^*,\sof(2n')]_{\mathbb{Z}_3},[I_{n''},\suf(n'')]
\right\}
\eeq
where $n=n'+n''$. Recall that the $[I_n^*,\sof(2n)]_{\mathbb{Z}_3}$ is identified as a $SU(2)+nF$ with a gauged $\Z_3$ automorphism \cite{Martone:2021ixp}. Then the mass deformation above can be interpreted as giving a common mass to $n''$ fundamental hypers and therefore make the prediction that the $\Z_3$ automorphism allows for such a decoupling. The fact that the CB stratification at the bottom of figure \ref{fig:CBthree} ($b$) has a partially unspecified stratum $[I_\star^*,\sof(6)]$ is due to the fact that no IR-free $SU(2)$ gauge theory with global symmetry $\sof(6)$ and which is connected by mass deformation to the ones above, is currently known to the authors. The natural guess would be $SU(2)+3F$ but this is asymptotically free. This is a puzzle whose resolution we leave for future work.

%%%%%%%%%%%%%%%%%%%%%%%%%%%%%%%%%%%%%%%%%%%%%%%%%%%%%%%

\acknowledgments We would like to thank Philip Argyres, Antoine Bourget, Cyril Closset, Julius Grimminger, Shlomo Razamat and Sakura Schafer-Nameki for helpful discussions and comments. MM is supported by NSF grants PHY-1151392 and PHY-1620610 and GZ is supported in part by the ERC-STG grant 637844-HBQFTNCER and by the INFN.

\appendix

\section{Matching flavor symmetry along the Higgs branch}\label{app:Higgs}

Here we collect all the information regarding the way that the mass deformation of the SCFT at the superconformal fixed point are matched along the HB and in particular on the bottom symplectic leaves. The main results are reproduced in table \ref{Higgs:one}, \ref{Higgs:two} and \ref{Higgs:three}. Here we use the notation $\gf$ to denote both a lie algebra $\gf$ and its corresponding minimal nilpotent orbit. Since much of the data discussed here rely on properties of such orbits, for the convenience of the reader we summarized them in table \ref{NilOrbits}.

\begin{table}
\begin{adjustbox}{center,max width=.9\textwidth}
\renewcommand{\arraystretch}{1.3}
$\begin{array}{|c|c|c|c|c|}
\hline
\hline
\multicolumn{5}{|c|}{\text{\bf Minimal nilpotent orbit of Lie algebras}}\\
\hline
\hline
\ff & {\rm dim}_\H &\ff^\natural & \pi_{\bRf}&I_{\ff^\natural\hookrightarrow\ff}\\
\hline
\ \ \af_N\ \  & N & \af_{N-2}\oplus \C &\ \  ({\bf N-2})_+\oplus (\bar{{\bf N-2}})_-\ \ &\ (1,1) \ \\
\bff_N & 2N-2 & \af_1\oplus \bff_{N-2} & ({\bf2},{\bf2N-3})&\ (1,1) \ \\
\cf_N & N & \cf_{N-1} & {\bf 2(N-1)}&\ (1,1) \ \\
\df_N &\ \  2N-3\ \  &\ \ \af_1\oplus \df_{N-2}\ \ & ({\bf2},{\bf 2N-4})&\ (1,1) \ \\
\ef_6 & 11 & \af_5 & {\bf 20}&1\\
\ef_7 & 17 & \df_6 & {\bf 32}&1\\
\ef_8 & 29 & \ef_7 & {\bf 56}&1\\
\gf_2 & 3 & \af_1 & {\bf 4}&3\\
\ff_4 & 8 & \cf_3 & {\bf 14}' &1\\
\hline
\hline
\end{array}$
\caption{\label{NilOrbits}Properties of minimal nilpotent orbits of classical and exceptional Lie algebras. The first column indicates the lie algebra $\ff$, the second the quaternionic dimension of the orbit, the third the stabilizer group of a generic non-singular point of the orbit $\ff^\natural$. Finally the last two columns indicate the represenation of the goldstone bosons under $\ff^\natural$ and the index of embedding of $\ff^\natural\hookrightarrow\ff$. This table is almost verbatim taken from \cite{Beem:2019tfp}.}
\end{adjustbox}
\end{table}

Let's describe in detail what the columns of these very busy tables mean:
\begin{itemize}
    \item[$1^{\rm st}$] Numbering of each theory matching what with the ones used in \cite{Martone:2021ixp}.
    \item[$2^{\rm nd}$] Global flavor symmetry, with levels, of each SCFT.
    \item[$3^{\rm rd}-5^{\rm th}$] Information about the \emph{unknotted} $u$ bottom symplectic stratum, \emph{i.e.} the one accessible by turning on Higgs moduli of the SCFT supported on the $u=0$ CB stratum (here $u$ and $v$ are the globally defined CB coordinates). Specifically we report the type of symplectic leaf $(\bSf_u)$, the unbroken flavor symmetry on its generic point along with its index of embedding and the representation of the goldstone bosons associated with the given pattern of symmetry breaking. See \cite{Argyres:2020wmq,Martone:2021ixp} for more details.
    \item[$6^{\rm th}-7^{\rm th}$] Information about the bottom theory supported on the $\bSf_u$ leaf which is indicated by $\bTf_u$. We also specify the subgroup of the flavor symmetry of $\bTf_u$ which is matched along the $\bSf_u$.
    \item[$8^{\rm th}-10^{\rm th}$] Information about the \emph{unknotted} $v$ bottom symplectic stratum now denoted by $\bSf_v$.
    \item[$11^{\rm th}-12^{\rm th}$] Information about the bottom theory supported on the $\bSf_v$ leaf which is indicated by $\bTf_v$.
\end{itemize}

To further understand how to read the tables, it is helpful to work out some examples:

\paragraph{$D_1^{20}(E_8)$} Let's start from a simple case. For this theory there is a single simple factor, $\ef_8$, and the stratum associated with it is its minimal nilpotent orbit, $\bar{\blue{\Sf}}_{\ef_8}\equiv \ef_8$. From the properties of nilpotent orbits we readily obtain that $\ff^\natural=\ef_7$ which will have $k_{\ef_7}=20$ and the goldstone bosons transform in the ${\bf 56}$ of $\ef_7$ with $T_2({\bf 56})=12$, as in fact is reported in the corresponding entry in table \ref{Higgs:one}. The index of embedding of $\ef_7$ inside $\ef_8$ is one as it can be computed by the decomposition of any $\ef_8$ irreducible representations in $\ef_7\oplus\suf(2)$ ones (for example ${\bf 248}\to({\bf 133,1})\oplus ({\bf 1,3})\oplus ({\bf 56,2})$). The theory supported on this stratum is $\bTf_{\ef_8}\equiv\cT^{(1)}_{E_7,1}$ and thus $k_{\ff_{\rm IR}}=8$. This information is enough for the reader to check that \eqref{kmatch} is indeed satisfied.

\paragraph{$\sof(20)_{16}$} Let's go back to an example previously discussed where the matching of the level is not always as simple as the previous example. Consider now another theory of the $\ef_8-\sof(20)$ series, namely $\sof(20)_{16}$. As discussed above, the HB stratum associated to the simple flavor factor is the minimal nilpotent orbit of $\sof(20):\bar{\blue{\Sf}}_{\sof(20)}\equiv\df_{10}$. Again properties of nilpotent orbits tell us that $\ff^\natural=\sof(16){\times}\suf(2)$, in particular $\ff^\natural$ is semi-simple, and that the goldstone bosons transform in the $({\bf2},{\bf16})$. On the other hand the theory supported on this stratum is $\bTf_{\sof(20)}\equiv \cT^{(1)}_{E_8,1}$ which has flavor symmetry $\ef_8$ and we noticed that $\sof(16)$ is a maximal subalgebra of $\ef_8$, thus $\ff_{\rm IR}=\sof(16)$ at level $k_{\sof(16)}=12$. Using the fact that $I_{\sof(16)\hookrightarrow \ef_8}=1$ and that in our normalization $T_2({\bf 16})=2$, we can immediately reproduce the result we are after: $k_{\sof(20)}=16$. But what about the extra $\suf(2)$? It arises from the breaking of the initial $\sof(20)$, so we expect the level of this $\suf(2)$ to be also 16. Since we have ``used up'' all the moment maps of the rank-1 SCFT on the stratum, our only hope is that the goldstone bosons alone can make up for that. It is a nice surprise to notice that indeed the goldstone transform as 16 copies of the fundamental of this $\suf(2)$ and we are working in a normalization such that $T_2({\bf 2})=1$.

In a similar manner, the data in tables \ref{Higgs:one}, \ref{Higgs:two} and \ref{Higgs:three} can be leveraged to show the consistency of \eqref{kmatch} of all theories discussed in this paper.

\begin{landscape}
\begin{table}[ht]
\begin{adjustbox}{center,max width=.64\textwidth}
$\def\arraystretch{1.0}
\begin{array}{r|c|ccc:cc|ccc:cc|}
\multicolumn{12}{c}{\Large\textsc{Rank-2:\ Higgs\ branch\ data\ I}}\\
\hline
\hline
\#&\ff&
\bSf_u&([\ff^\natural]_{k^\natural},I_{\ff^\natural\hookrightarrow\ff_{\rm UV}})&\pi_{\bRf}&\bTf_u
&([\ff_{\rm IR}]_{k_{\rm IR}},I_{\ff^\natural\hookrightarrow\ff_{\rm IR}}) 
&\blue{\Sf}_v&( [\ff^\natural]_{k^\natural},I_{\ff^\natural\hookrightarrow\ff_{\rm UV}})&\pi_{\bRf}&\bTf_v
&([\ff_{\rm IR}]_{k_{\rm IR}},I_{\ff^\natural\hookrightarrow\ff_{\rm IR}}) 
\\
\hline\hline

\multicolumn{12}{c}{\ef_8-\sof(20)\ \text{series}}\\
\hline

%Beginning of the E8-SO(20) series

1.&[\ef_8]_{24}{\times}\suf(2)_{13}
&\varnothing&\text{-}&\text{-}&\text{-}&\text{-}
&\ef_8&([\ef_7]_{24}{\times}\suf(2)_{13},(1,1))&({\bf56},{\bf1})&\cT^{(1)}_{E_8,1}{\times}\H&([\ef_7]_{12}{\times}\suf(2)_{12}{\times}\suf(2)_1,(1,1,1))
\\
2.&\sof(20)_{16}
&\varnothing&\text{-}&\text{-}&\text{-}&\text{-}
&\df_{10}&(\suf(2)_{16}{\times}\sof(16)_{16},(1,1))&({\bf2},{\bf16})&\cT^{(1)}_{E_8,1}&(\sof(16)_{12},1)
\\
3.&[\ef_8]_{20}
&\varnothing&\text{-}&\text{-}&\text{-}&\text{-}
&\ef_8&([\ef_7]_{20},1)&{\bf56}&\cT^{(1)}_{E_7,1}&([\ef_7]_8,1)
\\
4.&[\ef_7]_{16}{\times}\suf(2)_9
&\varnothing&\text{-}&\text{-}&\text{-}&\text{-}
&\ef_7&(\sof(12)_{16}{\times}\suf(2)_9,(1,1))&({\bf 32},{\bf1})&\cT^{(1)}_{E_7,1}{\times}\H&(\sof(12)_8{\times}\suf(2)_8{\times}\suf(2)_1,(1,1,1))
\\
5.&
 \suf(2)_8{\times}\sof(16)_{12}
&\af_1 & (\sof(16)_{12},1) & - & \cT^{(1)}_{E_8,1}&(\sof(16)_{12},1)&
\df_8& (\suf(2)_8{\times}\suf(2)_{12}{\times}\sof(12)_{12},(1,1,1)) &  ({\bf 2},{\bf12}) & \cT^{(1)}_{E_7,1} & (\suf(2)_8{\times}\sof(12)_8,(1,1)) 
\\
6.&\
\suf(10)_{10}
&\varnothing&\text{-}&\text{-}&\text{-}&\text{-}
&\af_9&(\suf(8)_{10}{\times}\uf(1),1)&{\bf8}\oplus\bar{{\bf8}}&\cT^{(1)}_{E_7,1}&(\suf(8)_8,1)
\\
7.&
[\ef_6]_{12}{\times}\suf(2)_7
&\varnothing&\text{-}&\text{-}&\text{-}&\text{-}
&\ef_6&(\suf(6)_{12}{\times}\suf(2)_7,(1,1))&({\bf20},{\bf1})&\cT^{(1)}_{E_6,1}{\times}\H&(\suf(6)_6{\times}\suf(2)_6{\times}\suf(2)_1,(1,1,1))
\\
8.&
\sof(14)_{10}{\times}\uf(1)
&\varnothing&\text{-}&\text{-}&\text{-}&\text{-}
&\df_7&(\suf(2)_{10}{\times}\sof(10)_{10},(1,1))&({\bf2},{\bf10})&\cT^{(1)}_{E_6,1}&(\sof(10)_6{\times}\uf(1),1)
\\
9.&
\suf(2)_6{\times}\suf(8)_8&
\af_1&(\suf(8)_8,1)&-&\cT^{(1)}_{E_7,1}&(\suf(8)_8,1)
&\af_7&(\suf(6)_8{\times}\suf(2)_6{\times}\uf(1),(1,1))&{\bf6}\oplus\bar{{\bf6}}&\cT^{(1)}_{E_6,1}&(\suf(6)_6{\times}\suf(2)_6,(1,1))
\\
\rcy10.&
\sof(12)_8
&\varnothing&\text{-}&\text{-}&\text{-}&\text{-}
&\df_6&(\suf(2)_8{\times}\sof(8)_8,(1,1))&({\bf2},{\bf8})&\cT^{(1)}_{D_4,1}&(\sof(8)_4,1)
\\
\rcy11.&
\sof(8)_8{\times}\suf(2)_5
&\varnothing&\text{-}&\text{-}&\text{-}&\text{-}
&\df_4&(\suf(2)^3_8{\times}\suf(2)_5,(1,1))&({\bf 2}^3,{\bf1})&\cT^{(1)}_{D_4,1}\times \H&(\suf(2)^3_4{\times}\suf(2)_1,(1,1))
\\
\rcy12.&
\uf(6)_6
&\varnothing&\text{-}&\text{-}&\text{-}&\text{-}
&\af_5&(\suf(4)_6{\times}\uf(1),1)&{\bf4}\oplus\bar{{\bf4}}&\cT^{(1)}_{D_4,1}&(\suf(4)_5{\times}\uf(1),1)
\\
\rcy13.&
\suf(2)^5_4&
\af_1&(\suf(2)^4_5,1)&-&\cT^{(1)}_{D_4,1}&(\suf(2)^4_4,1)
&\varnothing&\text{-}&\text{-}&\text{-}&\text{-}
\\

\cdashline{1-12}
%End of the E8-SO(20) series -- beginning of the Sp(12)-Sp(8)-F4 series

\multicolumn{12}{c}{\spf(12)-\spf(8)-\ff_4\ \text{series}}\\
\hline

22.&
\spf(12)_8&
\cf_6&(\spf(10)_8,1)&{\bf10}&\cS^{(1)}_{E_6,2}&(\spf(10)_7,1)
&\varnothing&\text{-}&\text{-}&\text{-}&\text{-}
\\
23.&
 \spf(4)_7{\times}\spf(8)_8
&\cf_4 & (\spf(4)_7{\times}\spf(6)_8,(1,1)) & {\bf6} & \cS^{(1)}_{E_6,2}&(\spf(4)_7{\times}\spf(6)_7,(1,1))&
\cf_2& (\suf(2)_7{\times}\spf(8)_8,(1,1)) &  {\bf 2} & \hyperref[sec:t9]{{\rm Th. 9}} & (\suf(2)_6{\times}\spf(8)_8,(1,1)) 
\\
24.&
[\ff_4]_{12}{\times}\suf(2)_7^2
&\ff_4&(\spf(6){\times}\suf(2)^2,(1,1))&({\bf14}',{\bf1})&\cS^{(1)}_{E_6,2}&(\spf(6)_7{\times}\suf(2)_7^2,(1,1))
&\varnothing&\text{-}&\text{-}&\text{-}&\text{-}
\\
25.&
\suf(2)_8{\times}\spf(8)_6&
\cf_4&(\spf(6)_6{\times}\suf(2)_8,(1,1))&{\bf6}&\cS^{(1)}_{D_4,2}&(\spf(6)_5{\times}\suf(2)_8,(1,1))&
\af_1&(\spf(8)_6,1)&{\bf1}&\cT^{(1)}_{E_6,1}&(\spf(8)_6,1)
\\
26.&
\suf(2)_5{\times}\spf(6)_6{\times}\uf(1)&
\cf_3&(\spf(4)_6{\times}\suf(2)_5{\times}\uf(1),(1,1))&{\bf4}&\cS^{(1)}_{D_4,2}&(\spf(4)_5{\times}\suf(2)_5{\times}\uf(1),(1,1))&
\af_1&(\spf(6)_6,1)&-&\hyperref[sec:SU3Nf6]{\suf(3)+6F}&(\spf(6)_6,1)
\\
27.&
\sof(7)_8{\times}\suf(2)^2_5
&\bff_3&(\suf(2){\times}\suf(2){\times}\suf(2)^2,(1,2,1))&({\bf 2},{\bf3},{\bf1})&\cS^{(1)}_{D_4,2}&(\suf(2)_5^3{\times}\suf(2)_8,(1,1))
&\varnothing&\text{-}&\text{-}&\text{-}&\text{-}
\\
28.&
[\ff_4]_{10}{\times}\uf(1)
&\varnothing&\text{-}&\text{-}&\text{-}&\text{-}
&\ff_4&(\spf(6)_{10}{\times}\uf(1),1)&{\bf14}'&\cS^{(1)}_{D_4,2}&(\spf(6)_5{\times}\uf(1),1)
\\
29.&
\spf(6)_5{\times}\uf(1)&
\cf_3&(\spf(4)_5{\times}\uf(1),1)&{\bf4}&\cS^{(1)}_{A_2,2}&(\spf(4)_4,1)
&\varnothing&\text{-}&\text{-}&\text{-}&\text{-}
\\
30.&
\suf(3)_6{\times}\suf(2)^2_4
&\af_2&(\uf(1){\times}\suf(2)^2,(-,1))&{\bf1}_+\oplus{\bf1}_-&\cS^{(1)}_{A_2,2}&(\suf(2)_4^2{\times}\uf(1),(1,\text{-}))
&\varnothing&\text{-}&\text{-}&\text{-}&\text{-}
\\
\rcy31.&
\spf(4)_4&
\cf_2&(\suf(2)_4,1)&{\bf2}&\blue{\cS^{(1)}_{\varnothing,2}}&(\suf(2)_3,1)
&\varnothing&\text{-}&\text{-}&\text{-}&\text{-}
\\
\rcb32.&
 \suf(2)_6&
\af_1&(\suf(2)_3,1)&-&\blue{\cS^{(1)}_{\varnothing,2}}&(\suf(2)_3,1)
&\varnothing&\text{-}&\text{-}&\text{-}&\text{-}
\\
\cdashline{1-12}
\multicolumn{12}{c}{\suf(6)\ \text{series}}\\
\hline
33.&\red{\suf(6)_{16}{\times}\suf(2)_9}
&\varnothing&\text{-}&\text{-}&\text{-}&\text{-}
&\af_1&(\suf(6)_{16},1)&\text{-}&\hyperref[sec:t22]{\text{Th. 22}}&(\suf(6)_8,2)
\\
34.&\red{\suf(4)_{12}{\times}\suf(2)_7{\times}\uf(1)}
&\red{?}&\red{?}&\red{?}&\red{?}&\red{?}
&\af_1&(\suf(4)_{12}{\times}\uf(1),1)&\text{-}&\hyperref[sec:t25]{\text{Th. 25}}&(\suf(2)_8{\times}\suf(4)_6,(1,2))
\\
35.&\red{\suf(3)_{10}{\times}\suf(3)_{10}{\times}\uf(1)}
&\varnothing&\text{-}&\text{-}&\text{-}&\text{-}
&\bar{h}_{2,3}&\red{\suf(3)_{10}{\times}\suf(2)_{10}}&\red{{\bf2}\oplus\bar{\bf 2}} &\cS^{(1)}_{D_4,2}&(\suf(3)_5{\times}\suf(2)_8,(2,1))
\\
36.&\red{\suf(3)_{10}{\times}\suf(2)_6{\times}\uf(1)}
&\varnothing&\text{-}&\text{-}&\text{-}&\text{-}
&\af_1&(\suf(3)_{10}{\times}\uf(1),1)&\text{-}&\hyperref[sec:tTE62]{\widetilde{\cT}_{E_6,2}}&(\suf(3)_5{\times}\uf(1),2)
\\
37.&\red{\suf(2)_{8}{\times}\suf(2)_8{\times}\uf(1)^2}
&A_3&(\suf(2)_8{\times}\suf(2)_8,(1,1))&\text{-}&\cT^{(1)}_{D_4,1}&(\suf(2)_4{\times}\suf(2)_4,(2,2))
&\bar{h}_{2,2}&\red{\suf(2)_8{\times}\uf(1)^3}&\text{-}&\cS^{(1)}_{A_2,2}&(\suf(2)_4{\times}\uf(1),2)
\\
\rcy38.&\uf(1){\times}\uf(1)
&\varnothing&\text{-}&\text{-}&\text{-}&\text{-}
&\af_1&(\varnothing,\text{-})&\text{-}&\blue{\cS^{(1)}_{\varnothing,2}}&\uf(1)
\\
%% End of the Sp(12)-Sp(8)-F4 series

\hline\hline
\end{array}$
\caption{\label{Higgs:one}{\footnotesize This is the first of three tables summarizing the HB data for rank-2 theories. The second column lists the flavor symmetry of the SCFT, while the rest lists the information of the higgsing of the flavor symmetry realized on bottom symplectic leaves which is indicated under the column $\bSf_{u/v}$. $\uf(1)$ factors in this table will be mostly omitted.}}
\end{adjustbox}
\end{table}
\end{landscape}

\begin{landscape}
\begin{table}[ht]
\begin{adjustbox}{center,max width=.65\textwidth}
$\def\arraystretch{1.0}
\begin{array}{r|c|ccc:cc|ccc:cc|}
\multicolumn{12}{c}{\Large\textsc{Rank-2:\ Higgs\ branch\ data\ II}}\\
\hline
\hline
\#&\ff&
\bSf_u&([\ff^\natural]_{k^\natural},I_{\ff^\natural\hookrightarrow\ff_{\rm UV}})&\pi_{\bRf}&\bTf_u
&([\ff_{\rm IR}]_{k_{\rm IR}},I_{\ff^\natural\hookrightarrow\ff_{\rm IR}}) 
&\blue{\Sf}_v&( [\ff^\natural]_{k^\natural},I_{\ff^\natural\hookrightarrow\ff_{\rm UV}})&\pi_{\bRf}&\bTf_v
&([\ff_{\rm IR}]_{k_{\rm IR}},I_{\ff^\natural\hookrightarrow\ff_{\rm IR}}) 
\\
\hline\hline

%%%End SU(6) series beginning of the Sp(14)

\multicolumn{12}{c}{\spf(14)\ \text{series}}\\
\hline
39.& \spf(14)_9
&\varnothing&\text{-}&\text{-}&\text{-}&\text{-}
&\cf_7&(\spf(12)_9,1)&{\bf12}&\hyperref[sec:t22]{\rm Th. 22}&(\spf(12)_8,1)
\\
40.&\suf(2)_8{\times}\spf(10)_7&
\af_1&(\spf(10)_7,1)&-&\cS^{(1)}_{E_6,2}&(\spf(10)_7,1)
&\cf_5&(\spf(8)_7{\times}\suf(2)_8,(1,1))&{\bf8}&\hyperref[sec:t25]{{\rm Th. 25}}&(\spf(8)_6{\times}\suf(2)_8,(1,1)
\\
41.&\suf(2)_5{\times}\spf(8)_7&
\af_1&(\spf(8)_7,1)&-&\hyperref[sec:G247]{\gf_2+4F}&(\spf(8)_7,1)
&\cf_4&(\suf(2)_5{\times}\spf(6)_7,(1,1))&{\bf6}&\hyperref[sec:t26]{\rm Th.\ 26}&(\suf(2)_5{\times}\spf(6)_6,(1,1))
\\
42.&\spf(8)_6{\times}\uf(1)
&\varnothing&\text{-}&\text{-}&\text{-}&\text{-}
&\cf_4&(\spf(6)_6{\times}\uf(1),1)&{\bf6}&\hyperref[sec:tTE62]{\widetilde{\cT}_{E_6,2}}&(\spf(6)_5{\times}\uf(1),1)
\\
\rcy43.&\spf(6)_5
&\varnothing&\text{-}&\text{-}&\text{-}&\text{-}
&\cf_3&(\spf(4)_5,1)&{\bf4}&\hyperref[sec:SU22a]{\suf(2)\text{-}\suf(2)}&(\spf(4)_4,1)
\\

\cdashline{1-12}

%%%End Sp(14) series beginning of the SU(5)

\multicolumn{12}{c}{\suf(5)\ \text{series}}\\
\hline

44.&\suf(5)_{16}
&\varnothing&\text{-}&\text{-}&\text{-}&\text{-}
&\bar{h}_{5,3}&\suf(4)_{16}&{\bf 4}\oplus\bar{\bf4}&\cS^{(1)}_{D_4,3}&(\suf(4)_{14},1)
\\
45.&\suf(3)_{12}{\times}\uf(1)
&\red{?}&\red{?}&\text{-}&\cS^{(1)}_{A_2,4}&(\suf(3)_{14},1)
&\bar{h}_{3,3}&(\suf(2)_{12}{\times}\uf(1),1)&{\bf 2}\oplus\bar{\bf2}&\cS^{(1)}_{A_1,3}&(\suf(2)_{10}{\times}\uf(1),1)
\\
46.&\suf(2)_{10}\uf(1)
&\varnothing&\text{-}&\text{-}&\text{-}&\text{-}
&\bar{h}_{5,3}&\suf(4)_{16}&{\bf 4}\oplus\bar{\bf4}&\cS^{(1)}_{D_4,3}&(\suf(4)_{14},1)
\\

\cdashline{1-12}

%%%End SU(5) series beginning of the Sp(12)

\multicolumn{12}{c}{\spf(12)\ \text{series}}\\
\hline

47.&
\spf(12)_{11}
&\varnothing&\text{-}&\text{-}&\text{-}&\text{-}
&\cf_6&(\spf(10)_{11},1)&{\bf10}&\hyperref[sec:S5]{S_5}&(\spf(10)_{10},1)
\\
\rcy48.&
\spf(4)_5{\times}\sof(4)_4
&\varnothing&\text{-}&\text{-}&\text{-}&\text{-}
&\cf_2&(\suf(2)_5{\times}\sof(4)_4,(1,1))&{\bf2}&\hyperref[sec:SU22b]{2F+\suf(2)-\suf(2)+F}&(\suf(2)_4{\times}\sof(4)_4{\times}\uf(1)^2,(1,1))
\\
\rcy49.&
\spf(8)_7
&\varnothing&\text{-}&\text{-}&\text{-}&\text{-}
&\cf_4&(\spf(6)_7,1)&{\bf6}&\hyperref[sec:SU3Nf6]{\suf(3)+6F}&(\spf(6)_6,1)
\\

\cdashline{1-12}

%%%End Sp(12) series beginning of the Sp(8)-SU(2)^2

\multicolumn{12}{c}{\spf(8)-\suf(2)^2\ \text{series}}\\
\hline
50.&
\spf(8)_{13}{\times}\suf(2)_{26}
&\varnothing&\text{-}&\text{-}&\text{-}&\text{-}
&\cf_4&(\spf(6){\times}\suf(2),(1,1))&({\bf 6},{\bf1})&\cT^{(2)}_{E_6,2}&(\spf(6)_{12}{\times}\suf(2)_{12}{\times}\suf(2)_7^2,(1,1,1))
\\
51.&
\spf(4)_9{\times}\suf(2)_{16}{\times}\suf(2)_{18}
&\varnothing&\text{-}&\text{-}&\text{-}&\text{-}
&\cf_2&(\suf(2){\times}\suf(2),(1,1))&({\bf2},{\bf1})&\cT^{(2)}_{D_4,2}&(\suf(2)_{8}{\times}\suf(2)^2_{8}{\times}\suf(2)_5^2,(2,1,1))
\\
52.&
\suf(2)_7{\times}\suf(2)_{14}\times \uf(1)
&\varnothing&\text{-}&\text{-}&\text{-}&\text{-}
&\af_1&(\suf(2),1)&{\bf1}&\cT^{(2)}_{A_2,2}&(\suf(2)_6{\times}\suf(2)_4^2{\times}\uf(1),(1,1,\text{-})
\\
53.&\suf(2)_6{\times}\suf(2)_8
&\af_1&(\suf(2)_8,1)&{\bf1}&\cS^{(1)}_{A_2,4}&\uf(1){\times}\uf(1)
&\varnothing&\text{-}&\text{-}&\text{-}&\text{-}
\\
54.&\suf(2)_5&
\af_1&(\varnothing,\text{-})&-&\green{\cS^{(1)}_{\varnothing,4}}&(\uf(1))
&\varnothing&\text{-}&\text{-}&\text{-}&\text{-}
\\
\rcb55.&\suf(2)_{10}
&\varnothing&\text{-}&\text{-}&\text{-}&\text{-}
&\af_1&(\suf(2)_3,1)&\text{-}&\blue{\cS^{(1)}_{\varnothing,2}}{\times}\uf(1)&(\suf(2)_3,1)
\\

\cdashline{1-12}

%%%End Sp(8)-SU(2)^2 series beginning of the G2

\multicolumn{12}{c}{\gf_2\ \text{series}}\\
\hline

56.&[\gf_2]_8{\times}\suf(2)_{14}
&\gf_2&(\suf(2){\times}\suf(2),(3,1))&({\bf 4},{\bf 1})&\cS^{(1)}_{D_4,3}&(\suf(2)_{14}{\times}\suf(2)_{14}{\times}\uf(1),(1,1))
&\varnothing&\text{-}&\text{-}&\text{-}&\text{-}
\\
57.&\suf(2)_{\frac{16}3}{\times}\suf(2)_{10}
&\af_1&(\suf(2),1)&{\bf1}&\cS^{(1)}_{A_1,3}&(\suf(2)_{10}{\times}\uf(1),1)
&\varnothing&\text{-}&\text{-}&\text{-}&\text{-}
\\
58.&[\gf_2]_{\frac{20}3}&
\gf_2&(\suf(2)_{\frac{20}3},3)&{\bf4}&\cS^{(1)}_{A_1,3}&(\suf(2)_{10},1)
&\varnothing&\text{-}&\text{-}&\text{-}&\text{-}
\\
\rcb59.&\suf(2)_8
&\varnothing&\text{-}&\text{-}&\text{-}&\text{-}
&\af_1&(\suf(2)_3,1)&\text{-}&\blue{\cS^{(1)}_{\varnothing,2}}{\times}\uf(1)&(\suf(2)_3,1)
\\

\cdashline{1-12}

%%%End G2 series beginning of the SU(3)

\multicolumn{12}{c}{\suf(3)\ \text{series}}\\
\hline

60.&\suf(3)_{26}{\times}\uf(1)
&\varnothing&\text{-}&\text{-}&\text{-}&\text{-}
&h_{2,4}&(\suf(2)_{26},1)&{\bf2}\oplus\bar{\bf 2}&\cT^{(2)}_{D_4,3}&(\suf(2)_8{\times}\suf(2)_{14},(3,1))
\\
61.&\uf(1)^2
&\varnothing&\text{-}&\text{-}&\text{-}&\text{-}
&h_{2,3}&\uf(1)&\text{-}&\cT^{(2)}_{A_1,3}&(\suf(2)_{\frac{16}3}{\times}\suf(2)_{10},(1,1))
\\
\rcg62.&\uf(1)
&\varnothing&\text{-}&\text{-}&\text{-}&\text{-}
&A_4&\text{-}&\text{-}&\blue{\cS^{(1)}_{\varnothing,2}}{\times}\uf(1)&(\suf(2)_3,1)
\\

\cdashline{1-12}

%%%End SU(3) series beginning of the SU(2)

\multicolumn{12}{c}{\suf(2)\ \text{series}}\\
\hline

63.&\suf(2)_{16}{\times}\uf(1)
&\varnothing&&\text{-}&\text{-}&\text{-}
&h_{3,4}&\uf(1)^2&\text{-}&\cT^{(2)}_{A_2,4}&(\suf(2)_6{\times}\suf(2)_8,(1,1))
\\
\rcg64.&\uf(1)
&\varnothing&\text{-}&\text{-}&\text{-}&\text{-}
&A_5&\text{-}&\text{-}&\blue{\cS^{(1)}_{\varnothing,2}}{\times}\uf(1)&(\suf(2)_3,1)\\
\hline\hline
\end{array}$
\caption{\label{Higgs:two}{\footnotesize This is the first of three tables summarizing the HB data for rank-2 theories. The second column lists the flavor symmetry of the SCFT, while the rest lists the information of the higgsing of the flavor symmetry realized on bottom symplectic leaves which is indicated under the column $\bSf_{u/v}$. $\uf(1)$ factors in this table will be mostly omitted.}}
\end{adjustbox}
\end{table}
\end{landscape}

\begin{table}[ht]
\begin{adjustbox}{center,max width=.7\textwidth}
$\def\arraystretch{1.0}
\begin{array}{r|c|ccc:cc|ccc:cc|}
\multicolumn{12}{c}{\Large\textsc{Rank-2:\ Higgs\ branch\ data\ III\ (Isolated)}}\\
\hline
\hline
\#&\ff&
\bSf_u&([\ff^\natural]_{k^\natural},I_{\ff^\natural\hookrightarrow\ff_{\rm UV}})&\pi_{\bRf}&\bTf_u
&([\ff_{\rm IR}]_{k_{\rm IR}},I_{\ff^\natural\hookrightarrow\ff_{\rm IR}}) 
&\blue{\Sf}_v&( [\ff^\natural]_{k^\natural},I_{\ff^\natural\hookrightarrow\ff_{\rm UV}})&\pi_{\bRf}&\bTf_v
&([\ff_{\rm IR}]_{k_{\rm IR}},I_{\ff^\natural\hookrightarrow\ff_{\rm IR}}) 
\\
\hline\hline

\cdashline{1-12}
65.&\spf(4)_{14}{\times}\suf(2)_8&
\af_1&(\spf(4)_{14},1)&-&\cS^{(1)}_{D_4,3}&(\spf(4)_{14},1)
&\varnothing&\text{-}&\text{-}&\text{-}&\text{-}
\\
\cdashline{1-12}
66.&\suf(2)_{14}
&\af_1&\text{-}&\text{-}&\cT^{(2)}_{\varnothing,1}&\text{-}
&\varnothing&\text{-}&\text{-}&\text{-}&\text{-}
\\
\cdashline{1-12}
\rcb67.&\suf(2)_{14}
&\varnothing&\text{-}&\text{-}&\text{-}&\text{-}
&\af_1&(\suf(2)_3,1)&\text{-}&\blue{\cS^{(1)}_{\varnothing,2}}{\times}\uf(1)&(\suf(2)_3,1)
\\

\cdashline{1-12}
\multicolumn{12}{c}{\text{Theory with no known string theory realization}}\\
\hline
\rcy68.&\varnothing
&\varnothing&\text{-}&\text{-}&\text{-}&\text{-}
&\varnothing&\text{-}&\text{-}&\text{-}&\text{-}
\\
\hline\hline
\end{array}$
\caption{\label{Higgs:three}{\small This is the first of three tables summarizing the HB data for rank-2 theories. The second column lists the flavor symmetry of the SCFT, while the rest lists the information of the higgsing of the flavor symmetry realized on bottom symplectic leaves which is indicated under the column $\bSf_{u/v}$. $\uf(1)$ factors in this table will be mostly omitted.}}
\end{adjustbox}
\end{table}

\bibliographystyle{JHEP}

\end{document}